\documentclass[11pt,letterpaper]{article}
\pdfoutput=1
\usepackage{amsmath,amssymb,amsfonts}
\usepackage[T1]{fontenc}
\usepackage{palatino}
\def\author{Luca Ciambelli}
\def\title{From Asymptotic Symmetries to  the Corner Proposal} 
\usepackage{graphicx}
\usepackage[colorlinks=true, pdfstartview=FitV, linkcolor=black, citecolor=black, urlcolor=blue]{hyperref}
\usepackage{cancel}
\usepackage{mdframed,framed}
\usepackage[usenames,dvipsnames]{xcolor}
\usepackage{mathpazo}
\usepackage{layout}
\usepackage[nosort]{cite}

\newcommand{\D}{\text{d}}
\newcommand{\beq}{\begin{equation}}
\newcommand{\eeq}{\end{equation}}
\newcommand{\beqn}{\begin{eqnarray}}
\newcommand{\eeqn}{\end{eqnarray}}
\newcommand{\pa}{\partial}

\newcommand{\RR}{\mathbb{R}}
\newcommand{\cL}{{\cal L}}
\newcommand{\fL}{\mathfrak{L}}
\newcommand{\he}{ \ \hat{=} \ }
\newcommand{\Vx}{{V_\xi}}
\newcommand{\bigslant}[2]{{\raisebox{.5em}{$#1$}\left/\raisebox{-.5em}{$#2$}\right.}}
\newcommand{\dS}{\mathfrak{diff}(S)}
\newcommand{\ucs}{\mathfrak{ucs}}
\newcommand{\ecs}{\mathfrak{ecs}}
\newcommand{\acs}{\mathfrak{acs}}
\newcommand{\g}{\mathfrak{g}}
\newcommand{\h}{\mathfrak{h}}
\newcommand{\T}{\tilde{t}}

\usepackage{tikz-cd}
\usepackage{pict2e}

\makeatletter
\DeclareRobustCommand{\loplus}{\mathbin{\mathpalette\dog@lsemi{+}}}
\DeclareRobustCommand{\lotimes}{\mathbin{\mathpalette\dog@lsemi{\times}}}
\DeclareRobustCommand{\roplus}{\mathbin{\mathpalette\dog@rsemi{+}}}
\DeclareRobustCommand{\rotimes}{\mathbin{\mathpalette\dog@rsemi{\times}}}

\newcommand{\dog@rsemi}[2]{\dog@semi{#1}{#2}{-90,90}}
\newcommand{\dog@lsemi}[2]{\dog@semi{#1}{#2}{270,90}}
\newcommand{\dog@semi}[3]{%
  \begingroup
  \sbox\z@{$\m@th#1#2$}%
  \setlength{\unitlength}{\dimexpr\ht\z@+\dp\z@\relax}%
  \makebox[\wd\z@]{\raisebox{-\dp\z@}{%
    \begin{picture}(1,1)
    \linethickness{\variable@rule{#1}}
    \roundcap
    \put(0.5,0.5){\makebox(0,0){\raisebox{\dp\z@}{$\m@th#1#2$}}}
    \put(0.5,0.5){\arc[#3]{0.5}}
    \end{picture}%
  }}%
  \endgroup
}
\newcommand{\variable@rule}[1]{%
  \fontdimen8  
  \ifx#1\displaystyle\textfont3\else
    \ifx#1\textstyle\textfont3\else
      \ifx#1\scriptstyle\scriptfont3\else
        \scriptscriptfont3\relax
  \fi\fi\fi
}

\usepackage{mathtools}
\newcommand{\defeq}{\vcentcolon=}

\usepackage{tocloft}
\begin{document}

{\centering
 \vspace*{1cm}
\textbf{\LARGE{\title{}}}
\vspace{0.5cm}
\begin{center}
\author{}\\
\vspace{0.5cm}
\textit{Perimeter Institute for Theoretical Physics,\\
31 Caroline Street North, Waterloo, Ontario, Canada N2L 2Y5}
\end{center}
\vspace{1cm}
\begin{abstract}
\vspace{0.5cm}
These notes are a transcript of lectures given by the author in the XVIII Modave summer school in mathematical physics. The introduction is devoted to a detailed review of the  literature on asymptotic symmetries, flat holography, and the corner proposal. It covers much more material than needed, for it is meant as a lamppost to help the reader in navigating the vast existing literature. The notes then consist of three main parts. The first is devoted to Noether's theorems and their underlying framework, the covariant phase space formalism, with special focus on gauge theories. The surface-charges algebra is shown to projectively represent the asymptotic symmetry algebra. Issues arising in the gravitational case, such as conservation, finiteness, and integrability, are addressed. In the second part, we introduce the geometric concept of corners, and show the existence of a universal asymptotic symmetry group at corners. A careful treatment of corner embeddings provides a resolution to the issue of integrability, by extending the phase space. In the last part we bridge asymptotic symmetries and corners by formulating the corner proposal. In essence, the latter focuses on the central question of extracting from classical gravity universal results that are expected to hold in the quantum realm. After reviewing the coadjoint orbit method and Atiyah Lie algebroids, we apply these concepts to the corner proposal. Exercises are solved in the notes, to elucidate the arguments exposed.
\end{abstract}}

\vspace{5cm}

\begin{center}
\textit{Please write to} ciambelli.luca@gmail.com \textit{for corrections, typos, and literature suggestions.}
\end{center}

\thispagestyle{empty}

\newpage
\tableofcontents
\thispagestyle{empty}
\newpage
\clearpage
\pagenumbering{arabic} 


\section{Introduction}

The theory of asymptotic symmetries is a thriving area of investigation. It aims at understanding the real notion of  observables in a physical system, especially in the presence of gauge symmetries. This theory made us appreciate how to characterize degrees of freedom of gauge theories, at the classical level. In these notes, we approach this theory from a geometric viewpoint. This approach inevitably has two consequences. The first is that it brings it closer to holography, and thus opens the door to an entire community working on the holographic nature of gravity. The second is that it is now well-suited  to be reformulated in terms of corners, namely codimension-$2$ surfaces in the spacetime, on which Noether charges have support, believed to be the atomic constituents of any gauge theory. Consequently, these notes aim at the task of creating a bridge between these various approaches to symmetries in gauge theories. In the author's opinion, these lectures are therefore an original take that brings a fresh perspective to the community. Given the impressive amount of works in recent years on asymptotic symmetries, holography, and corners, this introduction has the goal of guiding the reader through the literature. It is meant as a contextualization of these notes, and contains much more material than needed. In particular, it touches upon topics that are only marginally discussed in the main body. It can be read on its own or, if the reader wants to directly access the main body of these notes, it can be skipped at a first stage.

Symmetries are one of the most important tool we have in physics, and arguably science altogether. In his lectures at Cornell \cite{Feynman1964}, Feynman attributes to  Weyl (see \cite{Weyl1952a}) the common sense definition: \textit{A thing is symmetrical if there is something that you can do to it, so that after you finish doing it, it looks exactly the same as it did before.}
This definition is not far from what it is meant by symmetry in physics, module the replacement \textit{thing $\leftrightarrow$ physical law / state / field / \dots}. 

In the same lectures \cite{Feynman1964}, Feynman presents the famous riddle proposed by Gardner (known as the Ozma problem \cite{Gardner1990}): ``how can you tell to a Martian, describing a human being, that the heart is on the left side?'' Clearly, given the definition above, if bilateral symmetry is an exact symmetry of the universe, it is impossible. However, at the time of Feynman lectures,\footnote{The experiment proving the non-conservation of parity in weak interactions was conducted by Chien-Shiung Wu et al. in 1956 \cite{Wu1957}.} it was already clear that bilateral symmetry is broken in certain processes, such as $\beta$-decays $N\to p+e^-+\bar\nu$, where the emitted electron always spins to the left. This is the way we can tell to Martians in which side of our body the heart is. With his typical humour, Feynman then observes that, if the Martian has only anti-matter, then the Martian would conclude that the heart is to the right. So if one day we will meet Martians, greet them with the right hand, and they would offer to us their left hand, then we should be careful: it means that they are made of anti-matter and would annihilate us!

The mathematical description of symmetries that we are interested in follows the seminal works of Lagrange \cite{Lagrange1811, Lagrange1815}, Hamilton \cite{Hamilton1835, Hamilton1834}, and Jacobi \cite{Jacobi1866}. The turning point is due to Noether, that formulated in 1918 the celebrated Noether's theorems \cite{Noether1918}. A long-standing debate followed, permeating both the physics and philosophy community, on the notion of physical symmetry and observables, as opposed to gauge (trivial) symmetry.  Notably, Einstein exchanged opinions with Klein about this, in the context of gravity \cite{Einstein1988} (see also \cite{LaurentFreidel2021}). This debate is relevant to philosophers, still nowadays, and has already produced many interesting perspectives and argumentations \cite{Norton1993, Morton1993, Brown1995, Brading2000, Brown2002, Brading2003, Brading2004, Smolin2006, Lange2007, Smith2008, Pooley2010, Greaves2011, KosmannSchwarzbach2011, Teh2016, Pooley2017, Read2018, Olver2018, Rowe2019, Brown2020, KosmannSchwarzbach2020, Roberts2021, DeHaro2021, Gomes2021, Gomes2021a}, see in particular \cite{Read2022}. One of the main modern reformulations of this debate in physics is through the covariant phase space formalism, which was discovered in the seventies \cite{GawEdzki1972, Kijowski1973, Kijowski1976}, was refined in the late eighties \cite{Witten1986, Crnkovic1986, Crnkovic1988,Ashtekar1991, Hayward1993}, and reached its final form with the works of Wald et al. \cite{Lee:1990nz,Wald:1993nt,Iyer:1994ys,Iyer1995,Wald:1999wa}.\footnote{Some relevant subsequent works based on this formalism are \cite{Julia2002, Hollands2005, Harlow:2019yfa, Kirklin2019, Freidel:2021cjp, Chandrasekaran2022, Grant2022, Odak2022}, see also the review \cite{Gieres2021}.}

In this formalism, the aforementioned debate is addressed by constructing the Noether charge associated to the symmetry in question, and stipulating that a symmetry with vanishing associated charge is a trivial gauge symmetry. The construction of charges associated to symmetries comes from Noether's theorems. While the first Noether theorem deals with global symmetries, that is, symmetries associated to transformations applied simultaneously and equally at all points of spacetime, Noether herself remarked that for a gauge symmetry -- defined as a local symmetry acting differently at different spacetime points -- her first theorem gives a current associated to the symmetry that necessarily vanishes on-shell, up to total derivatives. These total derivatives imply that Noether charges for gauge symmetries must be defined as surface integrals of $d-2$ forms, if $d$ is the spacetime dimension, i.e., as surface charges. This is the core of Noether's second theorem. This result should not be surprising, for instance, it is well-known that the total electric charge of a system in electromagnetism is indeed defined as a surface integral, thanks to Gauss' law.

The first discussion of the role of gauge symmetries and their surface charges appeared using the Hamiltonian formalism in the seminal work of Regge and Teitelboim \cite{Regge:1974zd}, see also \cite{Benguria1977}.\footnote{A discussion about the total mass of the system as surface charge appeared previously in the ADM formalism \cite{Arnowitt1960}.} Many years later, Barnich and Brandt \cite{Barnich:2001jy} (see also \cite{Barnich2008}) formulated a general relation between asymptotically conserved $d-2$ forms and gauge symmetries in Lagrangian gauge theories. The term ``asymptotically'' refers here to the fact that these conserved forms are computed near the boundary (not necessarily at infinite distance) on which the surface integral is evaluated, for specific boundary conditions imposed on the fields. The mathematical foundation behind these results resides in the variational bicomplex, constructed independently by Vinogradov \cite{Vinogradov1977, Vinogradov1978,Vinogradov1984, Vinogradov1984a}  and Tulczyjew \cite{Tulczyjew1980}, refined by Anderson \cite{MR1188434}, and applied to conservation laws by Anderson and Torre \cite{Anderson:1996sc} (see also \cite{Khudaverdian2002}). The relationship between this formalism and the covariant phase space formalism of Wald et al. is carefully reviewed e.g. by Comp\`ere \cite{Compere2019b}. Other useful reviews and theses on asymptotic symmetries are \cite{Oblak:2016eij,Ruzziconi2019,Ruzziconi2020,Fiorucci2021}.

As we will carefully discuss in the main body of these notes, the theory of asymptotic symmetries states that, given a choice of dynamics in the bulk of a spacetime with certain falloffs for the dynamical fields near the boundary, only certain gauge transformations (called ``allowed'') preserve the falloffs of the fields. These allowed gauge symmetries can be separated in two categories at the boundary, by evaluating the associated surface charges. If the charge vanishes, then the associated symmetry is still a local one (called trivial), representing even in the presence of the boundary a gauge redundancy of the theory. Contrarily, if the associated surface charge is non-vanishing, then the associated symmetry is a physical transformation of the field space of the theory, mapping a field configuration into an inequivalent one. These are sometimes called non-trivial or large gauge symmetries. Non-vanishing surface charges then satisfy a potentially centrally extended charge algebra, since their algebra is a projective representation of the algebra of asymptotic gauge symmetries \cite{Brown1986,Brown:1986nw}, as we will prove in these notes. This procedure of finding which symmetry is physical and which is trivial is the way physicists address the aforementioned debate. A crucial ingredient is thus the definition of charges. In the absence of fluxes, that is, in a non-dissipative system, charges are defined in a canonical way. For dissipative systems, different notions of charges appeared in the literature, and the guiding principle is focussed on physical assumptions rather than a mathematical prescription. In these notes we will analyse the tremendous progress that has been made in this topic in recent years. 

The theory of asymptotic symmetries has been widely used in many dynamical theories, for a variety of boundary conditions, obtaining more and more general results over the years. An enlightening precursor is $3$-dimensional anti-de Sitter (AdS$_3$) Einstein gravity, where, imposing Dirichlet boundary conditions at the conformal boundary, one recovers the well-known Virasoro algebra with Brown-Henneaux central extension \cite{Brown:1986nw}. In these notes, we will derive the asymptotic symmetry algebra and find the Brown-Henneaux central charge as an exercise. Another notable example is asymptotically flat Einstein gravity. The original motivation to study asymptotic symmetries in this context was to understand the properties of gravitational waves far away from the strong gravity regime, i.e., at asymptotic infinity \cite{BONDI1960,Sachs1961,doi:10.1098/rspa.1962.0161,Sachs:1962wk,Sachs1962a,Bondi1964,Blanchet1986}. As per the AdS$_3$ case mentioned above, one of the difficulties in doing so is that the  boundary is located at conformally infinite distance from the bulk. The treatment of conformal boundaries goes back to Penrose \cite{Penrose:1962ij,Penrose:1964ge} (see also \cite{Newman1962a,Geroch1977}), and it is a standard procedure. Nonetheless, an a priori surprising result is that the group of asymptotic physical symmetries, even on empty Minkowski space, is not Poincar\'e but rather the infinite-dimensional Bondi–van der Burg–Metzner–Sachs ($BMS$) group at null conformal infinity \cite{doi:10.1098/rspa.1962.0161,Sachs:1962wk,Newman1966} (see also the modern review \cite{Alessio2018}), where Abelian translations are enhanced to infinite-dimensional Abelian supertranslations. This opened the door to many interesting developments and results on asymptotic symmetries \cite{Tamburino1966, Newman1968, Ashtekar1978, Ashtekar1981, Geroch1981, Hogan1985, Christodoulou1991
, Brown1993, Ashtekar1997, Adamo2009a, Barnich2013, Compere2016b, Flanagan2017, Maedler2016, Barnich2017a, Dolan2018, Bunster2018, Prabhu2019, Bunster2020, Blanchet2021, Prabhu2022, Compere2022}. In  recent times, more and more relaxed boundary conditions and asymptotic symmetry groups have been found, that generalize the $BMS$ group. The first series of papers in this direction are by Barnich and Troessaert \cite{Barnich:2009se,Barnich:2010eb,Barnich2010,Barnich:2011mi,Barnich2013b} in which, among other things, it is realized that one can consider conformal Killing vectors of the boundary structure that are only locally invertible. Further generalizations followed \cite{Campiglia:2014yka, Hollands2017, Compere:2018ylh, Flanagan:2019vbl, Campiglia:2020qvc, Freidel:2021fxf, Geiller:2022vto},\footnote{A whole tower of dual magnetic asymptotic charges has been recently derived, see \cite{Godazgar2019, Kol2019,  Huang2020, Godazgar2020, Godazgar2020a, Kol2020, Oliveri2020a,  Ciambelli2021, Seraj2021, Freidel2022a, Godazgar2022a, Geiller2022, Long2022}.} building bigger and bigger symmetry groups and leading to an appreciation of infinite-dimensional asymptotic symmetry groups also in $4$-dimensional AdS Einstein gravity \cite{Compere2008, Poole2019, Compere:2019bua, Compere2020}.

This quest of finding enlarged  asymptotic symmetry groups spread in our community, and many papers relaxed boundary conditions, applied the theory of asymptotic symmetries to other boundaries, or studied them in other theories of gravity. For instance, efforts in $3$-dimensional asymptotically flat Einstein gravity include \cite{Barnich2007,Barnich2012,Barnich2013c,Detournay2014, Barnich:2014kra, Barnich:2015uva, Andrade2015,  Detournay2017, Compere2017, Grumiller2017, Fuentealba2018, Adami2020, Geiller2021a, Geiller2021}, while $3$-dimensional AdS Einstein gravity is treated in \cite{Coussaert1995, Compere2013, Troessaert:2013fma, Bagchi2016a, Compere2016a, Donnay2016, Perez2016, Grumiller:2016pqb, Ojeda2019, Alessio:2020ioh, Campoleoni2022}. Asymptotic symmetries are also studied in the Newman-Penrose formalism, \cite{Newman1962}, in \cite{Barnich2012b, Barnich2016, Paoli2017, Barnich2019, Godazgar2019a, Liu2022}, and in the BV-BFV formalism in \cite{Rejzner2021}. In $3$-dimensions, Einstein gravity is a topological theory. On the other hand, one can construct theories of gravity that include Einstein gravity but possess a propagating massive graviton. Known examples are Topologically Massive Gravity (TMG) \cite{Deser1982, Deser1982a} (see as well \cite{Kraus2006, Solodukhin2006, Carlip2008, Carlip2009, Skenderis2009} for TMG with negative cosmological constant and holographic applications), and New Massive Gravity (NMG) \cite{Bergshoeff2009}. Studying asymptotic symmetries in these theories allowed us to discover a rich symmetry structure, as discussed e.g. in \cite{Bouchareb2007, Anninos2009, Blagojevic2009,
 Compere2009b, Henneaux2010, Anninos2010, Henneaux2011, Detournay2012, Donnay2015, Donnay2015a, 
Detournay2016, Adami2019, Ciambelli2020, Aggarwal2020, Adami2021, Aggarwal2022}. In addition to asymptotic symmetries associated to bulk diffeomorphisms, a whole new class of physical symmetries can be derived using the first-order formulation of  gravity, as done e.g. in \cite{Ashtekar2008, Corichi2014, Jacobson2015, Freidel:2015gpa, Corichi2016, Barnich2016a, Frodden2018, Paoli2018, Bonder2018, Cattaneo2018, Oliveri2020, Canepa2021, Godazgar2022, Canepa2022}. 
As mentioned, the same procedure can also be applied to other boundaries (and various boundary conditions, \cite{Odak2021}). Indeed, enlarged asymptotic symmetry groups have been found at spatial infinity in \cite{Ashtekar1978, Campiglia2017b, Troessaert2018, Henneaux2018, Henneaux2018a, Henneaux2018b, Bakhoda2019, Henneaux2019, Fuentealba2022}. 
Another important example is a generic null hypersurface \cite{Torre1986, Goldberg1992, Mars1993, Booth2001, Gourgoulhon2006,  Alexandrov2015, Hopfmuller:2016scf, Wieland2017, Hopfmuller:2018fni, Ciambelli:2019lap, Speranza:2019hkr, Adami2022, SheikhJabbari2022} and null boundary \cite{Ashtekar2014, Chandrasekaran:2018aop, Chandrasekaran:2020wwn, Chandrasekaran:2021hxc, Herfray2021} (on a similar direction, see also \cite{Ananth2021, Ananth2021a}), with particular focus on isolated horizons such as the black hole horizon \cite{Ashtekar2000a, Koga2001, Hotta2001, Ashtekar2004, Bagchi2013b, Donnay:2015abr, Afshar2016a, Penna2016, Blau2016, Eling2016, Donnay:2016ejv, Penna2017, Carlip2018, Carlip:2019dbu, Donnay:2019jiz, Chen2020, Adami2020a, Geiller2021b, Danielson2022, RedondoYuste2022}. It is in this context that Hawking, Perry and Strominger proposed to use asymptotic symmetries to tackle the information paradox \cite{Hawking2015, Hawking2016, Hawking2017, Strominger2020, Haco2018}. The latter is based on Hawking's old observation \cite{Hawking1976}
that there would be loss of predictability of the final state if the black hole evaporated completely, because one could not measure the quantum state of what fell into the black hole. How does the information of the quantum state of infalling particles re-emerge in the outgoing radiation? Hawking, Perry and Strominger proposed that the information is stored in a specific $BMS$ transformation (a supertranslation, called soft hair) associated with the shift of the horizon caused by ingoing particles. This has opened the avenue to the study of black hole soft hairs \cite{Averin2016, Afshar2016, Mirbabayi2016, Afshar2017, Bousso2017, Bousso2017a, Afshar2017a, Gomez2018, Grumiller2018, Donnay2018, Grumiller:2019fmp, Rahman2020}. 
The understanding of the information paradox is an extremely active subject of research nowadays, recently addressed using quantum extremal islands  \cite{Faulkner2013, Engelhardt2015, 
Penington2020, Almheiri2019, 
Almheiri2020, Geng2020, Chen2020a}, see as well \cite{Laddha2021, Raju2022} for a different perspective.

With all these new insights, enhancement of symmetries, and relaxation of boundary conditions, some problems arise in the theory of asymptotic symmetries. There are three essential features of the Noether charges that are not a priori guaranteed. First, the charges one obtains may not be conserved. The reason for this is the presence of fluxes at the boundary. Our perspective is that this is a feature, rather than an issue, but it should be stressed that this feature questions the interpretation of these quantities as charges used to label states, in the first place. Indeed, one should rather associate to non-conserved charges the meaning of observables, which can evolve in time, and become conserved only in the vacuum of the theory. 

Secondly, and especially when dealing with boundaries at conformal infinite distance, charges may diverge. Counterterms can be added to the bulk action (or directly to the pre-symplectic potential) to cure these divergences. In AdS, counterterms are understood in the framework of AdS/CFT holographic renormalization \cite{Balasubramanian1999, Emparan1999, Skenderis2001, Haro2001, Skenderis2002, Bianchi2002, Papadimitriou2005, Hollands2005a, Mansi2009a, Mansi2009, Papadimitriou2010, Ciambelli:2019bzz, Anastasiou2020, Fiorucci:2020xto}. A similar construction applies to the flat case, and the search of a universal renormalization scheme is a very active avenue of investigation \cite{Mann2006, Parattu2016, Lehner2016, Hartong2016, Jubb2017, Chandrasekaran2022}. In this quest, the relationship between holographic anomalies  \cite{Henningson1998,Imbimbo2000, Schwimmer2000, Rooman2001a, Rooman2001, Schwimmer2008} (see also \cite{Ciambelli:2019bzz,Jia2021}) and asymptotic symmetries has been discovered in \cite{Alessio:2020ioh, Fiorucci:2020xto, Campoleoni2022}. In the covariant phase space formalism \`a la Wald, the definition of charges comes from the Lagrangian. The former are codimension-$2$ forms while the latter is a top form. It is therefore not surprising that the notion of charges comes with built-in ambiguities, \cite{Jacobson1994, Iyer:1994ys, Julia2002, Harlow:2019yfa, MargalefBentabol2021, J.FernandoBarbero2021, Chandrasekaran2022}. It is the choice of these ambiguities that allows us to properly renormalize divergent charges.\footnote{There is no mathematical theorem proving this statement, but one can show that it holds case by case.} Rather than a top-down mathematical prescription, ambiguities can be resolved using physical inputs on the system, and thus a variety of ambiguity-resolution procedures have been developed in the literature. On the other hand, in Anderson's variational bicomplex, charges are derived from the equations of motion, with a unambiguous procedure \cite{MR1188434, Anderson:1996sc, Barnich:2001jy}, see also \cite{Freidel:2020xyx, Freidel:2020svx, Freidel:2020ayo}. One can rephrase this result saying that Anderson's procedure selects the ambiguities. The discussion on ambiguities is still an on-going interesting debate. To reiterate, the divergency of charges is the second issue, that we are learning  how to cure better and better. 

The last issue is that charges may not be integrable, meaning that one cannot integrate the  phase-space variation on a path to obtain the charges. If this is the case, then the charge algebra is not represented by the Poisson bracket of charges, and the system is not closed. This issue is more challenging, and still a source of active discussions. To begin with, charges may still be integrable if one chooses the correct field-space variables. In lower-dimensional Einstein gravity, it has been shown that it is always possible to render charges integrable, which is compatible with the fact that the bulk theory is topological, \cite{Adami:2021nnf, Ruzziconi2021}. Applying this logic to $4$ dimensions and higher, one can then isolate the physical non-integrable part.
The next important question is: is this an issue to be cured, or a feature to be understood? For some authors, this is just a feature that indicates that the system is open, and there is leakage. In this direction, Barnich and Troessaert proposed a new notion of brackets \cite{Barnich:2011mi} among the integrable parts of the charges, see also \cite{Troessaert2016, Chandrasekaran:2020wwn,  Wieland2022}. This bracket then retains most of the properties of the Poisson bracket, and allows us to understand the non-integrable part as being related to fluxes. This is a prescription to understand non-integrable systems, but we will adopt a different perspective here, which is to look for an enlarged phase space on which charges are all integrable, so a bigger symplectic structure on which charges are again canonical. This can be achieved introducing a particular set of extra dynamical fields \cite{Donnelly:2016auv, Ciambelli:2021nmv, Freidel:2021dxw, Speranza:2022lxr}, and thus working on a specific extended phase space. These extra phase-space variables come in gravitational theories from the embedding of corners, and are known as edge modes \cite{Donnelly:2016auv, Gomes:2016mwl, Speranza:2017gxd, Geiller:2017xad, Geiller:2017whh, Freidel2019, Donnelly2022}, which are the core of a novel, thriving, area of investigation \cite{Donnelly2015, Donnelly2016, Gomes2018, Gomes2019,  Takayanagi2020, Geiller2020c, Freidel:2020xyx, Freidel:2020svx, Freidel:2020ayo, Donnelly:2020xgu, Francois2021, Riello2021, Wolf2021, Carrozza2022a, Carrozza2022, Kabel2022}. One of the merits of this method to treat integrability is that the charge algebra is then realized by the standard Poisson bracket. As stated, we will mostly focus on this method, as edge modes are  important elements of the corner proposal, and Poisson brackets will be crucial for the moment map. Eventually, independently of the perspective on the topic, the two approaches must be two different ways of understanding the same underlying physics. Nevertheless, a complete link between these two approaches has yet to be done. 

Asymptotic symmetries are the underlying symmetry structure of holography. In the author's view, a way to understand that gravity is holographic is readily written in Noether's theorems: charges associated to gauge symmetries are codimension-$2$ quantities, that can be reinterpreted as global symmetries of a theory living on the codimension-$1$ boundary (references are given below in the context of AdS/CFT). This is arguably the strongest evidence that gravity (and all gauge theories, actually) is holographic. 

Despite so, the word ``holography'' deserves more explanation. Indeed, it is here intended in a loose sense as the possibility of recasting gravity on a boundary theory. This however does not imply that from one boundary and finitely many degrees of freedom one can exactly reconstruct the full bulk gravitational phase space. This happens in AdS/CFT, when imposing Dirichlet boundary conditions on the conformal boundary, but one should not expect this to hold for all boundaries and all boundary conditions. Instead, one can (and should) expect that it is possible to recast asymptotic symmetries and the phase space of gravity near a boundary in terms of a theory fully intrinsic to the boundary.
As mentioned above, the AdS$_3$ asymptotic-symmetry result fits nicely within the AdS/CFT correspondence \cite{Maldacena:1997re, Witten1998, Gubser1998, Klebanov1999, Aharony2000, Bousso2002}, see also the early works \cite{Hooft1993, Susskind1995}. This correspondence states that (quantum) gravity on a $d$-dimensional asymptotically AdS spacetime is dual to a CFT in one dimension less, living on the boundary of AdS.\footnote{From a geometric viewpoint, it is crucial that there exists a gauge for the bulk metric such that the latter is foliated by timelike hypersurfaces \cite{Fefferman1985,Fefferman2011}, and Einstein equations are a Hamiltonian problem in the holographic coordinate.} The asymptotic symmetries of AdS then become the global symmetries of the CFT on the boundary. A spin-off of the AdS/CFT correspondence is the fluid/gravity duality\cite{Policastro2001, Policastro2002, Policastro2002a, Kovtun2003, Kovtun2005, Bhattacharyya2008, Loganayagam2008, Baier2008, Haack2008, Bhattacharyya2008a, Bhattacharyya2008b, Rangamani2009, Banerjee2011, Hubeny2012, Skenderis2017, Ciambelli2017}, in which the hydrodynamic limit is taken in the boundary field theory. In spite of the drawback that we lose microscopic information on the boundary theory, this limit is useful because it allows us to gain a more geometric control on the theory and its conservation laws.

One can naturally wonder if something similar to the AdS/CFT correspondence exists for an asymptotically flat spacetime. There are various ways to address this question, which are tackling the same problem from different angles. The first one is from the symmetries point of view. In the asymptotically flat case, the bulk asymptotic symmetry group is $BMS$, so the dual field theory should be a $BMS$-field theory. A crucial result is that the $BMS_d$ group is isomorphic to the level-$2$ conformal Carroll group\footnote{The Carroll group was discovered by Levy-Leblond and Sen Gupta in \cite{LevyLeblond1965, Gupta1966} as a particular $c\to 0$ contraction of the Poincar\'e group, see also \cite{Henneaux1979}.} $CCarr_{d-1}$, when the spatial non-degenerate metric is a $(d-2)$-sphere \cite{Duval1991, Duval2014, Duval2014a, Duval2014d}. This opened the door to the study of $BMS$- or Carroll-field theories, as putative holographic duals to gravity in asymptotically flat spacetimes. Early works in this direction are \cite{Arcioni:2003xx, Arcioni:2003td, Dappiaggi2004, Dappiaggi:2005ci}, see also \cite{deBoer:2003vf} for a different viewpoint. More recently, field-theoretical efforts include \cite{Bagchi2012, Fareghbal2014, Fareghbal2015, Bagchi2015, Fareghbal2017, Bagchi2019, Bagchi2020, Gupta2021, Bagchi2021, Henneaux2021, Donnay:2022aba, Bagchi2022, Bagchi2022b, RiveraBetancour2022, Baiguera2022, Donnay2022a}, while a more group-theoretical take on the topic is \cite{Hartong:2015xda}. The second approach to asymptotically flat holography is based on the flat limit of hydrodynamics and the fluid/gravity correspondence \cite{Ciambelli:2018wre, Ciambelli:2018ojf, Campoleoni:2018ltl, Ciambelli:2020eba, Ciambelli2020a, Freidel2022d}, see also \cite{Harmark2017,Ball2019, Hijano2019, Hijano2020} for non-hydrodynamic approaches to the flat and other interesting limits of AdS/CFT. This requires the understanding of Carrollian fluids \cite{Boer2018, Ciambelli2018, Boer2020, Boer2022, Petkou2022, Freidel2022}. As for the AdS case, the advantage of this approach is the higher control one obtains on the geometric aspects. In a loose sense, rather than Carrollian hydrodynamics one should refer to this holographic setup as Carrollian geometro-hydrodynamics, because the degrees of freedom composing the dual fluids are dictated by the boundary Carrollian geometry. 

The third and last approach to asymptotically flat holography is what is known as celestial holography (see the reviews \cite{Strominger2017, Raclariu2021, Pasterski2021}, and \cite{Pasterski:2021raf}). This approach is also field-theoretical, but exploits the fact that the dual theory can be seen as a CFT living on the codimension$-2$ space at infinity, colloquially known as the celestial sphere (and thus the CFT called celestial CFT, CCFT). This framework was built off the observation of 
Strominger et al. that gravity and gauge theories in asymptotically flat spacetimes are governed in the infrared by a triangular equivalence: soft theorems can be recast as conservation laws associated to non-trivial bulk gauge symmetries (interpreted as boundary Ward identities), while memory effects are observable signatures of these asymptotic symmetries. Tremendous advancements have been gathered in recent years, in each corner of the triangle. Asymptotic symmetries in this context have been studied e.g. in \cite{Strominger2014, Strominger:2013jfa, He2014, Campiglia2015a, Kapec2015, Campiglia2015b, Oblak2015, Avery2016, Avery2016a, Strominger2017a, Kapec2017, Laddha2018a, Pasterski2019, He2019, He2019a, Donnay:2020guq, Donnay:2021wrk, Nguyen2021, Freidel2022c}. They act on the celestial sphere generating conserved charges at this specific corner, and thus they are readily interpreted as corner symmetries, see below. Soft theorems, introduced in \cite{Low1958, Weinberg1965}, have been linked to asymptotic symmetries and memory effects in \cite{He2015, Cachazo2014, Lysov2014, Campiglia:2014yka, Kapec:2015vwa, Strominger2016a, Campiglia2017a, Campiglia2016, Campiglia2017,  Campoleoni:2017mbt, Nande2018, Laddha2017, Pate:2017fgt, Francia:2018jtb, Laddha2018, Donnay2019, Aggarwal:2018ilg, Himwich2019a, Himwich2019, Banerjee2020b,  Pate2019, Puhm2020, Banerjee2020c, Guevara2019, Campoleoni:2019ptc, Campoleoni:2020ejn,    Sahoo2022, Melton2022}. Lastly, memory effects, originally proposed in \cite{Zeldovich1974, Braginsky1985, Braginsky1987, Bieri2013, Susskind2015, Nichols2017}, are analysed in this context in \cite{Strominger2016, Pasterski2016, Pasterski2017, Pate2017, Satishchandran2019, Seraj2021a, Prabhu2022a, Seraj2022, Godazgar2022b}. Independently from repercussions on the infrared triangle, there have been a lot of intrinsic developments in CCFT and the understanding of the $S$-matrix, see \cite{Kapec2014, He2016, Kapec:2016jld, Cheung:2016iub, Pasterski2017a, He2017, Lam2018, Kapec2018, Fotopoulos2019, Pate:2019lpp, Fotopoulos:2019vac, Banerjee2020a, Fan2020, Ebert2021, Guevara2021a, Guevara2021, Atanasov2021, Crawley2021, Pasterski2021b, Pasterski2021c, Kapec2022, Kapec2022a, Adamo2022, Freidel2022e}. Furthermore, the structure of IR divergences in celestial holography has been analysed in \cite{Gomez2016, Kapec2017a, Choi2018, Magnea2021, Gonzalez2021, Nguyen2021a, Kalyanapuram2021a, Nastase2022}.

Fascinatingly, celestial holography is slowly filling an important gap between asymptotic symmetries and scattering amplitudes in asymptotically flat spacetimes, thanks to the introduction of a new observable, the celestial amplitude. The scattering amplitude community is thriving, and this link made by the celestial holography proposal is opening the door to a deep exchange of knowledge, with fruitful results for both sides, see the non-exhaustive list of works in this direction \cite{Campiglia:2015yka, Pasterski:2016qvg, Pasterski:2017kqt, Schreiber2018, Stieberger2019, Adamo2019, Fan2019, Nandan2019, Law2020a, Law2020, Himwich2020, ArkaniHamed2021, Liu2021, Fan2021, GarciaSepulveda2022}. Loop corrections to the celestial amplitude have been studied in \cite{Banerjee2018, Gonzalez2020, Albayrak2020, Pasterski2022, Donnay2022}, while double-copy techniques have been applied to this problem in  \cite{Casali2021, Casali2020, Kalyanapuram2021, Pasterski2021d}.
The infrared triangle is just the visible tip of a much deeper iceberg. A whole new algebra of symmetries recently emerged in CCFT, the w$_{1+\infty}$ algebra, \cite{Strominger2021, Himwich2021, Adamo2022a, Jiang2022, Mago2021, Ball2022, Hu2022},
 related to asymptotic symmetries in \cite{Freidel2022b}. This new discovery indicates that the boundary theory retains profound knowledge of the bulk, perhaps deeper than expected, and thus there is yet a lot to be unravelled and understood.

The celestial sphere is just a specific instance of a codimension$-2$ surface, on which Noether charges associated to gauge symmetries have support. Recently, there has been an appreciation of these surfaces from a purely geometric point of view, which led to the formulation of the corner proposal \cite{Donnelly:2016auv, Speranza:2017gxd, Geiller:2017whh, Freidel:2020xyx, Freidel:2020svx, Freidel:2020ayo, Donnelly:2020xgu, Ciambelli:2021vnn, Freidel:2021cjp, Ciambelli:2021nmv, Ciambelli2022a, Donnelly2022}, see also \cite{Canepa2022}. In this setup, the term ``corner'' indicates a generic codimension-$2$ surface on which surface charges are evaluated. The non-trivial gauge symmetry algebra is then an essential tool to reformulate geometric observables in an algebraic language, amenable to quantization. This proposal is based on the concept of local holography. In this context, the idea is that the geometry is included into the charge algebra, the quantum kinematics into its representations, and the quantum dynamics into fusion properties. It is thus a shift of paradigm that focuses on universal quantities that are expected to survive in a quantum theory of gravity. Restated in other words, classical dynamics, a spacetime, and gauge symmetries, are not expected to survive in the quantum realm of gravity. Nonetheless, corners and their charge algebra could survive, and be an organising principle for quantum observables. This is the main underlying idea of the corner proposal: a refocusing on corners and charges, rather than spacetime and classical dynamics. 

We will spend considerable time in these notes explaining this proposal. In practice, many questions must be addressed in order to make the latter concrete. A main question is whether gravity provides a maximal group of symmetries at generic corners, so that we can focus on its representations only.  We recently found a maximal non-trivial asymptotic symmetry algebra at isolated corners \cite{Ciambelli:2021vnn, Freidel:2021cjp, Ciambelli2022a}, thus providing a universal corner symmetry group ($UCS$). Various sub-groups give rise to known asymptotic symmetry groups found in the literature, both for asymptotic corners (such as the celestial sphere), and finite distance corners (such as codimension$-2$ surfaces on the black hole horizon). In these notes we will solve as an exercise the reformulation of the near horizon symmetries \cite{Donnay2015a} in terms of corner symmetries, in a gauge-free formalism. Another important question we recently addressed is whether there exists a well-defined Poisson bracket representing the charge algebra at corners in full generality. The answer to this question is positive, thanks to the introduction of edge modes, and it is ultimately related to the issue of integrability we previously discussed \cite{Ciambelli:2021nmv, Freidel:2021dxw}, see also \cite{Speranza:2022lxr, Carrozza2022a, Carrozza2022, Kabel2022}. The existence of a universal symmetry group, canonically realized on the extended phase space, is a promising starting point for the corner proposal.

The next natural step is then to study representations of the $UCS$. We will discuss this in the last part of these notes. The $UCS$ is a complicated group, and little is known about its representations. One tool is the so-called coadjoint orbit method, proposed by Kirillov \cite{Kirillov_1962, Kirillov1976ElementsOT, Kirillov_Merits, Kirillov1990, kirillov2004lectures}, and subsequently analysed in \cite{Kostant_2006, ginz, Duistermaat:1982aa, Alekseev1988, ALEKSEEV1989719, Alekseev1990, wildberger_1990, DELIUS1990, Aratyn1990, Brylinski_1994, Alekseev2022}. This method instructs us to study symplectic leaves in the dual algebra, which has a natural Poisson structure. On the symplectic leaves there is then a symplectic form, called the Kirillov-Konstant-Souriau form, \cite{souriau1970structure, Kostant_2006, Kirillov1976ElementsOT, Kostant2009}, and thus there is a natural Poisson bracket. This is also the starting point of the theory of geometric quantization (proposed in \cite{souriau1970structure, Kostant_2006, Guillemin1977, Kostant1987}, see also \cite{Gotay1980, Sniatycki1980, Guillemin1982, Vaisman1983, Ashtekar1986, Blau1988, Duval1990, Kirillov1990, Axelrod1991, FigueroaOFarrill1991, Tuynman1992, Leon1997, Woodhouse1997, Bates1997a, EcheverriaEnriquez1998, Sardanashvily2001, Gukov2009, Cattaneo2022}), and applying the latter to corner symmetries is part of our  agenda. The coadjoint orbit method is a powerful tool that found physical applications in \cite{Witten:1987ty, Balog:1997zz, Barnich:2014zoa, Barnich:2014kra, Barnich:2015uva, Oblak2015a, Cotler2019, Donnelly:2020xgu, Barnich:2021dta, Lahlali2021, Ciambelli2022a, Bergshoeff:2022eog, Riello:2022din}, see also \cite{Rai1990, Barnich2017b, Merbis2020, Henneaux2020, Barnich2022} for examples of geometric action constructions. We will show in these notes how we can use this method to make a step further in the corner proposal. Although it is still an active research avenue, progress has been made in this direction in \cite{Ciambelli2022a}, by thinking of the $UCS$ in terms of the automorphisms of an Atiyah Lie algebroid. Atiyah Lie algebroids, introduced in \cite{Atiyah:1957, Pradines1966, Pradines1967, Atiyah1978} (see also \cite{mackenzie_1987, mackenzie_2005}), are geometric structures that can be used to mathematically formulate gauge theories \cite{Lazzarini_2012, Fournel:2012uv, Francois2014, Carow-Watamura:2016lob, Kotov:2016lpx, Attard2020, Ciambelli:2021ujl}, and a recasting of the corner proposal in this framework is expected to have far-reaching consequences. In particular, representations are then given by bundles associated to the algebroid. 

The notes are organised as follows. Sec. \ref{sec2} is devoted to symmetries. We start in Subsec. \ref{sec2.1} with an introduction to the covariant phase space formalism, the backbone of Noether's theorems, that we then state in Subsec. \ref{sec2.2}. There, the Poisson bracket among charges is shown to projectively represents the vector fields Lie bracket. We then propose an exercise in Subsec. \ref{sec2.3}, on how to compute Noether charges using the covariant phase space formalism in electromagnetism. In particular, we discuss the total electric charge of the system in these terms. Subsequently, we state in Subsec. \ref{sec2.4} the procedure to go from a gauge theory to its Noether charges, the core of the asymptotic symmetries theory. When applied to gravity, this procedure produces charges that could be non-conserved, divergent, or non-integrable. We discuss this, and possible resolutions, in Subsec. \ref{sec2.5}. We conclude this Section applying these results to AdS$_3$ with Dirichlet boundary conditions, recovering the well-known Virasoro algebra with Brown-Henneaux central extension \cite{Brown:1986nw}, in Subsec. \ref{sec2.6}. 

In Sec. \ref{sec3} we turn our attention to the geometric structure underlying Noether charges, the corners. We show in Subsec. \ref{sec3.1} how embeddings play an important role. Afterwards, we demonstrate in Subsec. \ref{sec3.2} the existence of a maximal finitely-generated sub-algebra of the bulk diffeomorphisms at corners, the so-called Universal Corner Symmetry ($UCS$). Applying this to finite distance corners, we discuss in Subsec. \ref{sec3.3} how  Noether charges give a faithful representation of the Extended Corner Symmetry group ($ECS$), and how the latter is  nested in the $UCS$. At generic finite-distance corners there is leakage and the covariant phase space produces non-integrable charges. However, one can introduce edge modes to render all charges integrable, by working in an extended phase space, as we prove in Subsec. \ref{sec3.4}. As mentioned previously, we then propose as an exercise in Subsec. \ref{sec3.5} the reformulation of the near horizon symmetries  \cite{Donnay2015a} in terms of corner symmetries. This raises interesting questions concerning the gauge-fixing approach. 

Eventually, we state in Sec. \ref{sec4} the corner proposal. The main idea, explained in Subsec. \ref{sec4.1}, is to refocus our attention on charges and corners, as the fundamental ingredients of gauge theories. The fact that there exists a universal symmetry group is very useful, but in practice only certain sub-algebras are dynamically realised, as we argue in Subsec. \ref{sec4.2}. Nonetheless, it is the $UCS$ that one needs to study, and finding representations is a challenging task. We then devote the rest of this Section to an explanation of various ideas and instruments that we have to tackle the study of representations. We present the coadjoint orbit method in Subsec. \ref{sec4.3}, discussing how it is one of the best tools at our disposal to achieve our goal. The $UCS$ contains diffeomorphisms of the corner, which are complicated to treat. In $2$ bulk dimensions, or at a single point on the codimension$-2$ corner, we apply the coadjoint orbit method in Subsec. \ref{sec4.4} and find the Kirillov-Kostant-Souriau two-form on orbits. Reintroducing corner diffeomorphisms, in Subsec. \ref{sec4.5} we recognize in the Atiyah Lie algebroid the correct geometric framework to study orbits and representations. From the algebroid perspective, this is achieved by  introducing associated bundles. We show how an associated rank$-2$ affine bundle gives rise to two normal coordinates, and thus how the associated representation can be thought of as probing the  classical spacetime in the proximity of the corner. We conclude the section with Subsec. \ref{sec4.6}, where we introduce the concept of moment maps, and provide explicit examples of the latter as an exercise. In Sec. \ref{sec5}, we offer a summary of the material treated and a digression on open questions and future directions.

\section{Symmetries}\label{sec2}

There is a lot of recent activity focussed on understanding new symmetries in physics and  revisiting old results. Non-invertible, generalized, higher-form symmetries are some examples of recent avenues of investigation in the topic. To enumerate all recent results is obviously far from the scope of these lectures; we will rather focus on global and local symmetries firstly, and later on local symmetries only, as they are the main actor in the theory of asymptotic symmetries and the corner proposal. 

The naive textbook definition of global and local symmetries is that the latter are generated by spacetime-dependent parameters while the former are generated by constant-in-spacetime ones. Then, global symmetries are physical, acting non-trivially on the system, whereas local symmetries are redundancies, i.e. gauge symmetries, simply expressing our ignorance in defining the physical variables of said system. As we will discuss in detail, this textbook definition is too naive, because there are local symmetries that become physical in the presence of boundaries. The method one should use to distinguish trivial symmetries from physical ones is not whether they are generated by constant parameters or not, but rather whether they have vanishing associated charges or not. The tools to address this question, the celebrated Noether's theorems, are ubiquitous in theoretical and mathematical physics. 

There are two main frameworks that we can use to study Noether's theorems, each coming with some advantages and drawbacks
\begin{itemize}
\item Hamiltonian approach \cite{Hamilton1835, Hamilton1834, Regge:1974zd, Benguria1977}: in this framework, time and space are split, and one studies the evolution of trajectories in phase space. The pros are: it is a robust, unambiguous formulation, and it is the closest approach of classical physics to quantum physics. The cons are mainly two. First it can become quickly a technical and heavy machinery. Second, and more importantly for the discussion here, it explicitly breaks spacetime covariance. 
\item Lagrangian approach \cite{Lagrange1811, Lagrange1815, Noether1918}: instead of an evolutionary problem, the phase space here is thought of as the set of solutions to the equations of motion of the theory. The latter can be shown to be in one to one correspondence with the space of evolution of trajectories. The pros are that it is easier to manipulate, and retains spacetime covariance. The cons are that it is more ambiguous, and sometimes it is harder to gather physical intuition with it.
\end{itemize}
It is in the Lagrangian approach that we will express Noether's theorems. In this approach, the so-called covariant phase space formalism applies, which is the framework we will use for the rest of these lectures. It stems from Noether's works, but it reached its final form thanks to the works of Wald et al. \cite{Lee:1990nz,Wald:1993nt,Iyer:1994ys,Iyer1995,Wald:1999wa}. We refer to the Introduction for a complete list of references.

\subsection{Covariant Phase Space}\label{sec2.1}

The idea is to develop the calculus both in spacetime and in field space.

\paragraph{Spacetime calculus} Consider a differentiable manifold $M$. Prior to a metric structure, one can introduce the de Rham calculus on the space of forms. Given a vector field $\xi\in TM$, a one form is an application from $TM$ to $C^\infty(M)$, that is, an element of the dual bundle $T^*M$. A generic $p-$form is an element of $\wedge^p T^*M$, where the symbol $\wedge$ stands for the wedge (skew-symmetric) product. The space of all forms constitutes the de Rham complex
\beq
\Omega^\bullet (M,\RR)\defeq \bigoplus_{n=0}^{d}\wedge^n T^*M,
\eeq 
where $0-$forms are scalars, and the maximal degree of a form is the dimension of the manifold, $d=$dim(M). We will denote by $\D$ and $i$ the exterior derivative and interior product on this complex, respectively. The exterior derivative increases the form degree by one, while the interior product decreases it by one. The exterior derivative is assumed to be a co-boundary on the complex, which means that it is a nilpotent operation $\D^2=0$. The Lie derivative compares these two operations when applied in different orders,
\beq
\cL_{\xi}\defeq \D i_\xi+i_\xi \D,
\eeq
and thus it does not change the form degree. This formula is sometimes referred to as Cartan's magic formula. In the following, we will repeatedly use that two interior products anti-commute, that $i_\xi$ of a scalar is zero, and $\D$ of a spacetime top-form vanishes.

\paragraph{Variational calculus} One can reproduce these results on the field space. A field space $\Gamma$ is defined as the space of all possible field configurations. It is assumed to be a differentiable manifold. One can thus introduce a calculus on the space of forms on such manifold, called variational calculus. Given a vector field $V\in T\Gamma$, a one form is an application from $T\Gamma$ to the space of functionals, $F=C^\infty(\Gamma)$. A generic $p-$form is an element of $\wedge^p T^*\Gamma$, and the space of all forms constitutes the variational complex
\beq
\Omega^\bullet (\Gamma,F)\defeq \bigoplus_{n=0}^{\text{dim$\Gamma$}}\wedge^n T^*\Gamma,
\eeq 
where $0-$forms are now functionals. We will denote by $\delta$ and $I$ the exterior derivative and interior product on this complex, respectively, with $\delta^2=0$. The exterior derivative increases the form degree by one, and it is what we commonly refer to as a field variation, while the interior product decreases it by one, and we usually think of it as a field contraction. Cartan's magica formula then gives the Lie derivative
\beq
\fL_{V}\defeq \delta I_V+I_V \delta.
\eeq
One can put together spacetime and field-space forms to form the variational bicomplex, defined on $(M,\Gamma)$. This is the basic ingredient of Anderson's theory \cite{MR1188434, Anderson:1996sc}, although we will not explore this theory here, but just confine to the covariant phase space. We will use the notation ``$(p,q)-$form'' to refer to a form on the bicomplex which is a spacetime $p-$form and a field-space $q-$form. On the bicomplex, one can construct an exterior derivative which is $\D+\delta$. The latter is nilpotent if $\D\delta=-\delta \D$. In the following, we will repeatedly use that $I_V$ of a functional is zero, and that two interior products anti-commute. Unless stated otherwise, we will also assume that vector fields in $TM$ are field-independent, such that $\delta\xi=0$, for all $\xi\in 
TM$.

\paragraph{Lagrangian theory} Let us apply this formalism to a Lagrangian theory. The action reads
\beq
S=\int_M L.
\eeq
The Lagrangian $L$ is thus a top form in spacetime, and does not contain field variations, so it is a functional of the fields $\varphi\in\Gamma$. Given that $d$=dim(M), the Lagrangian is a $(d,0)-$form, in the language just established. 

Under an arbitrary field variation $\varphi\to\varphi+\delta\varphi$, it is always possible to rewrite the Lagrangian as a term linear in the variations, and a total derivative
\beq\label{dL}
\delta L=\text{EOM}\delta\varphi+\D \theta.
\eeq
This is an important equality. The first term on the right-hand side of \eqref{dL} gives the equations of motion of the theory, while $\theta$ is the local pre-symplectic potential (we will explain shortly the ``pre-''). It contains a field variation and appears with a total derivative in spacetime, so it is a $(d-1,1)-$form. Suppose $M$ has a boundary $\pa M$. Denoting by $\he$ an equality that holds only on shell of the equations of motion, we then have
\beq
\delta S=\int_M\delta L=\int_M(\text{EOM}\delta \varphi+\D\theta)\he \int_{\pa M}\theta.
\eeq
The local pre-symplectic form is defined as $\omega \defeq \delta\theta$, which is a $(d-1,2)-$form. This local expression can be integrated on a Cauchy slice $\Sigma$ to give the so-called pre-symplectic $2-$form
\beq\label{defo}
\Omega\defeq  \int_\Sigma \omega,
\eeq
which is thus a $(0,2)-$form. This quantity is a crucial ingredient of the theory, because it carries the Poisson bracket, as we will see.

Consider a vector field $V\in T\Gamma$ which is an isometry of $\omega$, that is, $\fL_V\omega=0$. Using that $\delta\omega=\delta^2\theta=0$, we get
\beq
\fL_V \omega=(I_V\delta+\delta I_V)\omega=\delta(I_V\omega)=0.
\eeq
Assuming trivial cohomology in the space of $1-$forms on $\Gamma$ (or else assuming $\fL_{V}\theta=0$), we derive
\beq
\delta I_V\omega=0 \quad \Longrightarrow \quad I_V\omega=-\delta J_V,
\eeq
for a $(d-1,0)-$form $J_V$. Notice that we will disregard embeddings in this first part, to connect to older parts of the literature, and thus we will perform manipulations such as
\beq\label{dO}
\delta\Omega=\delta\int_\Sigma\omega=\int_\Sigma\delta\omega=0.
\eeq
This kind of manipulations is not harmless, and it is by understanding it carefully that the theory of embeddings in the covariant phase space emerged. This will be the core of Sec. \ref{sec3}. Assuming \eqref{dO} holds, one defines the global functional $H_V$ via
\beq\label{Ivo}
I_V\Omega=-\delta H_V, \qquad H_V\defeq \int_\Sigma J_V.
\eeq
A vector field $V$ in $T\Gamma$ that satisfies $\fL_V\Omega=0$ is called a symplectomorphism. A vector field satisfying \eqref{Ivo} is called a Hamiltonian vector field. For trivial cohomology, no distinction is required, for all symplectomorphisms are Hamiltonian. 

We have introduced all the ingredients necessary to formulate Noether's theorems. We reiterate that we are here considering a setup where there are no issues coming from symplectic or spacetime fluxes, and the charges we will construct are assumed to be integrable, conserved, and finite, even without a careful treatment of embeddings (assuming \eqref{dO}). In this setup, charges come directly from contracting the pre-symplectic $2-$form. These are called canonical charges, and are unanimously defined in the literature to be the physical charges. We will review later what happens when these assumptions fail, and the possible resolutions. In particular, the assumptions made here are too stringent to describe a general gravitational system (or more precisely to describe all asymptotic symmetries in such system), where charges cannot be constructed canonically from the pre-symplectic $2-$form, due to fluxes. There are many proposals addressing this problem, as we will review in the following. 

\subsection{Noether's Theorems and Charge Algebra}\label{sec2.2}

The quantity $J_V$ introduced above is called the local Noether current. Its expression depends on the nature of the symmetry. 

\paragraph{Internal symmetry} We directly compute
\beq
-\delta J_V=I_V\omega=\fL_V\theta-\delta I_V\theta.
\eeq
If one assumes that $\fL_V\theta=0$, then
\beq\label{gj}
J_V=I_V\theta,
\eeq
modulo $\delta-$exact terms. Note that we have not imposed the equations of motion at this stage.

\paragraph{Spacetime symmetry} To distinguish this case from the previous one, we refer to the vector field in field space associated to a spacetime vector field $\xi$ as $\Vx$. We compute as before
\beq
-\delta J_\Vx=I_\Vx\omega=\fL_\Vx\theta-\delta I_\Vx\theta.
\eeq
Now, however, one has that $\fL_{V_\xi}\theta=\cL_\xi \theta$, because we are dealing with spacetime transformations (this is background independence, in the usual language, $\delta_\Vx=\cL_\xi$). Then we get
\beq
-\delta J_\Vx=\cL_\xi\theta-\delta I_\Vx\theta =  i_\xi \D \theta +\D i_\xi \theta-\delta I_\Vx \theta.
\eeq
If we assume that $\D i_\xi \theta$ pulls back to zero at the boundary under scrutiny,\footnote{It is this assumption that is problematic in general, with fluxes, and led to different appreciations of the Noether current in the literature, see below.} and we go on-shell, we obtain
\beq\label{noe}
J_\Vx \he I_\Vx \theta - i_\xi  L,
\eeq
again modulo $\delta-$exact terms. In contrast to the previous derivation, we here assumed the equations of motion. This current is called the local weakly-vanishing Noether current because, as we show hereafter, it is identically zero on-shell, modulo $\D-$exact terms.

\paragraph{Noether's first theorem} This theorem states that given a global symmetry, there is an associated codimension$-1$ conserved quantity. To show this, we evaluate the field-space Lie derivative of the action under the symmetry in question
\beq
\fL_V S=\int_M(I_V\delta+\delta I_V)L=\int_M I_V\delta L=\int_M \left(I_V (\text{EOM} \delta\varphi)+\D I_V\theta\right).
\eeq
If $V$ is a global symmetry, then $\fL_V S=0$. Using \eqref{gj}, we get
\beq
\D J_V=-I_V (\text{EOM} \delta\varphi)\he 0.
\eeq
So the local Noether current is conserved on-shell. This implies that the global functional
\beq
H_V\defeq \int_\Sigma J_V,
\eeq
is the integral of a codimension$-1$ conserved global charge on-shell, and it is called the Noether charge.

\paragraph{Noether's second theorem} This theorem states that given a gauge symmetry, there is an associated codimension$-2$ conserved quantity. This means that\footnote{We generically refer to a vector field in field space as $V$. When some properties apply only to a $V$ associated to a spacetime diffeomorphism, we use $\Vx$. We will call the charge $H_\xi$ instead of $H_\Vx$, simply to be compatible with the common notation in the literature.}
\beq\label{N2}
J_V \he \D Q_V.
\eeq
Conservation then follows straightforwardly from $\D^2=0$. Let us show that this current is at best a $\D-$exact term on-shell.
To show this, we evaluate the field-space Lie derivative of the action under the symmetry in question
\beq\label{Ls}
\fL_V S=\int_M(I_V\delta+\delta I_V)L=\int_M \left(I_V (\text{EOM} \delta\varphi)+\D I_V\theta\right)=\int_M I_V (\text{EOM} \delta\varphi)+\int_{\pa M} I_V\theta.
\eeq
If $V$ is an internal gauge symmetry, then we  have $\fL_V S=0$ and we proceed as before. If, on the other hand, $V=\Vx$ is associated to a spacetime symmetry, then we have $\fL_\Vx S=\cL_\xi S$. So we compute
\beq\label{lS}
\cL_\xi S=\int_M \cL_\xi L=\int_M \D i_\xi L=\int_{\pa M}i_\xi L.
\eeq
Equating \eqref{Ls} with \eqref{lS} we gather
\beq
\int_{\pa M}I_\Vx\theta \he \int_{\pa M}i_\xi L \quad \Longrightarrow \quad  J_\Vx\he\D Q_\xi.
\eeq
Proceeding as before, we define the global functional
\beq
H_\xi\defeq  \int_\Sigma J_\Vx.
\eeq
However, this becomes a codimension$-2$ conserved charge on-shell. Indeed, if the hypersurface $\Sigma$ has a codimension$-2$ surface $S=\pa \Sigma$ at its boundary, we obtain
\beq
H_\xi \he \int_\Sigma \D Q_\xi=\int_S Q_\xi.
\eeq
This is again called the Noether charge, but for gauge symmetries. Given its codimension$-2$ nature, it is sometimes referred to as surface charge. The surface $S$ is generically called a corner, and so this charge is also called a corner charge, in the recent literature. To link to the discussion in the Introduction, it is this theorem that determines whether a gauge symmetry, in the presence of a boundary, is still a pure gauge transformation or becomes physical. In fact, this distinction is encoded in whether the associated Noether charge vanishes or not, respectively. We will discuss this  in Subsec. \ref{sec2.4}.

\paragraph{Poisson bracket} A symplectic $2-$form has two properties, it is closed in field space, and it is non-degenerate:
\beq
\delta \Omega=0, \qquad I_V\Omega=0 \Leftrightarrow V=0.
\eeq
The reason why we used the expression ``pre-''symplectic so far is that non-degeneracy is  not guaranteed, and indeed there are in general zero modes, that is, non-vanishing symmetries that annihilate $\Omega$. Once these modes are removed (by quotienting these symmetries out), the symplectic $2-$form carries a Poisson bracket representation of the charge algebra. Consider $V,W\in T\Gamma$, then we define the Poisson bracket between symplectomorphisms as
\beq\label{defO}
\{H_V,H_W\}=\fL_W H_V.
\eeq
Using this definition and the fact that $V,W$ are symplectomorphisms, one easily proves that the bracket is skew symmetric
\beq
\{H_V,H_W\}=I_W\delta H_V=-I_W I_V \Omega=I_V I_W \Omega=-I_V\delta H_W=-\{H_W,H_V\}.
\eeq
The Poisson bracket is a fundamental ingredient in the theory of asymptotic symmetries, as it gives rise to the symmetry principle organising observables of a given theory. Along the way, we have also shown that
\beq
\{H_V,H_W\}=I_V I_W \Omega,
\eeq
which demonstrates that the symplectic $2-$form determines the Poisson bracket of the theory.

Consider now spacetime symmetries, where the definition \eqref{defO}, given two spacetime vector fields $\xi,\zeta\in TM$, reads
\beq\label{HH}
\{H_\xi,H_\zeta\}=\fL_{V_\zeta}H_\xi.
\eeq
On the other hand, we also have in this case the Lie bracket of vector fields in $TM$
\beq
[\xi,\zeta]=\cL_\xi \zeta.
\eeq
How is the Poisson bracket of charges (and thus the charge algebra) related to the Lie bracket  of vector fields (and thus the symmetry algebra)? To answer this question, we need to demonstrate the following result. Given two symplectomorphisms $V_\xi$ and $V_\zeta$, one has
\beq\label{t1}
I_{V_{[\xi,\zeta]}}\Omega=\delta \{H_\xi,H_\zeta\}.
\eeq
Let us prove this. Using $I_{[V_\xi,V_\zeta]}= I_{V_{[\xi,\zeta]}}$, and recalling the calculus identity\footnote{The reader unfamiliar with this result should prove it before continuing, as it will be used systematically.}
\beq
I_{[V_\xi,V_\zeta]}=[\fL_\Vx,I_{V_\zeta}],
\eeq
we get
\beq
I_{V_{[\xi,\zeta]}}\Omega=[\fL_\Vx,I_{V_\zeta}]\Omega=\fL_\Vx I_{V_\zeta}\Omega-I_{V_\zeta}\fL_\Vx\Omega.
\eeq
The last term vanishes, for $\Vx$ is a symplectomorphism, and thus
\beq
I_{V_{[\xi,\zeta]}}\Omega=\delta I_\Vx I_{V_\zeta}\Omega+ I_\Vx \delta I_{V_\zeta}\Omega.
\eeq
Using $\delta\Omega=0$, we can rewrite $I_\Vx \delta I_{V_\zeta}\Omega=I_\Vx \fL_{V_\zeta} \Omega=0$, because also $V_\zeta$ is a symplectomorphisms, such that we eventually gather
\beq
I_{V_{[\xi,\zeta]}}\Omega=\delta I_\Vx I_{V_\zeta}\Omega=\delta \{H_\xi,H_\zeta\},
\eeq
which is exactly what we wanted to prove, eq. \eqref{t1}. Using similar manipulations, we leave as an exercise to prove that the Lie bracket Jacobi identity induces the Poisson bracket Jacobi identity. 

We wish to establish the relationship between the charges and the vector fields algebras. First we observe that the identity $I_\Vx \Omega=-\delta H_\xi$ sets a correspondence between the field-space vector field $\Vx$ and the functional $H_\xi$. This correspondence is cohomological: two charges $H_\xi$ and $H_\xi+\kappa$ with $\delta \kappa=0$ correspond to the same vector field $V_\xi$. From \eqref{t1}, it similarly follows that  $\{H_\xi,H_\zeta\}$ and $\{H_\xi,H_\zeta\}+\kappa_{\xi,\zeta}$ with $\delta \kappa_{\xi,\zeta}=0$ correspond to the same vector field $V_{[\xi,\zeta]}$. Consider thus
\beq
-\delta H_{[\xi,\zeta]}=I_{V_{[\xi,\zeta]}}\Omega=\delta \{H_\xi,H_\zeta\}.
\eeq
From the cohomological argument above, we then deduce
\beq\label{pro}
\{H_\xi,H_\zeta\}=-H_{[\xi,\zeta]}+\kappa_{\xi,\zeta},
\eeq
as long as $\delta \kappa_{\xi,\zeta}=0$.\footnote{Mathematically, this central extension is a Gelfand-Fucks $2-$cocycle \cite{Gefand1972}.} 
What we have just shown is that \textit{the Poisson bracket of charges represents the Lie bracket of symmetries projectively.} A constant in field space (like $\kappa$) appearing in the Poisson bracket is known as a central extension, where the word ``central'' means that it commutes with all generators of the algebra ($\delta\kappa=0$). So another way to restate this result is that the charge algebra represents the symmetry algebra modulo central extensions. This is a general and important result. We will later show an example of this in AdS$_3$ gravity, where the central extension will be proportional to the well-known Brown-Henneaux central charge.

Before presenting the general theory of asymptotic symmetries, we  familiarize with the  construction of conserved quantities for gauge symmetries in a simple example. 

\subsection{Exercise: Electromagnetism}\label{sec2.3}

In this exercise, we will compute the Noether charges in electrodynamics, to get acquainted with the formal discussion of the previous Subsections. 

Consider classical electrodynamics in $4$ spacetime dimensions, on a flat background. The top-form Lagrangian, in suitable units, reads 
\beq
L=-{1\over 4}F^{\mu\nu}F_{\mu\nu}Vol_M,
\eeq
where $Vol_M$ is the volume form on $M$, that is, $Vol_M=\star 1$, with $\star$ denoting Hodge duality. We define the field strength 
\beq
F=\D A={1\over 2}F_{\mu\nu}\D x^\mu \wedge \D x^\nu, \qquad F_{\mu\nu}=\pa_\mu A_
\nu-\pa_\nu A_\mu,
\eeq
with $A=A_\mu\D x^\mu$ the $U(1)$ Abelian gauge connection.

\begin{enumerate}
\item[Q1)] \textbf{Prove that $S=-{1\over 2}\int_M F\wedge \star F$} 

We first familiarise with manipulations and conventions involving $Vol_M$. Starting from the definition of the Hodge star
\beqn
Vol_M=\star 1={1\over 4!}\epsilon_{\alpha\beta\gamma\delta}\D x^\alpha \wedge \D x^\beta\wedge \D x^\gamma\wedge \D x^\delta,
\eeqn
where $\epsilon_{\alpha\beta\gamma\delta}$ is the $4-$dimensional Levi-Civita symbol (equal to the Levi-Civita tensor on flat space),
and using the identity
\beq
\D x^\alpha\wedge \D x^\beta\wedge \D x^\gamma\wedge \D x^\delta=-\epsilon^{\alpha\beta\gamma\delta} \D x^0 \wedge \D x^1\wedge \D x^2\wedge \D x^3,
\eeq
we derive the familiar expression
\beq
Vol_M=-{1\over 4!}\epsilon_{\alpha\beta\gamma\delta}\epsilon^{\alpha\beta\gamma\delta}\D x^0\wedge \D x^1\wedge \D x^2\wedge \D x^3=\D x^0\wedge \D x^1\wedge \D x^2\wedge \D x^3,
\eeq
where we used $\epsilon_{\alpha\beta\gamma\delta}\epsilon^{\alpha\beta\gamma\delta}=-4!$.

We then evaluate explicitly, using these relations and the skew-symmetry of $F^{\mu\nu}$,
\beqn
F\wedge \star F&=& {1\over 2}F_{\mu\nu}\D x^\mu\wedge \D x^\nu \wedge \left({1\over 4}\epsilon_{\alpha\beta\gamma\delta}F^{\gamma\delta}\D x^\alpha\wedge \D x^\beta\right)\\
&=& -{1\over 8}F_{\mu\nu}\epsilon_{\alpha\beta\gamma\delta}F^{\gamma\delta}\epsilon^{\mu\nu\alpha\beta} Vol_M\\
&=& {1\over 4}F_{\mu\nu}F^{\gamma\delta}\left(\delta^\mu_\gamma\delta^\nu_\delta-\delta^\mu_\delta\delta^\nu_\gamma\right)Vol_M\\
&=&{1\over 2}F_{\mu\nu}F^{\mu\nu}Vol_M.
\eeqn
Consequently
\beq
S=-{1\over 2}\int_M F\wedge \star F=-{1\over 4}\int_M F^{\mu\nu}F_{\mu\nu}Vol_M,
\eeq
as desired.

\item[Q2)] \textbf{Compute the local pre-symplectic potential $\theta$ and the equations of motion}

The gauge connection $A$ is the dynamical field of the theory in question, and thus it is part of the field space $\Gamma$. We will derive $\theta$ working exclusively in form language. We explicitly evaluate (recall $\D \delta=-\delta \D$)
\beqn
\delta S = -{1\over 2}\int_M \delta(F\wedge \star F) = \int_M \D(\delta A)\wedge \star F=\int_M \D(\delta A\wedge\star F)-\int_M\D \star F\wedge \delta A.
\eeqn
Comparing this result with \eqref{dL}, we obtain
\beq
\text{EOM:} \quad \D\star F=0, \qquad \theta=\delta A\wedge \star F.
\eeq
As expected, the equations of motion $\D\star F=0$ are Maxwell equations (together with the Bianchi identity $\D F=\D^2 A=0$), and we read off the local pre-symplectic potential.

\item[Q3)] \textbf{Show that} $\delta_\lambda A\defeq I_{V_\lambda} \delta A =\D \lambda$ \textbf{is a gauge transformation, and compute the Noether current. Does Noether's second theorem apply?}

To prove that $\delta_\lambda A$ is a gauge transformation, we have to show that the field strength is invariant
\beq
I_{V_\lambda} \delta F=-I_{V_\lambda}\D\delta A=-\D^2 \lambda=0.
\eeq

Then, to compute the Noether current, we must evaluate (note that $\delta\lambda=0$)
\beq
I_{V_\lambda}\Omega=\int_\Sigma I_{V_\lambda}\omega=\int_\Sigma I_{V_\lambda}\delta \theta=\int_\Sigma I_{V_\lambda}(\delta A\wedge \star  \D\delta A)=\int_\Sigma (\D\lambda\wedge \star  \D\delta A)=-\int_\Sigma\delta(\D\lambda\wedge \star \D A).
\eeq
From the general expression \eqref{Ivo}, we read off the Noether current
\beq
H_{\lambda}=\int_\Sigma \D\lambda\wedge \star \D A, \qquad J_{V_\lambda}=\D\lambda\wedge \star \D A.
\eeq

Since the symmetry in question is gauge, Noether's second theorem must hold, that is, it should be possible to rewrite $J_{V_\lambda}$ as an on-shell total derivative, \eqref{N2}. This is straightforward to prove:
\beq
J_{V_\lambda}=\D\lambda\wedge \star \D A=\D (\lambda  \star \D A)-\lambda \D \star F \he \D (\lambda  \star \D A), \qquad Q_\lambda=\lambda \star F.
\eeq

\item[Q4)] \textbf{Compute the Noether charge, and find the simplest value for which it is conserved. What is the physical quantity associated?}

We compute
\beq
H_\lambda=\int_\Sigma J_{V_\lambda}\he\int_\Sigma \D (\lambda \star F)=\int_{S}\lambda \star F.
\eeq
Here we have been generic concerning the Cauchy surface $\Sigma$ and its boundary $S$. For simplicity, and concreteness, we can assume that $S$ is a $2-$sphere at fixed value of the radius $r$ and at fixed time. Then, we observe that $H_\lambda$ is trivially conserved if $\lambda$ is a constant $\lambda_0$. The conserved quantity in question is thus
\beq
H_{\lambda_0}=\lambda_0 \int_{S} \star F,
\eeq
which is the total electric charge of the system. Note that this is not the only conserved quantities. Depending on the codimension$-2$ surface location, there can be other values of $\lambda$ that lead to conserved quantities. Note also that in this simple example we are essentially using the Gauss law of electrodynamics.
\end{enumerate}
The aim of this exercise was to familiarize with Noether's second theorem, and the fact that there are conserved charges associated to gauge symmetries, which have support on codimension$-2$ surfaces. We now continue and formulate the general theory of asymptotic symmetries. 

\subsection{Asymptotic Symmetries}\label{sec2.4}

Consider a manifold $M$ (the bulk) with boundary $B$ and a field space $\Gamma$. A classical dynamical theory is defined by the following steps\footnote{We stressed that this defines a classical theory, because most of the following are not the defining principles of a quantum theory. For instance, imposing boundary conditions, or fixing gauge, are  classical manipulations.}
\begin{enumerate}
\item[S1.] \textit{Dynamics} on $M$.
\item[S2.] \textit{Boundary conditions} $\Gamma\vert_B$, that is, the behaviour (falloffs) of the bulk fields near the boundary.
\item[S3.] In case of a gauge theory, a \textit{gauge fixing} for the bulk fields. 
\end{enumerate}
The last point is subtle, and it is a feature that is often abused in the literature. In principle, one should compute the conserved charges of the theory without gauge fixing, and show that the gauge symmetry needed to fix the gauge is indeed a trivial symmetry transformation, with vanishing associated charges. In practice, however, it is sometimes impossible to compute Noether charges without some restrictions on $\Gamma$, both for physical and technical reasons. It has been one of the main goals in the asymptotic symmetry community to relax boundary and gauge conditions, and obtain more and more general results. 

Once steps S1-S3 specified, we have a theory. Then, there is a systematic procedure to obtain Noether charges. The next steps are
\begin{enumerate}
\item[S4.] \textit{Symmetries}
\begin{enumerate}
\item[S4a-] \textit{Residual symmetries}. These are the symplectomorphisms preserving points S2 and S3 above. They are also referred to as allowed or gauge transformations. At this stage, there are still residual symmetries that give rise to zero modes of the pre-symplectic $2-$form. 
\item[S4b-] Compute the charge associated to a given residual symmetry. If the charge vanishes, then it is a \textit{trivial transformation}, a true redundancy of the system. These residual symmetries are also called proper or small gauge transformation. If the associated charge is instead non-vanishing, then the symmetry in question is an \textit{asymptotic symmetry}, that is, a physical transformation that acts non-trivially on the field space, and sends therefore the system to a different inequivalent configuration. These are also called improper or large transformations.
\item[S4c-] The \textit{asymptotic symmetry group} is then defined quotienting the residual symmetries by the trivial transformations
\beq
\text{Asymptotic Symmetries}= \bigslant{\text{Residual Symmetries}}{ \text{Trivial Symmetries}}
\eeq
This is possible because the trivial symmetry group is an ideal inside the residual symmetry group. Then, the pre-symplectic $2-$form becomes invertible if restricted to the asymptotic symmetries only. This procedure has thus removed the zero modes, which are the trivial transformations. 
\end{enumerate}
\item[S5.] \textit{Charge algebra}. Now that the symplectic $2-$form is invertible, the Poisson bracket of charges can be computed, and it gives rise to the algebra organizing the physical observables of the theory. 
\end{enumerate}

Some comments are in order. First, this discussion is formal, but we will shortly perform steps S1-S5 in a specific example, and there the reader will recognize familiar results. Second, there is no guarantee at this stage that the final charges are well-defined. This is because, in general, the boundary conditions chosen or the gauge fixing procedure may lead to pathologies. A conservative approach would define the final Noether charges associated to asymptotic symmetries to be well-defined if and only if the charges are:
\begin{itemize}
\item Integrable: $I_V \Omega=-\delta \int_S Q_V$. If this holds then the charges are called the canonical Noether charges.
\item Conserved: $H_V\vert_{S_2}-H_V\vert_{S_1}=\int_{S_1}^{S_2}\D Q_V\he 0$. This holds if the local Noether current is weakly vanishing.
\item Finite: as we approach the codimension$-2$ surface under scrutiny,  charges do not diverge.
\end{itemize}
Given points S1-S3 above, if one performs steps S4 and S5, and obtains integrable, conserved, and finite charges, then points S2 and S3 were not pathological, and the theory is well-defined. On the other hand, if charges do not satisfy these conditions, then one could go back and modify S2 and S3.\footnote{It is because of this that sometimes we refer to this procedure as ``the art of boundary conditions'', because there is not an a priori guarantee that they would lead to well-defined charges.} On the other hand, each of the features aforementioned, integrability, conservation, and finiteness, are questioned in gravity, and understanding how to cure these pathologies is a very far-reaching and active area of investigation.

Lastly, it is important to note that asymptotic symmetry groups are typically bigger than the full bulk symmetry groups, because they are generated by symmetries of the asymptotic-to-the-boundary field space only. The fundamental and historical example is Minkowski in $4$ dimensions, where the full bulk symmetries are given by the Poincar\'e group, but the asymptotic symmetry group at null infinity has been found to be the infinite-dimensional $BMS_4$ group \cite{BONDI1960,Sachs1961,doi:10.1098/rspa.1962.0161,Sachs:1962wk,Sachs1962a,Bondi1964}, where the $4$ bulk translations are enhanced to the so-called supertranslations, which form an infinite-dimensional Abelian algebra.

\subsection{Gravity}\label{sec2.5}

As mentioned, integrability, conservation, and finiteness of Noether charges have all been questioned in gravity. An important part of these lectures will be about integrability, but let us also mention how to deal with divergent and non-conserved charges. 

\paragraph{Finiteness} This is only guaranteed if the boundary is located at finite distance in the bulk. Asymptotic boundaries would instead typically lead to divergent charges. This is because the action itself could diverge, and thus all the derivation of charges is affected by it. The general method to cure this is called renormalization, and it is a whole enterprise. Without entering details, the general procedure is to add to the bulk action counterterms at each boundary
\beq
S\to S+\sum_j \int_{\pa M_j}\ell_j+\sum_i \int_{\pa\pa M_i}\tilde{\ell}_i,
\eeq
such that $\theta$ and $\omega$ become finite. Sometimes this is not possible, and only $\theta$ can be directly renormalized but not $S$. There is no a priori guarantee that this method always works, but we have gathered many evidences that it works on a case by case study. 
 
In AdS, this procedure is familiar in the context of AdS/CFT \cite{Balasubramanian1999, Emparan1999, Skenderis2001, Haro2001, Skenderis2002, Bianchi2002, Papadimitriou2005, Hollands2005a, Mansi2009a, Mansi2009, Papadimitriou2010, Ciambelli:2019bzz, Anastasiou2020, Fiorucci:2020xto}, and computations are facilitated by the fact that the asymptotic boundary is a timelike hypersurface. In asymptotically flat spacetimes, the asymptotic boundary is null infinity, which is a Carrollian manifold (see for instance \cite{Ciambelli:2018wre, Ciambelli:2018ojf, Campoleoni:2018ltl, Ciambelli:2020eba, Ciambelli2020a, Freidel2022d}), and computations are more challenging. Nonetheless, tremendous effort and progress has been made in the last decade. The reader can consult the Introduction, and \cite{Compere:2018ylh} and \cite{Chandrasekaran2022}, for more references and information on this thriving topic.

\paragraph{Conservation} There are two main reasons why the charges are not always conserved. Applied to diffeomorphisms, the conservation equation reads
\beq\label{concon}
H_{\xi}\vert_{S_2}-H_{\xi}\vert_{S_1}=\int_{S_1}^{S_2}\D Q_{\xi}=\int_{S_1}^{S_2}(I_{V_\xi}\theta-i_{\xi}L).
\eeq
The first reason is the presence of gravitational fluxes through the surface. Then non-conservation is due to the matter propagation, indicating that the subregion under scrutiny is not isolated from its complement. A discussion and a complete list of references on this argument is found in \cite{Fiorucci2021}. The second reason is the presence of anomalies, which results from a non-stationary variational problem for the bulk action. This has been discussed in \cite{Alessio:2020ioh, Fiorucci:2020xto, Campoleoni2022}.

\paragraph{Integrability} 

It is not always guaranteed that contracting the pre-symplectic $2-$form with a field-space vector field associated to a diffeomorphism gives a $\delta-$exact term. This has dramatic consequences, because the charge algebra does not close on itself, and thus the Poisson bracket cannot be used. One computes explicitly
\beq
I_{V_\xi}\Omega=\int_\Sigma I_{V_\xi}\delta \theta=\int_\Sigma\D i_\xi \theta-\delta \D Q_\xi \neq -\delta H_\xi,
\eeq
unless $\D i_\xi\theta$ vanishes. The physical interpretation is that the system is open, and thus there are physical degrees of freedom coming from the bulk and reaching the boundary. The evolution on the boundary seems therefore to be not predictable from a single surface, because there is dissipation. There are two main approaches to treat this situation
\begin{enumerate}
\item Interpret the non-integrable part as a symplectic flux
\beq
F_\xi= \int_\Sigma\D i_\xi \theta,
\eeq
and split the charges between integrable part and flux
\beq
I_{V_\xi}\Omega= -\cancel{\delta} H_\xi,\qquad \cancel{\delta} H_\xi=\delta H_\xi-F_\xi.
\eeq
This split is ambiguous, and there are various way to fix this ambiguity. One possibility is to introduce a bracket between the integrable parts, called the Barnich-Troessaert bracket \cite{Barnich:2011mi} (see also \cite{Troessaert2016, Chandrasekaran:2020wwn,  Wieland2022})
\beq
\{H_\xi,H_\zeta\}_{BT}\defeq \{H_\xi,H_\zeta\}+I_{V_\xi}F_\zeta.
\eeq
Then this bracket is skew-symmetric, satisfies Jacobi (it has to be postulated), and gives
\beq
\{H_\xi,H_\zeta\}_{BT}=- H_{[\xi,\zeta]}+\kappa_{\xi,\zeta},
\eeq
except that $\kappa$ is now field dependent. Another prescription is given in \cite{Freidel:2021cjp}, where the Noetherian split is imposed, such that the charges are posit to be the Noether charges. This prescription gives rise to a unique split, and calls for the deep question about the different (from canonical) notion of charges and their physical interpretation.  
\item Consider an extended field space, such that 
\beq
I_{V_\xi}\Omega^{\text{Ext}}=-\delta H_\xi.
\eeq
There is no a priori certainty that such an extension exists, but this is exactly where a careful treatment of codimension$-2$ surfaces comes to the rescue. In particular, one finds that there are extra degrees of freedom living on the surface, called edge modes, such that the  symplectic$-2$ form in which edge modes are dynamical fields gives rise to integrable charges for all diffeomorphisms. Clearly, dissipation is still present, but it is now associated to non-trivial edge modes variations, rather than non-integrability. In this approach, one can then use the standard Poisson bracket, and all charges are again canonical, on this extended field space \cite{Ciambelli:2021vnn, Ciambelli:2021nmv, Freidel:2021dxw}. 
\end{enumerate}

The second approach stems from the corner proposal, which is why we will adopt it and study  in details in the following. It leads to deep questions about symplectic systems in the presence of dissipation, and challenges the existing literature with a new, radically different, perspective. Indeed, this approach suggests that the evolution is still predictable, by keeping track of embeddings (edge modes) and their field-space variation. We will discuss this in Subsec. \ref{sec3.4}. As mentioned in the Introduction, a complete relationship between the two approaches is still lacking at present, and it is definitely worth pursuing.

\subsection{Exercise: \texorpdfstring{AdS$_3$}{}}\label{sec2.6}

In this exercise, we show how the general theory of asymptotic symmetries, that we enunciated in Subsec. \ref{sec2.4}, works in an explicit example. We will derive the AdS$_3$ asymptotic symmetries, with the final goal of finding the charge algebra, and the famous Brown-Henneaux central extension \cite{Brown:1986nw}. We will rigorously follow the steps of Subsec. \ref{sec2.4}: first we define the theory, and then we compute the symmetries. 

\begin{enumerate}
\item[S1.] The dynamics is Einstein gravity in $3$ dimensions with negative cosmological constant
\beq
S={1\over 2k}\int_M (R-2\Lambda)\sqrt{-g} \ \D^3 x,
\eeq
where $k=8\pi G$, $c=1$, $R$ is the Ricci scalar,  $\Lambda$ the cosmological constant, and $\sqrt{-g}$ the square root of the determinant of the (Lorentzian) metric. 
\item[S2.] The boundary we will study is the conformal timelike boundary at spatial infinity. Calling $\rho$ the inverse radial coordinate, the conformal boundary $B$ is reached by sending $\rho\to 0$. The falloff we impose is that the boundary appears with a pole of order $2$ in the radial coordinate, calling $x^\mu=(\rho, x^a)$ the bulk coordinates, 
\beq
\lim_{\rho\to 0}g_{ab}={g^{(0)}_{ab}(x)\over \rho^2}+O(\rho^0),
\eeq
where $g^{(0)}_{ab}$ is thus interpreted as the (conformal class of) boundary metric at the boundary.
We will then adopt the so-called Brown-Henneaux boundary conditions, and require that the leading order of the bulk metric as we approach the boundary is the $2-$dimensional Minkowski metric
\beq
g^{(0)}_{ab}=\eta_{ab}.
\eeq
\item[S3.] The gauge theory is gravity, and the gauge symmetries are diffeomorphisms in the bulk. We will use part of this bulk symmetries to fix the gauge, i.e., to bring the bulk metric to a specific form. The gauge we adopt is the Fefferman-Graham gauge \cite{Fefferman1985, Fefferman2011}, reached setting
\beq\label{gc}
g_{\rho\rho}={\ell^2\over \rho^2}\qquad g_{\rho a}=0,
\eeq
where we introduced the AdS radius $\ell^2=-{1\over \Lambda}$. That this gaige can always be achieved, is a mathematical theorem. That the diffeomorphisms required to bring the metric to this gauge are trivial (vanishing associated charges), is not obvious, and indeed there are indications in the literature that this is generically false. We will nevertheless proceed with this gauge fixing in the following.
\end{enumerate}
Even without Brown-Henneaux boundary conditions, points S1-S3 are enough to completely solve the bulk equations of motion, and find the line element (this can be proven as an exercise)
\beq\label{ds}
\D s^2={\ell^2\over \rho^2}\D\rho^2+\left({g^{(0)}_{ab}\over \rho^2}+g^{(2)}_{ab}+\rho^2 g^{(4)}_{ab}\right)\D x^a\D x^b,
\eeq
with
\beq
g^{(4)}_{ab}={1\over 4}g^{(2)}_{ac}g^{(0)cd}g^{(2)}_{db},
\eeq
and $g^{(0)ab}$ is the inverse of the boundary metric $g^{(0)}_{ab}$. This line element is on-shell if, defining \cite{Henningson1998} ($R^{(0)}$ is the boundary Ricci scalar of $g^{(0)}_{ab}$)
\beq
T_{ab}={1\over k \ell}\left(g^{(2)}_{ab}+{\ell^2\over 2}R^{(0)}g^{(0)}_{ab}\right),
\eeq
we have that
\beq\label{Tab}
D^{(0)}_a T^{a}{}_b=0 \qquad T^a{}_a={c\over 24\pi} R^{(0)}\qquad c={3\ell\over 2G}.
\eeq
In this expression, $D^{(0)}$ is the boundary Levi-Civita covariant derivative, and the indexes are raised and lowered using the boundary metric.

We have thus all the ingredients to find the residual symmetries, charges, and charge algebra. Details can be found for instance in \cite{Alessio:2020ioh}, and references therein. 

\begin{enumerate}
\item[Q1)]\textbf{The gauge-preserving diffeomorphisms must leave \eqref{gc} invariant, which requires $\cL_{\xi}g_{\rho\rho}=0=\cL_{\xi}g_{\rho a}$. Solve these equations.}

Given a vector field $\xi=\xi^\rho(x,\rho)\pa_\rho+\xi^a(x,\rho)\pa_a$, we compute
\beqn\label{s1}
0=\cL_\xi g_{\rho\rho}=\xi^\mu\pa_\mu g_{\rho\rho}+2g_{\rho \mu}\pa_\rho\xi^\mu=-2 {\ell^2\over \rho^3}\xi^\rho+2 {\ell^2\over \rho^2}\pa_\rho \xi^\rho,
\eeqn
where we used the definition of the Lie derivative of the metric, 
\beq
\cL_\xi g_{\mu\nu}=\xi^\alpha\pa_\alpha g_{\mu\nu}+g_{\mu\alpha}\pa_\nu\xi^\alpha+g_{\alpha\nu}\pa_\mu\xi^\alpha,
\eeq
and the fact that
\beq
g_{\rho\rho}={\ell^2\over \rho^2}\qquad g_{\rho a}=0
\eeq
in the Fefferman-Graham gauge. Then we can solve explicitly \eqref{s1}
\beq
\ln \xi^\rho=\ln\rho+\beta(x),
\eeq
where $\beta(x)$ is an integration constant. Calling $e^\beta=\sigma$, we eventually get
\beq\label{sigma}
\cL_\xi g_{\rho\rho}=0, \quad \Rightarrow \quad \xi^\rho=\rho \sigma(x).
\eeq

Similarly, using this result, we evaluate
\beq
0=\cL_\xi g_{\rho a}=g_{ab}\pa_\rho\xi^b+g_{\rho\rho}\pa_a\xi^\rho \quad \Rightarrow\quad
{\ell^2\over \rho}\pa_a\sigma=-g_{ab}\pa_\rho\xi^b.
\eeq
Multiplying this expression by $g^{ab}$,\footnote{This is a bulk inverse, that admits an expansion in powers of $\rho$, not to be confused with $g^{(0)ab}$, the inverse of the boundary metric.} the inverse of $g_{ab}$, we gather
\beq
\pa_\rho\xi^a=-{\ell^2\over \rho}g^{ab}\pa_b\sigma.
\eeq
This can now be integrated from the boundary to the bulk, and yields
\beq\label{x0}
\xi^a(x,\rho)=\xi_{(0)}^{a}(x)-\ell^2\pa_b\sigma\int_0^{\rho}{\D \rho^\prime\over \rho^\prime}g^{ba}(x,\rho^\prime),
\eeq
where $\xi_{(0)}^{a}(x)$ are integration constants.
\end{enumerate}
So we obtained the most general diffeomorphism preserving the Fefferman-Graham gauge conditions (step S3)\footnote{This vector is field-dependent, but only in  subleading orders, and thus $\delta\xi$ never contributes in this exercise.}
\beq
\xi=\rho\sigma(x)\pa_\rho+\xi_{(0)}^{a}(x)\pa_a-\ell^2\pa_b\sigma\int_0^{\rho}{\D \rho^\prime\over \rho^\prime}g^{ba}(x,\rho^\prime)\pa_a.
\eeq
This depends on three arbitrary functions on the boundary. The function $\sigma(x)$ generates Weyl rescaling of the boundary, whereas $\xi_{(0)}^{a}(x)$ generate boundary diffeomorphisms. This is explained for instance in \cite{Imbimbo2000, Ciambelli:2019bzz, Alessio:2020ioh}. 
\begin{enumerate}
\item[Q2)]\textbf{Impose Brown-Henneaux boundary conditions (point S2) to obtain the residual symmetries.}

The condition $g^{(0)}_{ab}=\eta_{ab}$ is preserved under bulk diffeomorphisms requiring that 
\beq\label{bh}
\delta_\xi g_{ab}\defeq \cL_\xi g_{ab}=O(\rho^0).
\eeq
For the $\rho$ expansion, we first notice that 
\beq
g^{ab}(x,\rho)=\rho^2 g^{(0)ab}+O(\rho^4),
\eeq
which implies
\beq
\int_0^{\rho}{\D \rho^\prime\over \rho^\prime}g^{ba}(x,\rho^\prime)=\int_0^{\rho}\D \rho^\prime(g^{(0)ba}\rho^{\prime}+O(\rho^{\prime 3}))={\rho^2\over 2}+O(\rho^4).
\eeq
Therefore, the second term in \eqref{x0} is subleading in the $\rho$ expansion, with respect to the first.

Using this, we evaluate
\beqn
\cL_\xi g_{ab}&=& \xi^\mu\pa_\mu g_{ab}+g_{ac}\pa_b\xi^c+g_{bc}\pa_a\xi^c\\
&=&\rho\sigma\pa_\rho \big({g^{(0)}_{ab}\over \rho^2}\big)+{\xi_{(0)}^{c}\over \rho^2}\pa_c g^{(0)}_{ab}+{g^{(0)}_{ac}\over \rho^2}\pa_b\xi_{(0)}^{c}+{g^{(0)}_{bc}\over \rho^2}\pa_a\xi_{(0)}^{c}+O(\rho^0)\\
&=& -2{\sigma\over\rho^2}g^{(0)}_{ab}+{1\over\rho^2}\cL_{\xi_{(0)}}g^{(0)}_{ab}+O(\rho^0),
\eeqn
where $\cL_{\xi_{(0)}}$ is the boundary Lie derivative along the vector field $\xi_{(0)}=\xi^a_{(0)}(x)\pa_a$. We now impose \eqref{bh} and gather
\beq
\cL_{\xi_{(0)}}g^{(0)}_{ab}=2 \sigma g^{(0)}_{ab}.
\eeq
This equation fixes $\sigma$ in terms of $\xi_{(0)}$. With Brown-Henneaux boundary conditions, this expression becomes
\beq
\cL_{\xi_{(0)}}\eta_{ab}=2 \sigma \eta_{ab}.
\eeq
This is the familiar conformal Killing equation. Indeed, taking the trace we obtain $\sigma={1\over 2}\pa_c\xi^c_{(0)}$, and thus
\beq
\cL_{\xi_{(0)}}\eta_{ab}=\pa_c\xi^c_{(0)} \eta_{ab}.
\eeq
Therefore, the Weyl factor is not independent but serves to compensate the conformal Killing diffeomorphism at the boundary.
\end{enumerate}
To recapitulate, the residual symmetries of the theory defined by steps S1-S3 above are parameterized by the conformal Killing vectors of the boundary metric. This is a very primitive and geometric way of unravelling the AdS/CFT correspondence, from the asymptotic symmetries viewpoint. And indeed, the AdS/CFT dictionary works exclusive with Brown-Henneaux boundary conditions, which are nothing but Dirichlet boundary conditions for the bulk metric.\footnote{In $3$ dimensions there is no radiation, but in higher dimensions these boundary conditions set the radiation to zero.} 
 
While everything has been carefully derived so far, we will now give the Noether charges, rather than derive them explicitly. This is due to two reasons: the computation is quite technical, and it is a standard result that can be found in numerous places in the literature (see e.g. \cite{Compere2019b}). Assuming that the boundary metric is $\eta_{ab}$, we can choose light-cone coordinates $x^a=x^\pm$ in the boundary. Then \eqref{Tab} can be solved explicitly, leading to two chiral functions $L^\pm(x^\pm)$, the integration constants of the equations of motion. The line element \eqref{ds} becomes the fully on-shell metric
\beq\label{dslc}
\D s^2={\ell^2\over \rho^2}\D\rho^2-\left({\D x^+\over\rho}-\rho\ell^2L^-(x^-)\D x^-\right)\left({\D x^-\over\rho}-\rho\ell^2L^+(x^+)\D x^+\right).
\eeq
The residual symmetries are consequently generated by the two chiral functions $\xi^\pm(x^\pm)$. The Noether charges, originally derived in \cite{Brown:1986nw} (see also \cite{Compere2019b, Alessio:2020ioh}) are found to be integrable, conserved, and finite. They are evaluated at the codimension$-2$ surface $S$ at $\rho\to 0$ and fixed time. The explicit expression is\footnote{Conventions: this expression -- and others in the following -- are two equations, one with only the plus signs (line above) and the other with only the minus signs (line below). It should be evaluated at fixed time $t$, related to $x^\pm$ via $x^\pm={t\over \ell}\pm \varphi$.}
\beq
H_{\xi^\pm}={\ell\over k}\int_0^{2\pi}\D\varphi L^\pm(x^\pm)\xi^\pm(x^\pm).
\eeq
We will compute their algebra in what follows.
\begin{enumerate}
\item[Q3)] \textbf{Using the modes decomposition
$\xi^\pm=\sum_{m=0}^\infty\alpha^{\pm}_m e^{i m x^\pm}$, and working only with the base generators $\xi^\pm_m(x^\pm)\defeq e^{im x^\pm}$, compute the residual symmetry algebra.}

Explicitly, we compute
\beq
[\xi^{\pm}_m\pa_\pm,\xi^\pm_n\pa_\pm]=[e^{imx^\pm}\pa_\pm,e^{inx^\pm}\pa_\pm]=i(n-m)e^{i(m+n)x^\pm}\pa_\pm=i(n-m)\xi^\pm_{m+n}\pa_\pm,
\eeq
while the mixed commutator vanishes trivially. Thus
\beq\label{lieal}
i[\xi^{\pm}_m\pa_\pm,\xi^\pm_n\pa_\pm]=(m-n)\xi^\pm_{m+n}\pa_\pm, \qquad i[\xi^{\pm}_m\pa_\pm,\xi^\mp_n\pa_\pm]=0.
\eeq
The modes algebra is a double copy Witt algebra.

\item[Q4)] \textbf{Evaluate $\delta_{\xi^\pm}L^\pm$.}

To find $\cL_{\xi^\pm} L^\pm$, we need to compute the next order in $\cL_\xi g_{ab}$. Explicitly
\beqn
\cL_\xi g_{ab}&=& \xi^\mu\pa_\mu g_{ab}+g_{ac}\pa_b\xi^c+g_{bc}\pa_a\xi^c\\
&=&\xi_{(0)}^{c}\pa_c g^{(2)}_{ab}+g^{(2)}_{ac}\pa_b\xi_{(0)}^{c}+g^{(2)}_{bc}\pa_a\xi_{(0)}^{c}\\
&&-{\ell^2\over\rho^2}\eta_{ac}\pa_b\big(\pa_d\sigma\int_0^\rho\D \rho^\prime \rho^\prime \eta^{dc}\big)-{\ell^2\over\rho^2}\eta_{bc}\pa_a\big(\pa_d\sigma\int_0^\rho\D \rho^\prime \rho^\prime \eta^{dc}\big)+O(\rho^2)\\
&=&\cL_{\xi_{(0)}}g^{(2)}_{ab}-\ell^2\pa_a\pa_b\sigma+O(\rho^2).
\eeqn
Comparing orders, and using $\cL_\xi=\delta_\xi$, we obtain 
\beq
\delta_\xi g^{(2)}_{ab}=\cL_{\xi_{(0)}}g^{(2)}_{ab}-\ell^2\pa_a\pa_b\sigma.
\eeq
Using light-cone coordinates, from \eqref{dslc} we read that $g^{(2)}_{\pm\pm}=\ell^2 L^\pm$ and $g^{(2)}_{\pm\mp}=0$. Then
\beq
\cL_{\xi^+} g^{(2)}_{++}=\ell^2 \delta_{\xi^+}L^+=\ell^2(\xi^+_{(0)}\pa_+L^++2 L^+\pa_+\xi^+_{(0)}-{1\over 2}\pa_+
^3\xi^+_{(0)}),
\eeq
and similarly for $g^{(2)}_{--}$, such that the two equations can be written as
\beq\label{delL}
 \delta_{\xi^\pm}L^\pm=\xi^\pm_{(0)}\pa_\pm L^\pm +2 L^\pm\pa_\pm\xi^\pm_{(0)}-{1\over 2}\pa_\pm
^3\xi^\pm_{(0)}\qquad \delta_{\xi^\mp}L^\pm=0.
\eeq
In a CFT, these are the transformation rules of two Virasoro currents. And indeed, as we have seen, the bulk asymptotic symmetries induce the boundary conformal symmetries underpinning the $2-$dimensional CFT, so the stress tensor modes transform as expected. The last term in \eqref{delL} indicates that the transformations are anomalous, that is, there is a non-trivial central charge in the dual CFT. This is what the charge algebra, that we study in the next step, will unveil.
\end{enumerate}
To evaluate the charge algebra, we decompose the Noether charges in modes:
\beq
H_{\xi^\pm}^m={\ell\over k}\int_0^{2\pi}\D \varphi e^{imx^\pm}L^\pm.
\eeq
\begin{enumerate}
\item[Q5)]\textbf{Compute $\{H^m_{\xi^\pm},H^n_{\xi^\pm}\}$ and $\{H^m_{\xi^\pm},H^n_{\xi^\mp}\}$. How is the charge algebra related to the vector fields Lie algebra?}

We start with the chiral/anti-chiral commutator. Since we are in a canonical setup (integrable, conserved, and finite Noether charges), the Poisson bracket generates the field space transformation. Thus, eq. \eqref{HH} gives
\beq
\{H^m_{\xi^+},H^n_{\xi^-}\}=\delta_{\xi^-_n}H^m_+={\ell\over k}\int^{2\pi}_0\D\varphi e^{imx^+}\delta_{\xi^-_n}L^+.
\eeq 
Using now \eqref{delL} we obtain
\beq
\{H^m_{\xi^\pm},H^n_{\xi^\mp}\}=0.
\eeq
This implies that the algebra is a direct sum of the chiral and anti-chiral sectors.

We then evaluate the chiral/chiral commutator. Using again \eqref{delL}, we obtain
\beqn
\{H^m_{\xi^+},H^n_{\xi^+}\}&=&\delta_{\xi^+_n}H^m_+\\
&=&{\ell\over k}\int^{2\pi}_0\D\varphi e^{imx^+}\delta_{\xi^+_n}L^+\\
&=&{\ell\over k}\int^{2\pi}_0\D\varphi e^{imx^+}\left(e^{inx^+}\pa_+L^++2i n L^+e^{inx^+}-{1\over 2}(in)^3e^{inx^+}\right)\\
&=&{\ell\over k}\int^{2\pi}_0\D\varphi e^{i(m+n)x^+}\left(\pa_+L^++2i n L^++{i\over 2}n^3\right).
\eeqn
Given $\int_0^{2\pi}\D \varphi e^{i(m+n)x^+}=2\pi\delta_{m+n,0}$, and  integrating by parts in $\pa_\pm$,\footnote{Integrals are performed at fixed time, and thus total derivatives vanish on the circle.} we find
\beqn
\{H^m_{\xi^+},H^n_{\xi^+}\}&=&{\ell\over k}\left[\int^{2\pi}_0\D\varphi e^{i(m+n)x^+}(-i(m+n)L^++2i n L^+)+{i\over 2}n^32\pi \delta_{m+n,0}\right]\\
&=&{\ell\over k}\int^{2\pi}_0\D\varphi e^{i(m+n)x^+}i(n-m)L^++{i n^3 \ell \over 8G} \delta_{m+n,0}\\
&=&i(n-m)H^{m+n}_{\xi^+}+{in^3\ell\over 8G}\delta_{m+n,0}.
\eeqn
This is our final result. For the anti-chiral/anti-chiral commutator a very similar computation holds, such that we can compactly present the charge algebra as
\beq
i\{H^m_{\xi^\pm},H^n_{\xi^\pm}\}=(m-n)H^{m+n}_{\xi^\pm}+{m^3\ell\over 8G}\delta_{m+n,0}.
\eeq
This algebra is a direct sum of two copies of the Virasoro algebra: $Vir\oplus Vir$. It represents the vector fields Lie algebra \eqref{lieal} projectively, because of the central term ${m^3\ell\over 8G}\delta_{m+n,0}$.\footnote{This is sometimes presented as ${m^2(m-1)\ell\over 8G}\delta_{m+n,0}$. The difference between these two expressions is the  value (offset) at the initial point in field space, which is assumed to be zero here.} So we have explicitly shown the general theorem \eqref{pro} in a specific example, and found the value of the central charge, dictated by the bulk. The physical central charge in the Virasoro algebra is the number $c$ in
\beq
i\{H^m_{\xi^\pm},H^n_{\xi^\pm}\}=(m-n)H^{m+n}_{\xi^\pm}+{c\over 12}m^3\delta_{m+n,0}.
\eeq
And thus we read off
\beq
c={3\ell\over 2G}.
\eeq
This is called the Brown-Henneaux central charge. It is a precursory holographic prediction, that is responsible for the conformal anomaly in the boundary CFT. Indeed, interpreting the bulk tensor $T_{ab}$ as (the expectation value of) the boundary stress tensor in the CFT, eq. \eqref{Tab} predicts an anomaly in the conformal Ward identity in the boundary, dictated by the central charge $c$, which depends on the bulk AdS radius and Newton's constant.
\end{enumerate}

So we have completely solved steps S4 and S5 of Subsec. \ref{sec2.4} for this theory. Without entering details, we want to conclude this exercise with a simplified discussion of the holographic dictionary, here explicitly realized. In AdS/CFT, the bulk asymptotic symmetries coming from diffeomorphisms give rise to codimension$-2$ Noether charges, because they follow Noether's second theorem. These are interpreted from a boundary theory, living on $B$, as codimension$-1$ global charges (and thus follow Noether's first theorem). The boundary theory is dictated to be a field theory on a fixed background metric (thanks to the bulk Dirichlet boundary conditions). Since we have found the bulk asymptotic symmetries to be the conformal Killing vectors on the boundary, the boundary theory is a CFT, and thus this is a AdS$_3$/CFT$_2$ holographic setup. The bulk charge algebra is exactly the double copy Virasoro algebra of the boundary Virasoro currents, and the bulk Einstein equations of motion dictate the value of the central charge, and thus give a holographic prediction on the CFT. While we will argue that a local holography nature permeates the corner theory, it is only in this setup (AdS with Dirichlet boundary conditions) that gravity is completely holographic, namely it is possible to fully reconstruct the bulk from the boundary. This is more subtle in general, especially when radiation is turned on, but tremendous progress has been made in recent years to understand better the holographic nature of gravity. The corner proposal is a step forward in this direction.

\section{Corners}\label{sec3}

Until now, we have deliberately not insisted on a crucial feature of Noether charges stemming from Noether's second theorem. Contrarily to global symmetries, Noether charges for gauge symmetries have support on codimension$-2$ surfaces. We have explicitly seen this in both  exercises. From now on, a generic codimension$-2$ surface will be called a corner. Corners are therefore the fundamental geometric ingredients of gauge theories. Consider a gauge theory without global symmetries, then all the observables and physical degrees of freedom are localized on corners. Thus, \textit{corners are the atomic constituents of gauge theories}. A better understanding of their properties is what led to the  corner proposal, through which we have gained better control on symmetries in classical and quantum gauge theories. 

The Introduction contains a review of the literature on the topic, and we refer to \cite{Donnelly:2016auv, Speranza:2017gxd, Geiller:2017whh, Freidel:2020xyx, Freidel:2020svx, Freidel:2020ayo, Donnelly:2020xgu, Ciambelli:2021vnn, Freidel:2021cjp, Ciambelli:2021nmv, Ciambelli2022a, Donnelly2022} for more details. Even though the ultimate goal of the corner proposal is to address open questions in quantum gravity, this whole Section is devoted to a study of corners in classical gravity. While the previous Section contained standard  material, this and the next Sections are focussed on more recent results. Consequently, we follow more closely certain papers of the author. This Section is mostly based on  \cite{Ciambelli:2021vnn, Ciambelli:2021nmv, Freidel:2021dxw}.

\subsection{Embeddings}\label{sec3.1}

As remarked in \eqref{dO}, we have been cavalier so far in integrating over manifolds like a boundary or a corner. Integrals on these manifolds require a pull-back, which can be non-trivial. A corner $S$ is a $(d-2)-$manifold, with its own atlas and coordinates system. One can then embed such a manifold into the $d-$dimensional bulk $M$. Using coordinates $\sigma^\alpha$ on $S$, ($\alpha=1,\dots,d-2$), and coordinates $y^M$ on $M$, ($M=0,\dots,d-1$), the embedding map $\phi$ is
\beq
\phi:S\to M, \qquad \phi\\
: \sigma^\alpha\mapsto y^M(\sigma).
\eeq
Without introducing a bulk metric, we can adapt the tangent bundle $TM$ to the tangential and normal coordinates. This is achieved splitting the tangent bundle in horizontal and vertical sub-bundles
\beq
TM=Hor\oplus Ver, \qquad \text{Rank}(Hor,Ver)=(d-2,2).
\eeq
We can introduce bulk coordinates $y^M=(u^a,x^i)$ with $(i=1,\dots,d-2)$ and $(a=0,d-1)$. 

Without entering too technical details on the theory of screening distributions, the two most general (modulo rescalings) one-forms that pull-back to zero on the corner are parameterized in these coordinates as
\beq
n^a=\D u^a-a^a_i(u,x)\D x^i.
\eeq
By requiring that $n^a$ annihilates elements of $Hor$, we find that the later is spanned by
\beq
Hor=\text{Span}(D_i\defeq \pa_i+a^a_i\pa_a).
\eeq
The sub-bundle $Hor$ is in general a non-integrable distribution inside $TM$, as long as the one forms $a^a_i\D x^i$ -- called Ehresmann connection -- have non-vanishing curvature. 

This framework in $TM$ allows us to better control embeddings. Indeed, using the bulk coordinates split, we have
\beq
\phi:\sigma^\alpha\mapsto (u^a(\sigma),x^i(\sigma)).
\eeq
Then, denoting the pull-back as
\beq
\phi^*:T^*M\to T^*S,
\eeq
the bulk one-forms $n^a$ span the  bundle $Ver^*$ dual to $Ver$ if and only if
\beq
\phi^*(Ver^*)=\phi^*(n^a)=0, \qquad \phi^*(Hor^*)=T^*S,
\eeq
where with a slightly abuse of notation we wrote pull-backs of dual bundles to indicate that it holds for all elements of the latter. It is important to stress that this construction does not involve any metric on $M$, and holds for any differentiable manifold. The simplest example of an embedding, which we call \textit{trivial embedding}, is given by
\beq
\phi_0:\sigma^\alpha\mapsto (u^a(\sigma)=0,x^i(\sigma)=
\delta^i_\alpha\sigma^\alpha).
\eeq
We will use this embedding in the following, but it is crucial to observe that embeddings are affected by bulk diffeomorphisms. In a passive interpretation of diffeomorphisms, the latter change coordinates in the bulks, and thus the image of the embedding map changes. Conversely, in the active interpretation, a bulk diffeomorphism changes the embedding map. Understanding this has dramatic effects on the covariant phase space.

\subsection{Universal Corner Symmetry}\label{sec3.2}

As we have proven, the charge algebra projectively represents via Poisson brackets the symmetry algebra. In gravity, we are interested in the algebra of diffeomorphisms, given by the Lie bracket. Therefore, if one understands the Lie bracket algebra around corners, and has a canonical covariant phase space theory at hand, then the Poisson bracket of charges gives the same algebra, modulo possible central extensions. 

We wish to study the Lie bracket of vector fields in our setup.
The Lie bracket is an operation defined on a generic differentiable manifold, and does not depend on a particular metric. Using the trivial embedding, an expansion around the embedded corner is thus an expansion in powers of $u^a$, because $u^a(\sigma)=0$. An analytic expansion in $u^a$ is not a priori guaranteed, but we will retain only the leading orders, where one can assume the same. Given $\xi,\zeta\in TM$, we can use the coordinates split and write\footnote{We could have used the adapted basis $(D_i,\pa_a)$, at the price that the vectors become field-dependent in the coordinate basis. It would be interesting to perform this analysis in that case.}
\beq
\xi=\xi^a(u,x)\pa_a+\xi^i(u,x)\pa_i,\qquad \zeta=\zeta^a(u,x)\pa_a+\zeta^i(u,x)\pa_i.
\eeq
Then, the components can be expanded in $u^a-$powers
\beqn
\xi^a(u,x)&=& \xi^a_{(0)}(x)+\xi^a_{(1)b}(x)u^b+\xi^a_{(2)bc}(x)u^bu^c+\dots\\
\xi^i(u,x)&=& \xi^i_{(0)}(x)+\xi^i_{(1)a}(x)u^a+\xi^i_{(2)ab}(x)u^au^b+\dots\\
\zeta^a(u,x)&=& \zeta^a_{(0)}(x)+\zeta^a_{(1)b}(x)u^b+\zeta^a_{(2)bc}(x)u^bu^c+\dots\\
\zeta^i(u,x)&=& \zeta^i_{(0)}(x)+\zeta^i_{(1)a}(x)u^a+\zeta^i_{(2)ab}(x)u^au^b+\dots,
\eeqn
where we assumed that the expansion starts at finite order. This is a working assumption, which turns out to be true in all examples, but it is not motivated by any first principles. In this expansion, each term $(\xi^a_{(n)b_1\dots b_n},\xi^i_{(m)b_1\dots b_m})$ and $(\zeta^a_{(n)b_1\dots b_n},\zeta^i_{(m)b_1\dots b_m})$ is a new symmetry generator, that can potentially give rise to a non-vanishing associated Noether charge on $S$.

Consider now the vector field
\beq
\eta=[\xi,\zeta]=\eta^a(u,x)\pa_a+\eta^i(u,x)\pa_i.
\eeq
It admits the same expansion, but each coefficient can now be computed from the Lie bracket. One can then wonder: given certain $(\xi^a_{(n)b_1\dots b_n},\xi^i_{(m)b_1\dots b_m})$ and $(\zeta^a_{(n)b_1\dots b_n},\zeta^i_{(m)b_1\dots b_m})$ turned on, which orders in $(\eta^a_{(n)b_1\dots b_n},\eta^i_{(m)b_1\dots b_m})$ are non-vanishing? If there are orders in $\eta$ turned on that were not present in $\xi$ and $\zeta$, then the Lie bracket algebra does not close, and by repeatedly acting with the Lie bracket, one generates an  algebra that has infinitely many orders in the expansions. On the other hand, if $\eta$ has the same orders of $\xi$ and $\zeta$ turned on, or less, then the algebra closes on itself, and the expansion truncates to these orders only.

We seek a finitely generated algebra, that is, a finite expansion in powers of $u^a$, that closes on itself, without generating higher orders. By explicitly computing the bracket, it is easy to show that the maximal finitely generated sub-algebra of diffeomorphisms is given by vector fields of the form
\beqn
\xi&=&\xi_{(0)}^i(x)\pa_i+\left(\xi^a_{(0)}(x)+\xi^a_{(1)b}(x)u^b\right)\pa_a,\label{xiucs}\\
\zeta&=&\zeta_{(0)}^i(x)\pa_i+\left(\zeta^a_{(0)}(x)+\zeta^a_{(1)b}(x)u^b\right)\pa_a.
\eeqn
One can check that including any higher order in the expansion would automatically enhance this algebra to an infinite expansion. 

We can then display the Lie bracket of two such vector fields, to show that it closes on itself, and find the explicit algebra. Calling
\beq
\hat\xi_{(0)}=\xi^i_{(0)}\pa_i,\qquad \hat\zeta_{(0)}=\zeta^i_{(0)}\pa_i,
\eeq
we obtain
\beqn\label{maxAl}
\underbrace{\big[\xi,\zeta\big]}_{\ucs}
&=&
\underbrace{\big[\hat\xi_{(0)},\hat\zeta_{(0)}\big]^j\pa_j}_\text{$\dS$}\label{fl1}
\\&&
+\Big[
\underbrace{\hat\xi_{(0)}(\zeta_{(0)}^a)
-\hat\zeta_{(0)}(\xi_{(0)}^a)}_\text{$\dS$ acts on $\RR^2$}
+\underbrace{\zeta^a_{(1)b}\xi^b_{(0)}
-\xi^a_{(1)b}\zeta^b_{(0)}}_\text{$\mathfrak{gl}(2,\RR)$ acts on $\RR^2$}
\Big]\pa_a\label{fl2}
\\&&
+u^b
\Big[-\underbrace{\big[\xi_{(1)},\zeta_{(1)}\big]^a{}_b}_\text{$\mathfrak{gl}(2,\RR)$}
+\underbrace{\hat\xi_{(0)}(\zeta^a_{(1)b})
-\hat\zeta_{(0)}(\xi^a_{(1)b})}_\text{$\dS$ acts on $\mathfrak{gl}(2,\RR)$}
\Big]\pa_a.\label{fl3}
\eeqn
The algebra we have found is
\beq\label{ucs}
\ucs=(\dS\loplus {\mathfrak{gl}}(2,\RR))\loplus \RR^2,
\eeq
where the generators are valued on the corner $S$, that is, are functions of the corner coordinates. Indeed, the first line \eqref{fl1} simply gives the Lie bracket on $S$, so it is just the algebra of diffeomorphisms of the corner (we are using the trivial embedding). In the other two lines the corner diffeomorphism acts semi-directly on $\xi^a_{(0)}(x)$ and $\xi^a_{(1)b}(x)$, which justifies the semi-direct sum appearing in the algebra. In line \eqref{fl2} we observe that $\xi^a_{(1)b}(x)$ acts on $\xi^a_{(0)}(x)$, but we do not have a $\xi^a_{(0)}(x)$, $\zeta^a_{(0)}(x)$ commutator, which implies that these two generators are Abelian, and acted upon by the rest of the algebra. This is the $\RR^2$ part. Finally, line \eqref{fl3} tells us that $\xi^a_{(1)b}(x)$ and $\zeta^a_{(1)b}(x)$ satisfy a $2x2$ matrix algebra, so a $\mathfrak{gl}(2,\RR)$, and they are acted upon by $\dS$. Putting all this together, we arrive to the algebra \eqref{ucs}.
Notice that one can then interpret the $\RR^2$ as the two normal (super)translations in the two normal bulk directions, and the $\mathfrak{gl}(2,\RR)$ as the most general transformation of the $2-$dimensional normal plane.

The algebra \eqref{ucs} is that of the group\footnote{As customary in Lie algebra theory, we indicate Lie groups and algebras with capital and gothic letters, respectively.}
\beq
UCS=(\text{Diff}(S)\ltimes \text{GL}(2,\RR))\ltimes \RR^2,
\eeq
where $UCS$ stands for Universal Corner Symmetry. We refer to this algebra as universal because it has been derived without introducing a metric or a specific dynamics. Clearly, not all metric configurations or dynamics will have non-vanishing Noether charges associated to all generators of this algebra, but we claim that this is the maximal one that can be sourced from a single corner. 

All possible field spaces of a gravitational theory with diffeomorphisms will have  a physical asymptotic symmetric group that is contained in the $UCS$. Reversing the logic, we see that corners and the $UCS$ are the fundamental ingredients of a gravitational theory. While classically there are restrictions coming from boundary conditions and gauge fixing, it is a promising feature that a maximal symmetry group has been found. Indeed, studying this group and its representations, and fusion rules, can lead to an appreciation of symmetries of observables that can be the guiding principle in quantum gravity, although we are yet at the very beginning of this journey. We will focus on this in Section \ref{sec4}. 

Before concluding, we would like to mention two relevant sub-algebras of the $\ucs$ that have appeared in the literature. The first one is the so-called Extended Corner Symmetry ($\ecs$), at finite distance corners:
\beq
\ecs=(\dS\loplus {\mathfrak{sl}}(2,\RR))\loplus \RR^2,
\eeq
which we will show in the next Subsection to be associated to Einstein-Hilbert Noether charges at a generic finite distance corner. This algebra was found in \cite{Donnelly:2016auv} and \cite{Speranza:2017gxd} and further discussed in the literature on corners. It is the biggest sub-algebra of the $\ucs$ for which we have found non-vanishing Noether charges. The other sub-algebra we want to mention is the $\mathfrak{bmsw}$ (W stands for Weyl) sub-algebra
\beq\label{bmsw}
\mathfrak{bmsw}=(\dS\loplus \RR)\loplus\RR,
\eeq
which has been show in \cite{Freidel:2021fxf} to be the algebra at asymptotic corners, for  asymptotically flat Einstein-Hilbert gravity. This algebra in turn contains the Generalized $\mathfrak{bms}$ ($\mathfrak{gbms}$) algebra found in \cite{Campiglia:2014yka},
\beq\label{gbms}
\mathfrak{gbms}=\dS\loplus\RR,
\eeq
the Extended $\mathfrak{bms}$ algebra of \cite{Barnich:2009se}, where the $\dS$ is restricted to the conformal Killing vectors of $S$, and the original $\mathfrak{bms}$ algebra \cite{BONDI1960,Sachs1961,doi:10.1098/rspa.1962.0161,Sachs:1962wk,Sachs1962a,Bondi1964}. We refer to \cite{Ciambelli:2021vnn} for more details on the relevant sub-algebras involved.

\subsection{Application to Finite Distance Corners}\label{sec3.3}

In this Subsection, we want to make a step further and introduce a bulk metric adapted to our tangent bundle split. Then, we can explicitly compute the metric variation under vector fields generating the $\ucs$. 

Any bulk $d-$dimensional metric can be parameterized as
\beq\label{fdc}
\D s^2=h_{ab}(u,x)n^a n^b+\gamma_{ij}(u,x)\D x^i \D x^j,
\eeq
where we remind that $n^a=\D u^a-a^a_i(u,x)\D x^i$ are the one-forms adapted to the $Hor\oplus Ver$ split of $TM$. This is a generic metric parameterization where the metric constituents transform as expected under the $\ucs$. Given that the pull-back of $n^a$ is zero, the leading order of $\gamma_{ij}$ gives the metric on the corner, while $h_{ab}$ is the metric on the normal fibres.

We assume that the corner under scrutiny is located at finite distance in the bulk. This implies that the metric constituents, $(a^a_i,h_{ab},\gamma_{ij})$, admit an expansion in powers of $u^a$, without singular poles. Using the trivial embedding $\phi_0$, we can expand
\beqn\label{fdcexp}
h_{ab}(u,x)&=&h^{(0)}_{ab}(x)+u^c h^{(1)}_{abc}(x) +u^cu^d h^{(2)}_{abcd}(x) +\dots\\
a^a_i(u,x)&=& a^a_{(0)i}(x)+u^ba^a_{(1)ib}(x)+u^bu^ca^a_{(2)ibc}(x)+\dots\\
\gamma_{ij}(x)&=& \gamma^{(0)}_{ij}(x)+u^a \gamma^{(1)}_{ija}(x) +u^au^b \gamma^{(2)}_{ijab}(x) +\dots.
\eeqn

\paragraph{Field variations}
Using $\delta_\xi=\cL_\xi$, we can compute the field variations of each metric constituent, under a vector field $\xi\in\ucs$, that is, \eqref{xiucs}. This is very similar to what we have done for AdS$_3$ in Exercise \ref{sec2.6}. A lengthy yet straightforward computation yields
\beqn
\delta_{{\xi}}h^{(0)}_{ab}&=&
\xi_{(0)}^j\pa_j h^{(0)}_{ab}
+\Big(h^{(0)}_{bc}\xi_{(1)a}^c+h^{(0)}_{ac}\xi_{(1)b}^c\Big)
+\xi_{(0)}^ch^{(1)}_{abc}\label{h0}
\\
\delta_{{\xi}}a_{(0)i}^a&=&
\Big(\xi_{(0)}^j\pa_ja_{(0)i}^a
+a_{(0)j}^a\pa_i\xi_{(0)}^j\Big)
-\xi_{(1)c}^aa_{(0)i}^c
+\Big(-\pa_i\xi_{(0)}^a+\xi_{(0)}^c a_{(1)ic}^a\Big)\label{a0}
\\
\delta_{{\xi}}a_{(1)ib}^a&=&
\Big(\xi_{(0)}^j\pa_ja_{(1)ib}^a
+a_{(1)jb}^a\pa_i\xi_{(0)}^j\Big)
+\Big(-\pa_i\xi_{(1)b}^a
+a_{(1)ic}^a\xi_{(1)b}^c
-\xi_{(1)c}^aa_{(1)ib}^c
\Big)
+\xi_{(0)}^c a_{(2)ibc}^a\label{a1}
\\
\delta_{{\xi}}\gamma^{(0)}_{ij}
&=&
\Big(\xi_{(0)}^k\pa_k \gamma^{(0)}_{ij}
+\gamma^{(0)}_{k j}\pa_i\xi_{(0)}^k
+\gamma^{(0)}_{ki}\pa_j\xi_{(0)}^k\Big)
+\xi_{(0)}^a \gamma^{(1)}_{ija}.\label{g0}
\eeqn
These equations seem complicated at first, but they can be rewritten in a compact form, and they unravel a systematic pattern. First notice that the $\dS$ action can be recast as the corner Lie derivative $\cL_{\hat\xi_{(0)}}$. Indeed, from the viewpoint of $S$, $h_{ab}$ is a scalar (and thus $\cL_{\hat{\xi}_{(0)}}h_{ab}=\xi^j_{(0)}\pa_j h_{ab}$), while $a^a_i$ are the components of a one form on $S$ (and thus $\cL_{\hat{\xi}_{(0)\\
}}a^a_{i}=(\xi_{(0)}^j\pa_ja_{i}^a
+a_{j}^a\pa_i\xi_{(0)}^j)$), for all orders in $u^a$. Lastly, $\gamma_{ij}$ is a rank$-(0,2)$ tensor from the point of view of $S$, so its Lie derivative is 
\beq
\cL_{\hat{\xi}_{(0)}}\gamma_{ij}
=
\xi_{(0)}^k\pa_k \gamma^{(0)}_{ij}
+\gamma^{(0)}_{k j}\pa_i\xi_{(0)}^k
+\gamma^{(0)}_{ki}\pa_j\xi_{(0)}^k,
\eeq
again to all orders in $u^a$. 

Then one can rewrite (\ref{h0}-\ref{g0}) as
\beqn\label{transform}
\delta_{{\xi}}h^{(0)}_{ab}&=&
\cL_{\hat{\xi}_{(0)}}h^{(0)}_{ab}
+\Big(h^{(0)}_{bc}\xi_{(1)a}^c+h^{(0)}_{ac}\xi_{(1)b}^c\Big)
+\xi_{(0)}^ch^{(1)}_{abc}
\\
\delta_{{\xi}}a_{(0)i}^a&=&
\cL_{\hat{\xi}_{(0)}}a_{(0)i}^a
-\xi_{(1)c}^aa_{(0)i}^c
+\Big(-\pa_i\xi_{(0)}^a+\xi_{(0)}^c a_{(1)ic}^a\Big)
\\
\delta_{{\xi}}a_{(1)ib}^a&=&
\cL_{\hat{\xi}_{(0)}}a_{(1)ib}^a
+\Big(-\pa_i\xi_{(1)b}^a
+a_{(1)ic}^a\xi_{(1)b}^c
-\xi_{(1)c}^aa_{(1)ib}^c
\Big)
+\xi_{(0)}^c a_{(2)ibc}^a
\\
\delta_{{\xi}}\gamma^{(0)}_{ij}
&=&
\cL_{\hat{\xi}_{(0)}}\gamma^{(0)}_{ij}
+\xi_{(0)}^a \gamma^{(1)}_{ija}.
\eeqn
The $\mathfrak{gl}(2,\RR)$ generator acts as expected by matrix multiplication, every time an index $(a,b,c,\dots)$ is free (with a plus sign for lower indexes and minus sign for upper indexes). The $\RR^2$ generators have been identified with the two normal translations of the corner. They thus couple a metric constituent (say ${\cal O}_{(n)}$) with the one one step further in the $u^a$ expansion (${\cal O}_{(n+1)}$), as one could have expected, since translations are affine transformations. 

There still two terms in these transformation laws that we have not explained. Suppose we are in a setup where $a^a_i$ is zero to all orders. Then, we have 
\beqn
\delta_{{\xi}}a_{(0)i}^a=
-\pa_i\xi_{(0)}^a
\qquad 
\delta_{{\xi}}a_{(1)ib}^a=
-\pa_i\xi_{(1)b}^a.
\eeqn
This means that $a^a_{(0)i}$ and $a^a_{(1)ib}$ transform non-linearly, as connections.
The interpretation is that $a_{(0)i}^a$ is the $\RR^2-$connection whereas $a_{(1)ib}^a$ is the $\mathfrak{gl}(2,\RR)-$connection. This is perfectly in line with the fact that $a^a_i$ is an Ehresmann connection, needed to preserve the $Hor\oplus Ver$ split in the bulk. Indeed, in the $\ucs$ the $\RR^2$ and $\mathfrak{gl}(2,\RR)$ transformations act non-trivially on $Hor$, and thus the connection shifts non-linearly. For the $\ucs$, higher orders in $a^a_i$ transform again tensorially, without shifts. 

So we have seen that the field space transforms as expected under the $\ucs$. In other words, in this bulk parameterization, every metric constituent is a tensor under the $\ucs$, with the exception of $a_{(0)i}^a$ and $a_{(1)ib}^a$, which are connections. We leave as an exercise to write down the transformation law of higher orders metric constituents. This should be done without computing the Lie derivative, simply applying the logic aforementioned. 

\paragraph{Charge computation} We wish to compute Noether charges in this setup, at finite-distance corners. When the charges are integrable, conserved, and finite, the canonical Noether charge associated to Einstein-Hilbert gravity is
\beq\label{n1}
H_\xi=\int_S\phi^*_0(Q_\xi)=
\int_S\phi^*_0(\star \D g(\xi,.)),
\eeq
where we assumed the trivial embedding, and we stress that there is a pull-back involved, $\phi^*_0$. In this expression, $g(\xi,.)$ is the one-form metric-dual to $\xi$, $g(.,.)$ is the bulk metric, and $\star$ the bulk Hodge duality, such that $\star \D g(\xi,.)$ is a $(d-2)-$form. In our setup, we  generically have that canonical charges at finite-distance corners are finite, but non-integrable and not conserved. However, even in this scenario, the Noether charge is \eqref{n1}, except that it is not canonical. We will assume that \eqref{n1} are our charges, and postpone to later to show on which symplectic structure these charges are canonical, even in the presence of fluxes. As we will see, this leads us to the extended phase space. If the reader found this discussion too technical, and not clear at this stage, one can proceed postulating that \eqref{n1} are the charges to compute, and wait until Subsec. \ref{sec3.4} for a proof of this fact. 

For the computation of \eqref{n1}, it is crucial to insert a general bulk vector field (not just a $\ucs$ generator). Indeed, we wish to find which generators contribute to the charges, and thus we cannot set some of them to zero a priori. This is a logic that we explained in the theory of asymptotic symmetries, Subsec. \ref{sec2.5}. Note that here, however, we are not assuming any boundary conditions nor any particular gauge fixing, and thus the residual symmetries are the whole diffeomorphisms on $M$. Using the $u^a$ expansion of the metric and of the vector fields, and using the trivial embedding $\phi_0$, we invite the reader to show as an exercise (proved in \cite{Ciambelli:2021vnn}) that the charges \eqref{n1} become
\beq\label{ch}
H_\xi=\int_S Vol_S \left(\xi_{(1)b}^aN^b{}_a+\xi^j_{(0)}b_j+\xi^a_{(0)}p_a\right),
\eeq
where $Vol_S={\sqrt{\det \gamma^{(0)}}\over (d-2)!}\epsilon_{\alpha_1\dots\alpha_{d-2}}\D\sigma^{\alpha_1}\wedge \dots
\wedge \D\sigma^{\alpha_{d-2}}$. The quantities $N^a{}_b$, $b_i$, and $p_a$ are called the $\mathfrak{gl}(2,\RR)$, $\dS$, and $\RR^2$ momentum, respectively. Their explicitly expressions in term of the metric constituents are
\beqn\label{momo}
N^b{}_a &=& \sqrt{-\det h^{(0)}}\ h_{(0)}^{bc}\epsilon_{ca}\\
b_j &=& -N^b{}_aa_{(1)jb}^a\\
p_d &=& \tfrac12 N^{a}{}_{c}h_{(0)}^{cb}(h^{(1)}_{dba}-h^{(1)}_{dab}),
\eeqn
with $\epsilon_{ab}$ the $2-$dimensional Levi-Civita symbol.  

We have shown that only vector fields of the form \eqref{xiucs} contribute to the charges. This is a very important result, because it proves in a specific example the statement made in the previous paragraph, that only elements of the $\ucs$ can contribute to the charges. Note nonetheless that only the traceless elements in $\xi^a_{(1)b}$ is present, because $N^a{}_b$ is traceless. This implies that only the $\mathfrak{sl}(2,\RR)$ sub-algebra of $\mathfrak{gl}(2,\RR)$ contributes. Therefore, the asymptotic symmetry algebra is the Extended Corner Symmetry discussed at the end of the previous Subsection, that is, 
\beq
\ecs=(\dS\loplus {\mathfrak{sl}}(2,\RR))\loplus \RR^2,
\eeq
which is compatible to the fact that this is the most general symmetry algebra for finite distance corners,  \cite{Donnelly:2016auv, Speranza:2017gxd}. This algebra is a central result of the corner's literature.

\paragraph{Charge algebra}

As already stressed, we have not derived the charges \eqref{n1} from first principles here, so we do not have a symplectic $2-$form at our disposal. Nevertheless, we have a set of charges and field variations. Thus we can compute the variation of the charges, $\delta_\xi H_\zeta$. Keeping track of the embedding we compute
\beq
\delta_\xi H_\zeta=\delta_\xi\int_S\phi^*_0
(Q_\zeta).
\eeq
Now the question is: does the field variation pass through the pull-back of the embedding? This is what we implicitly assumed so far, see for instance \eqref{dO}. However, we have seen that the embedding depends on the bulk coordinates, so a bulk diffeomorphism does not in general commute. In order to proceed carefully, we write
\beq
\delta_\xi H_\zeta=\delta_\xi\int_S\phi^*_0
(Q_\zeta)=\int_S[(\delta_\xi\phi^*_0
)(Q_\zeta)+\phi^*_0
(\delta_\xi Q_\zeta)].
\eeq
Then, when $\delta_\xi\phi^*_0=0$, our results coincide with Section \ref{sec2}. As we will see in the next Subsection, problems such as integrability arise exactly when $\delta_\xi\phi^*_0$ should not be disregarded. The contribution to the charge algebra coming from the variation of the embedding is crucial, and must be taken it into account.

We would like to compute how the embedding changes under a bulk diffeomorphism. This can be evaluated in full generality by interpreting the diffeomorphism actively or passively. An active interpretation is that the diffeomorphism changes the position of the embedded corner in the bulk, $\delta_\xi \phi_0$. A passive interpretation is that one can use the same embedding, but change the bulk coordinates. These two pictures must be equal and thus, infinitesimally, we have
\beq\label{dqd}
(\delta_\xi\phi^*_0)(Q_\zeta)=-\phi^*_0(\cL_\xi Q_\zeta).
\eeq
This is one of the most important  equations of this Subsection. Introducing a bracket notation,
\beq\label{hh}
\{H_\xi,H_\zeta\}\defeq \delta_\xi H_\zeta=\int_S\phi^*_0(\delta_\xi Q_\zeta-\cL_\xi Q_\zeta),
\eeq
we can explicitly evaluate the right-hand side. If we introduce the smeared charges
\beq
N_\xi\defeq\int_S Vol_S \xi^a_{(1)b}N^b{}_a\qquad b_\xi\defeq\int_S Vol_S \xi^j_{(0)}b_j\qquad p_\xi\defeq\int_S Vol_S \xi^a_{(0)}p_a,
\eeq
then, from \eqref{hh}, using the field variations (\ref{h0}-\ref{g0}), one proves that\footnote{This is a long and technical computation. The author decided not to report details here, in order to have a more fluent discussion, and because they are carefully and explicitly done in \cite{Ciambelli:2021vnn}.}
\beqn
\{b_\xi,b_\zeta\}=b_{[\xi,\zeta]}\quad &\{N_\xi,N_\zeta\}=N_{[\xi,\zeta]}&\quad \{p_\xi,p_\zeta\}=p_{[\xi,\zeta]}=0\\
\{b_\xi,N_\zeta\}=N_{[\xi,\zeta]}\quad &\{N_\xi,p_\zeta\}=p_{[\xi,\zeta]}&\\
\{b_\xi,p_\zeta\}=p_{[\xi,\zeta]}\quad &&
\eeqn
The charge algebra in this case is exactly the $\ecs$, and thus is a centreless (i.e., faithful, non-projective) representation of the asymptotic symmetry algebra. 

This application to finite-distance corners had the goal of clarifying the discussion, and seeing how embeddings are important for the charge algebra. There are two pending questions
\begin{itemize}
\item How can we account for $\delta_\xi\phi$ directly on the field space? In other words, can we find a variational calculus in which $\phi$ is part of the field space?
\item We assumed a definition of the Noether charges, and introduced a charge bracket. Are these coming from a symplectic $2-$form? That is, for which symplectic $2-$form are the brackets introduced here the Poisson brackets?
\end{itemize}
Both questions find an answer in the extended phase space.

\subsection{Integrability: Extended Phase Space}\label{sec3.4}

In gravitational theories, the field space is usually given by the bulk metric constituents (called here $g_{\mu\nu}$), with 
\beq
\fL_{V_\xi}g_{\mu\nu}=I_{V_\xi}\delta g_{\mu\nu}=\delta_\xi g_{\mu\nu}=\cL_{\xi}g_{\mu\nu}.
\eeq
As we have seen, to take into account embeddings, one has to assume that $\delta\phi\neq 0$. Thus, we have to consider an enlarged field space $(g_{\mu\nu},\phi)\in\Gamma$. The theory of variational calculus on this enlarged field space is called the \textit{extended phase space}.

\paragraph{The field $\chi$}
To mathematically implement embeddings on the field space, we introduce a vector field on $M$ which is also a one form on $\Gamma$ 
\beq
\chi\in TM\otimes T^*\Gamma,
\eeq
defined through the equation
\beq\label{defc}
I_{V_\xi}\chi=-\xi.
\eeq
This is how embeddings appear on the field space. Indeed, for any functional ${\cal F}$, we postulate that
\beq\label{car}
(\delta\phi^*)({\cal F})=\phi^*(\cL_{\chi}{\cal F}),
\eeq
where $\cL_{\chi}$ is the spacetime Lie derivative along $\chi$.
Then, contracting both sides with $I_{V_\xi}$, we obtain
\beq
(\delta_\xi\phi^*){\cal F}=-\phi^*(\cL_\xi {\cal F}),
\eeq
which is the generalization of \eqref{dqd}. 

The field $\chi$ is a unusual object. To familiarize with it, we will explicitly derive its variation. Consider a generic integrated functional
\beq
{\cal A}=\int_S\phi^*{\cal F},
\eeq
and take its variation
\beq
\delta {\cal A}=\int_S[(\delta\phi^*){\cal F}+\phi^*(\delta{\cal F})].
\eeq
Using \eqref{car}, this reads
\beq
\delta {\cal A}=\int_S\phi^*(\delta{\cal F}+\cL_\chi{\cal F}).
\eeq
Now take another variation. Since $\delta^2=0$, we gather
\beq
0=\delta^2 {\cal A}=\delta\int_S\phi^*(\delta{\cal F}+\cL_\chi{\cal F}).
\eeq 
To proceed further, we need to compute $\delta (\cL_\chi{\cal F})$. For this, one recalls that $\chi$ is a one form in field space, and thus
\beq
\delta (\cL_\chi{\cal F})=\cL_{\delta\chi}{\cal F}-\cL_\chi \delta{\cal F}.
\eeq
We will systematically use this kind of manipulations in the following. Thanks to this, we can compute
\beqn
0=\delta^2 {\cal A}&=&\delta\int_S\phi^*(\delta{\cal F}+\cL_\chi{\cal F})\\
&=&\int_S[(\delta\phi^*)(\delta{\cal F}+\cL_\chi{\cal F})+\phi^*(\delta^2{\cal F}+\delta(\cL_\chi{\cal F}))]\\
&=&\int_S\phi^*(\cL_\chi\delta{\cal F}+\cL_\chi\cL_\chi{\cal F}+\cL_{\delta\chi}{\cal F}-\cL_\chi \delta{\cal F})\\
&=&\int_S\phi^*((\cL_\chi\cL_\chi+\cL_{\delta\chi}){\cal F}).\label{wha}
\eeqn
The first term satisfies
\beq
\cL_\chi\cL_\chi={1\over 2}[\cL_\chi,\cL_\chi]=\cL_{{1\over 2}[\chi,\chi]},
\eeq
again because $\chi$ is a vector field on $M$, but also a one form on $\Gamma$. Therefore, \eqref{wha} gives
\beqn
0=\int_S\phi^*((\cL_{{1\over 2}[\chi,\chi]}+\cL_{\delta\chi}){\cal F}),
\eeqn
which eventually yields
\beq\label{dc}
\delta\chi=-{1\over 2}[\chi,\chi].
\eeq
This result is an important step toward the understanding of the extended phase space. Since the unfamiliar reader may find these computations difficult, let us write down in coordinates the $[\chi,\chi]$ commutator. In coordinates $y^M$ on $M$ and $\varphi^A$ on $\Gamma$, we can write
\beq
\chi=\chi^M_A \delta\varphi^A \pa_M.
\eeq
Then, we compute
\beq
[\chi,\chi]=[\chi^M_A \delta\varphi^A \pa_M,\chi^N_B \delta\varphi^B \pa_N]=\left(\chi^M_A\pa_M\chi^N_B-
\chi^M_B\pa_M\chi^N_A\right)\delta\varphi^A\wedge \delta \varphi^B \pa_N.
\eeq
The field $\chi$ satisfies a relationship reminiscent of the BRST transformation of a ghost. The latter can be seen as arising from the horizontality condition on the principal fibre bundle of gauge theories, where the ghosts act vertically, see \cite{ThierryMieg1980a}, and \cite{Ciambelli:2021ujl} for a reinterpretation of this result on Atiyah Lie algebroids. Similarly, the curvature of the field $\chi$ satisfies this horizontality condition, when seen as a field-space connection, as in \cite{Gomes:2016mwl}.

\paragraph{Extended $2-$form} To keep track of embeddings, we will consider a subregion $R$ of $M$. Then, using the notation $\phi^*_n$ for a $n-$dimensional embedding, the action for the subregion is
\beq
S_R=\int_R\phi^*_d(L).
\eeq
While one can at this stage assume that $R=M$ and the embedding is the identity map, we will keep track of it, for it plays an important role. This construction is very similar to  \cite{Donnelly:2016auv}, where a reference manifold is introduced, and indeed we will find that our results are related to the results found there via ambiguities. Embeddings are related to edge modes. This can be seen observing that their contribution to the bulk dynamics is only on the corner, as we will prove here-below. 

The variation of the action yields
\beq
\delta S_R=\int_R\phi^*_d(\delta L+\cL_\chi L)\he \int_R\phi^*_d(\D \theta+\D i_\chi L)=\int_{\pa R}\phi^*_{d-1}(\theta+i_\chi L).
\eeq
Given a codimension$-1$ Cauchy hypersurface $\Sigma$, the extended pre-symplectic potential is given by
\beq\label{text}
\Theta^{\text{ext}}\defeq \int_{\Sigma}\phi^*_{d-1}\theta^{\text{ext}}=\int_{\Sigma}\phi^*_{d-1}(\theta+i_\chi L),
\eeq
and reduces to the usual one when $\chi=0$.

From this, we evaluate the extended pre-symplectic $2-$form
\beqn\label{dt}
\Omega^{\text{ext}}=\delta\Theta^{\text{ext}}=\delta\int_{\Sigma}\phi^*_{d-1}(\theta+i_\chi L)
=\int_{\Sigma}\phi^*_{d-1}(\delta(\theta+i_\chi L)+\cL_\chi(\theta+i_\chi L)).
\eeqn
Using that $\cL_\chi\theta=\D i_\chi \theta+i_\chi \D\theta\he \D i_\chi \theta+i_\chi \delta L$, we compute
\beq\label{oex}
\Omega^{\text{ext}}\he\int_{\Sigma}\phi^*_{d-1}(\delta\theta+i_{\delta\chi} L-\cancel{i_\chi\delta L}+\D i_\chi \theta+\cancel{i_\chi \delta L}+\cL_\chi i_\chi L).
\eeq
The last term can be processed and, using the identity $i_{{[\xi,\zeta]}}=[\cL_\xi,i_{\zeta}]$, gives
\beq
\cL_\chi i_\chi L=[\cL_\chi,i_{\chi}]L-i_\chi\cL_\chi L=i_{[\chi,\chi]}L-i_\chi \D i_\chi L=i_{[\chi,\chi]}L-\cL_\chi i_\chi L+\D(i_\chi i_\chi L),
\eeq
which implies
\beq
\cL_\chi i_\chi L={1\over 2}i_{[\chi,\chi]}L+{1\over 2}\D(i_\chi i_\chi L).
\eeq
Injecting this in \eqref{oex} and utilizing \eqref{dc} we obtain
\beqn
\Omega^{\text{ext}}&\he &\int_{\Sigma}\phi^*_{d-1}(\delta\theta-{1\over 2}i_{[\chi,\chi]} L+\D i_\chi \theta+{1\over 2}i_{[\chi,\chi]}L+{1\over 2}\D(i_\chi i_\chi L))\\
&=&\int_{\Sigma}\phi^*_{d-1}(\delta\theta+\D (i_\chi \theta+{1\over 2}i_\chi i_\chi L)).
\eeqn
Given the definition of the non-extended $2-$form \eqref{defo}, we finally get
\beq\label{oext}
\Omega^{\text{ext}}=\Omega+\int_{\Sigma}\phi^*_{d-1}\D (i_\chi \theta+{1\over 2}i_\chi i_\chi L)=\Omega+\int_{\pa\Sigma}\phi^*_{d-2}(i_\chi \theta+{1\over 2}i_\chi i_\chi L)=\Omega+\Omega^\chi,
\eeq
where we defined
\beq
\Omega^\chi\defeq \int_{\pa\Sigma}\phi^*_{d-2}(i_\chi \theta+{1\over 2}i_\chi i_\chi L).
\eeq
This contribution is entirely due to the field $\chi$, so it is present only in the extended phase space. It is a corner -- or edge -- contribution, because it is a codimension$-2$ integral. Therefore, although we used embeddings like $\phi_d$ and $\phi_{d-1}$, it is only the corner embedding $\phi_{d-2}$ that contributes to the extended pre-symplectic $2-$form. The extended $2-$form is a central feature of the extended phase space, which will lead to integrable charges for all diffeomorphisms.

\paragraph{Integrability}

We wish to contract \eqref{oext} with an arbitrary diffeomorphism, to obtain the canonical Noether charges, as we did in Section \ref{sec2} for the non-extended case. 

This is a long computation, that we  report here in detail. It is wiser to contract $\Omega^{\text{ext}}$ as it appears in \eqref{dt}:
\beqn
I_{V_\xi}\Omega^{\text{ext}}&=&I_{V_\xi}\int_{\Sigma}\phi^*_{d-1}(\delta(\theta+i_\chi L)+\cL_\chi(\theta+i_\chi L))\\
&=&\int_{\Sigma}\phi^*_{d-1}(I_{V_\xi}\delta\theta+I_{V_\xi}\cL_\chi\theta+I_{V_\xi}\cL_\chi(i_\chi L)+I_{V_\xi}i_{\delta\chi}L-I_{V_\xi}i_\chi\delta L).\label{ivoext}
\eeqn
The first term can be processed into
\beq
I_{V_\xi}\delta\theta=\fL_{V_\xi}\theta-\delta I_{V_\xi}\theta=\cL_{\xi}\theta-\delta I_{V_\xi}\theta,
\eeq
where we used that $\fL_{V_\xi}\theta=\cL_{\xi}\theta$.\footnote{In this derivation, we assume that there are no anomalies, in the language of \cite{Hopfmuller:2018fni}, and subsequent works \cite{Chandrasekaran:2020wwn, Freidel:2021dxw}. We also assume that the vector fields are field-independent. This computation has been performed without these assumptions in \cite{Freidel:2021dxw, Speranza:2022lxr}.}
The fourth term in \eqref{ivoext} can also be reduced, using \eqref{defc} and \eqref{dc}, to
\beq
I_{V_\xi}i_{\delta\chi}L=-{1\over 2}I_{V_\xi}i_{[\chi,\chi]}L={1\over 2}i_{[\xi,\chi]}L-{1\over 2}i_{[\chi,\xi]}L=i_{[\xi,\chi]}L.
\eeq
Inserting these results into \eqref{ivoext}, repeatedly using that $I_{V_\xi}$ anti-commutes with $\chi$, and applying \eqref{defc}, we obtain
\beqn
I_{V_\xi}\Omega^{\text{ext}}
&=&\int_{\Sigma}\phi^*_{d-1}(I_{V_\xi}\delta\theta+I_{V_\xi}\cL_\chi\theta+I_{V_\xi}\cL_\chi(i_\chi L)+I_{V_\xi}i_{\delta\chi}L-I_{V_\xi}i_\chi\delta L)\\
&=&\int_{\Sigma}\phi^*_{d-1}(\cancel{\cL_{\xi}\theta}-\delta I_{V_\xi}\theta-\cancel{\cL_\xi\theta}-\cL_\chi I_{V_\xi}\theta+I_{V_\xi}\cL_\chi(i_\chi L)+i_{[\xi,\chi]}L-I_{V_\xi}i_\chi\delta L)\\
&=&\int_{\Sigma}\phi^*_{d-1}(-\delta I_{V_\xi}\theta-\cL_\chi I_{V_\xi}\theta-\cL_\xi(i_\chi L)+\cL_\chi(i_\xi L)+i_{[\xi,\chi]}L+i_\xi\delta L+i_\chi I_{V_\xi}\delta L).\label{step}
\eeqn
The third term in \eqref{step} can be further manipulated,
\beq
\cL_\xi(i_\chi L)=i_{[\xi,\chi]}L+i_\chi \cL_{\xi}L=i_{[\xi,\chi]}L+i_\chi \fL_{V_\xi}L,
\eeq
while the last term gives
\beq
i_\chi I_{V_\xi}\delta L=i_\chi \fL_{V_\xi} L.
\eeq
With these results, \eqref{step} becomes
\beqn
I_{V_\xi}\Omega^{\text{ext}}
&=&\int_{\Sigma}\phi^*_{d-1}(-\delta I_{V_\xi}\theta-\cL_\chi I_{V_\xi}\theta-\cL_\xi(i_\chi L)+\cL_\chi(i_\xi L)+i_{[\xi,\chi]}L+i_\xi\delta L+i_\chi I_{V_\xi}\delta L)\\
&=&\int_{\Sigma}\phi^*_{d-1}(-\delta I_{V_\xi}\theta-\cL_\chi I_{V_\xi}\theta-\cancel{i_{[\xi,\chi]}L}-\cancel{i_\chi \fL_{V_\xi}L}+\cL_\chi(i_\xi L)+\cancel{i_{[\xi,\chi]}L}+i_\xi\delta L+\cancel{i_\chi \fL_{V_\xi} L})\nonumber\\
&=&-\int_{\Sigma}\phi^*_{d-1}(\delta(I_{V_\xi}\theta-i_\xi L)+\cL_\chi(I_{V_\xi}\theta-i_\xi L)).
\eeqn
We have thus isolated the weakly-vanishing local Noether current $J_{V_\xi}$, see \eqref{noe}. This equation can then be rewritten as
\beq
I_{V_\xi}\Omega^{\text{ext}}
=-\int_{\Sigma}\phi^*_{d-1}(\delta J_{V_\xi}+\cL_\chi J_{V_\xi}).
\eeq
The second term in this expression is exactly what is needed to account for the variation of the embedding, \eqref{car}. Therefore, we have shown that
\beq\label{inte}
I_{V_\xi}\Omega^{\text{ext}}
=-\int_{\Sigma}\phi^*_{d-1}(\delta J_{V_\xi}+\cL_\chi J_{V_\xi})=-\delta\int_{\Sigma}\phi^*_{d-1} J_{V_\xi}\he -\delta\int_{\pa\Sigma}\phi^*_{d-2} Q_{\xi}=-\delta H_{\xi},
\eeq
where we used Noether's second theorem to show that this is entirely a corner contribution.
Equation \eqref{inte} is the most important result of this Section. It proves that all Noether charges associated to diffeomorphisms are integrable on the extended phase space. It further shows that the canonical charge is exactly the Noether charge stemming from Noether's second theorem. In order words, all diffeomorphisms are symplectomorphisms of the extended pre-symplectic $2-$form $\Omega^{\text{ext}}$. We are finally able to answer the questions at the end of Subsec. \ref{sec3.3}. For the first question, the variation of the embedding ($\delta_\xi\phi$) is accounted for in the field space thanks to $\chi$. Concerning the second question, the Noether charges that we postulated in that Subsection are the canonical charges on the extended phase space, and the bracket we used is nothing but the Poisson bracket for $\Omega^{\text{ext}}$. Thus, the whole $\ecs$ is canonically realized at finite-distance corners, using the extended phase space. 

Before concluding, we observe that the charges can be non-conserved, but this fact is fully accounted for by the extended pre-symplectic potential. Indeed, we compute as in \eqref{concon},
\beq
H_{\xi}\vert_{S_2}-H_{\xi}\vert_{S_1}=\int_{S_1}^{S_2}\phi^*_{d-1}\D Q_{\xi}=\int_{S_1}^{S_2}\phi^*_{d-1}(I_{V_\xi}\theta-i_{\xi}L)=\int_{S_1}^{S_2}\phi^*_{d-1}I_{V_\xi}(\theta+i_{\chi}L)=I_{V_\xi}\Theta_{S1,S2}^{\text{ext}},
\eeq
where we used \eqref{text}. So we have disentangled integrability from non-conservation. The former is always attained on the extended phase space, while the latter instructs us about physical fluxes in the spacetime. 

This is the framework that we advocated in Subsec. \ref{sec2.5} to be the resolution number $2$ to the problem of integrability. Clearly, the system at hand is still dissipative, but dissipation is now associated to non-trivial variations of the embedding. It is remarkable that embeddings can cure integrability. It seems that they know -- or can keep track of -- the dissipative degrees of freedom in the evolution of the corner. We here showed this fact mathematically, but we are still at a very primitive stage of understanding of the physics behind it. It is an extremely thriving and exciting avenue of investigation, that touches upon the physics of dissipative systems, and how to account for dissipation on the field space. 

 Given a classical system, the Poisson bracket algebra is the structure that is most amenable to quantization. Thus, the fact that the extended phase space provides a Poisson bracket even in the presence of dissipation is crucial for quantization, and it is one of reasons why the author believes that this is an interesting path to pursue. We return to this in Section \ref{sec4}, where we eventually formalize the corner proposal.

\subsection{Exercise: DGGP in our Approach}\label{sec3.5}

The general finite-distance discussion of Subsec. \ref{sec3.3} is applicable to asymptotic symmetries at the black hole horizon. We show this as an exercise, that will raise some interesting questions on the gauge-fixing approach. 

We follow closely the framework of the paper by Donnay, Giribet, Gonzalez, and Pino (DGGP), \cite{Donnay:2015abr}, where it was found that the near-horizon asymptotic symmetry group is $BMS$. For concreteness, we will confine our attention to $3-$dimensional Einstein gravity. We use Gaussian null coordinates, such that the line element, in coordinates $x^\mu=(\rho,v,\varphi)$, can be written as
\beq\label{dggp}
\D s^2=g_{\mu\nu} \D x^\mu\D x^\nu=f(\rho,v,\varphi)\D v^2+2 k(\rho,v,\varphi) \D v\D \rho+2 h(\rho,v,\varphi)\D v \D \varphi+R^2(\rho,v,\varphi)\D \varphi^2.
\eeq
The coordinate $\rho$ is the radial coordinate, $v$ is the null time, and $\varphi$ is the $2\pi-$periodic angular coordinate. 

To connect with the steps of Subsec. \ref{sec2.4}, the theory is $3-$dimensional Einstein gravity (S1), the line element \eqref{dggp} is in the Gaussian null gauge (S3), the boundary is located at $\rho\to 0$, and the falloffs (S2) are given by
\beqn\label{dggpexp}
f(\rho,v,\varphi)&=&-2\kappa\rho+O(\rho^2)\\ k(\rho,v,\varphi)&=& 1+O(\rho^2)\\
h(\rho,v,\varphi)&=& \theta(\varphi)\rho+O(\rho^2)\\
R^2(\rho,v,\varphi)&=& \gamma^2(\varphi)+\lambda(\varphi)\rho+O(\rho^2).
\eeqn
This completely specifies the theory (S1-S3) that we are dealing with.

\begin{enumerate}
\item[Q1)] \textbf{Compute the induced metric at $\rho\to 0$, and the norm of $\pa_v$.}

Using all the ingredients above, we get
\beq
\D s^2\vert_{\rho =0}=\gamma^2(\varphi)\D \varphi^2,
\eeq
for all values of $v$. Thus, the induced metric at $\rho\to 0$ is degenerate in time, which signals the presence of a $2-$dimensional Carrollian manifold -- a null hypersurface. Furthermore, we compute the norm of $\pa_v$
\beq
g(\pa_v,\pa_v)=f(\rho,v,\varphi)=-2\kappa\rho+O(\rho^2), \quad \Longrightarrow \quad \lim_{\rho\to 0}g(\pa_v,\pa_v)=0.
\eeq
Therefore $\pa_v$ is null on the Carrollian manifold. This is the structure of a black hole horizon, which is thus the boundary under scrutiny, at $\rho=0$. This boundary is at finite distance, because there are no non-removable poles in the radial coordinate. 
\end{enumerate}
Since this is a finite-distance configuration, we can use the general results of Subsec. \ref{sec3.3}.
\begin{enumerate}
\item[Q2)] \textbf{Comparing \eqref{dggp} with \eqref{fdc}, find which metric constituents are turned on in \eqref{fdcexp}.}

We have to adapt the notation of \eqref{fdc} to this $3-$dimensional setup. The corner that we are interested in is the circle on the horizon at fixed value of time $v$. Thus, the normal coordinates are $u^a=(\rho,v)$ while $x^i=\varphi$ is the coordinate on the $1-$dimensional embedded corner. Eq. \eqref{fdc} then reads
\beq
\D s^2=h_{ab}(\D u^a-a^a \D \varphi)(\D u^b-a^b \D \varphi)+\gamma_{\varphi\varphi}\D \varphi^2.
\eeq
Since $g_{\rho\rho}=0$, we get that $h_{\rho\rho}=0$, and thus
\beq
\D s^2=h_{vv}(\D v-a^v \D \varphi)(\D v-a^v \D \varphi)+2h_{\rho v}(\D \rho -a^\rho \D \varphi)(\D v-a^v \D \varphi)+\gamma_{\varphi\varphi}\D \varphi^2.
\eeq
Comparing with \eqref{dggp}, we  gather 
\beq
h_{vv}=f\qquad h_{\rho v}=k\qquad a^v=0\qquad a^\rho=-{h\over k}\qquad \gamma_{\varphi\varphi}=R^2.
\eeq
Expanding in powers of $\rho$ and $v$, as in \eqref{fdcexp}, and comparing with \eqref{dggpexp}, we read off that the non-vanishing metric constituents up to order $\rho$ are
\beq\label{dggpmc}
h^{(1)}_{vv\rho}=-2\kappa \qquad h^{(0)}_{v\rho}=1 \qquad a^\rho_{(1)\rho}=-\theta(\varphi) \qquad \gamma^{(0)}_{\varphi\varphi}=\gamma^2(\varphi)\qquad \gamma^{(1)}_{\varphi\varphi\rho}=\lambda(\varphi).
\eeq
This is a simple realization of the generic parameterization we presented in Subsec. \ref{sec3.3}.
\end{enumerate}
Although we will not explicitly solve it here, the reader is invited to show that the residual vector fields (S4) are of the form
\beq\label{bms3gf}
\xi=T(\varphi)\pa_v+{\theta(\varphi)\over 2\gamma^2(\varphi)}T^\prime(\varphi)\rho^2\pa_\rho+(Y(\varphi)-{1\over \gamma^2(\varphi)}T^\prime(\varphi)\rho+{\lambda(\varphi)\over 2\gamma^4(\varphi)}T^\prime(\varphi)\rho^2)\pa_\varphi+O(\rho^3).
\eeq
The explicitly form, and metric-dependency, of this vector field comes from requiring that it preserves the gauge. Indeed, the equation $\cL_\xi g=0$ inevitably makes the vector field metric-dependent. On the other hand, if one writes only the components and orders containing independent generators, one gets
\beq\label{bms3}
\xi=T(\varphi)\pa_v+Y(\varphi)\pa_\varphi.
\eeq
This is where the gauge-fixing approach clashes with Subsec. \ref{sec3.3}, in which the bulk metric is not gauge-fixed. Requiring to preserve a gauge leads to field-dependent vector fields, $\delta\xi\neq 0$. We know how to deal with this in the extended phase space (see \cite{Freidel:2021dxw}), but we prefer to continue following the analysis \ref{sec3.3}, and show that the charges turn out to be equal to those computed in \cite{Donnay:2015abr} in the gauge-fixing approach. This allows us to establish a general gauge-free framework.  

\begin{enumerate}
\item[Q3)] \textbf{Compare \eqref{bms3} with \eqref{xiucs}, what is the algebra of residual diffeomorphisms?}

In $3$ dimensions, \eqref{xiucs} reads
\beq
\xi=\xi^\varphi_{(0)}\pa_\varphi+\xi^a_{(1)b}u^b\pa_a+\xi^a_{(0)}\pa_a.
\eeq
Therefore, in order to obtain \eqref{bms3}, we have to set
\beq\label{ide}
\xi^\varphi_{(0)}=Y(\varphi)\qquad \xi^a_{(1)b}=0\qquad \xi^v_{(0)}=T(\varphi)\qquad \xi^\rho_{(0)}=0.
\eeq
Notice that we would have reached the same conclusion if we considered \eqref{bms3gf} instead, except that \eqref{bms3gf} contains other terms that are not in the $\ucs$, which nonetheless do not contain new independent generators. The algebra we obtain is thus $\dS\loplus \RR$, which is exactly the generalized $\mathfrak{bms}$ algebra, see \eqref{gbms}. The factor $Y(\varphi)$ generates $\dS$, while the function $T(\varphi)$ is a supertranslation in the null direction.

\item[Q3)] \textbf{Using the trivial embedding} $\varphi_0:\sigma\mapsto (u^a(\sigma)=0,\varphi(\sigma)=\sigma)$, \textbf{compute the charges.}

This is a simple application of \eqref{ch}. The corner is the $1-$dimensional circle spanned by the coordinate $\sigma$, and $Vol_S=\gamma(\sigma)\D \sigma$. Using \eqref{ide}, equation \eqref{ch} becomes
\beq
H_\xi=\int_0^{2\pi} \D \sigma \gamma(\sigma) \left(Y(\sigma)b_\varphi+T(\sigma)
p_v\right).
\eeq
Computing \eqref{momo} with \eqref{dggpmc}, we gather (conventions: $\epsilon_{v\rho}=-1$)
\beq
N^v{}_v=1\qquad N^\rho{}_\rho=-1\qquad b_\varphi=-\theta(\sigma)\qquad p_\rho=0\qquad p_v=2\kappa.
\eeq
Therefore, we obtain the charges
\beq
H_\xi=\int_0^{2\pi} \D \sigma \gamma(\sigma) \left(2\kappa T(\sigma)-Y(\sigma)\theta(\sigma)\right).
\eeq
These are exactly the charges found in \cite{Donnay:2015abr}, but they come here from the application of the general finite-distance analysis of Subsec. \ref{sec3.3}, rather than a top-down charge computation. The momentum in the $v$ direction, $\kappa$, is the black hole surface gravity. Its interpretation as the response of the system to a horizon supertranslation sheds light on the physics of near-horizon asymptotic symmetries. 
\end{enumerate}
We have found that the asymptotic symmetries are exactly given by \eqref{bms3}. The fact that our charges coincide with \cite{Donnay:2015abr} could have been anticipated, because we are here in a setup where charges are integrable, conserved, and finite, and thus the extended phase space calculations coincide with those of Section \ref{sec2}. To recapitulate, we have here re-derived the appearance of the $\mathfrak{bms}_3$ near-horizon symmetry using the extended phase space, in the gauge-free approach of Subsec. \ref{sec3.3}. Clearly, it would be interesting to apply our method to other constructions of asymptotic symmetries in the literature (see the Introduction for a complete list of examples). We expect that this could both clarify old results and allow us to gather a better understanding of the physics of embeddings, especially in contexts where integrability does not hold in the non-extended phase space, due to the presence of fluxes.
 
\section{The Proposal}\label{sec4}

We have gathered all the necessary ingredients to finally enunciate the corner proposal. The main idea has  already been discussed in the Introduction, and at various stages of Section \ref{sec3}. While so far we focussed on classical gravity, we we will explore in this Section potential outcomes of the corner proposal  in quantum gravity. 

The corner proposal is discussed in these papers \cite{Freidel:2015gpa, Donnelly:2016auv, Speranza:2017gxd, Geiller:2017whh, Freidel:2020xyx, Freidel:2020svx, Freidel:2020ayo, Donnelly:2020xgu, Ciambelli:2021vnn, Freidel:2021cjp, Ciambelli:2021nmv, Ciambelli2022a, Donnelly2022}, and this Section follows closely \cite{Ciambelli2022a}. Once the proposal enunciated, we will initiate a study of the $\ucs$ algebra. This is not meant to be a top-down analysis, but rather a survey of important mathematical tools that one needs to address the problem, such as the coadjoint orbit method and the theory of Atiyah Lie algebroids. Independently from applications to the corner proposal, these  are relevant tools that are permeating the theoretical physics community.

\subsection{The Main Idea}\label{sec4.1}

Classical gravity is not renormalizable, and direct attempts to quantize gravity lead to pathologies. As per the core of these lectures, we will use symmetries as a guiding principle. Gauge symmetries are classical redundancies of the system that are not expected to survive in quantum gravity. This applies in particular to diffeomorphisms that are pure gauge (trivial transformations in the language of \ref{sec2.4}). On the other hand, we have seen that there are diffeomorphisms that are asymptotic symmetries, carrying a non-vanishing Noether charge. In pure gravity, Noether charges are the only observables, organised according to the charge algebra. Therefore, the symmetries associated are not redundancy, they physically act on the field space. The corner proposal focuses on these symmetries, and posits that they survive in quantum gravity, and become the organising principle for quantum observables. This is clearly a long road, and we are merely starting to appreciate the far-reaching consequences of this proposal. Let us state it more rigorously. 

The direction we followed so far is schematically the following
\beq
\raisebox{.5em}{\text{Lagrangian Theory}}\searrow\raisebox{-.5em}{\text{Symplectic Structure at Boundaries}}\raisebox{-1em}{$\searrow$}\raisebox{-1.5em}{\text{Charges and Algebra at Corners}}\label{road}
\eeq
Along the way, we have established two fundamental results. The first is the existence of a universal symmetry group, the $UCS$, such that any classical Lagrangian theory sources one of its sub-groups at corners. The second is the extended phase space, such that there exists a symplectic structure on which all $UCS$ symmetries are canonically realized. This implies that for any Lagrangian theory, the road \eqref{road} leads to integrable charges and thus a well-defined charge algebra at corners. Given these two results, in the corner proposal we break ties with the Lagrangian theory and the symplectic structure at boundaries, and declare that\footnote{Forgetting the bulk in favour of a lower-dimensional geometry is a very primitive notion of holography.}
\begin{center}
Corner Proposal: \textit{
A gravitational theory is described by a set of charges and their algebra at corners.}
\end{center}
Loosing ties with the previous steps in  \eqref{road} automatically infers three important features. First, we get rid of trivial diffeomorphisms, thanks to the quotienting of Subsec. \ref{sec2.4}.\footnote{These diffeomorphisms must re-appear in the semi-classical bulk reconstruction, similar to what happens in the mean-field approximation in statistical mechanics.} Secondly, there is no classical metric to begin with, but only a set of charges. This is more reminiscent of quantum mechanics, where the fundamental ingredients are a Hamiltonian and the complete set of commuting observables. Charges at corners can indeed be promoted to quantum operators, whereas one cannot do so for a full bulk metric, because of pure gauge directions and fields, that are instead peeled off in the corner proposal. Third, and lastly,  corners can provide a discrete geometry in the normal planes. Indeed, given two corners and their charges, there is no a priori guarantee that the two corners are connected by a smooth $d-$dimensional space. This is a peculiar discretization program, because only the normal $2-$dimensional space is discrete. It is promising due to the unique features that characterize $2-$dimensional geometries.

In practice, the corner proposal focuses on charges and corners. How does spacetime emerge, as opposed to a quantum theory of charges? This question is at the core of nowadays investigations on corners, and it is addressed by studying the representations of the charge algebra. Ideally, a full understanding of the representation theory of the $\ucs$ will lead us to an appreciation of representations that are more suitable for a quantum theory:
\beqn
&\text{Charges and Algebra at Corners}\longrightarrow \text{Representations}\raisebox{1em}{$\nearrow$}\raisebox{1.5em}{\text{Classical Spacetime \ \ \ \ \ }}&\\
& \hspace{9.8cm}\raisebox{1.5em}{$\searrow$}\raisebox{1em}{\text{Quantum Representations?}}&\nonumber
\eeqn
A classical gravitational theory with a metric and diffeomorphisms is here a particular representation and a limiting procedure, as it should be for classical physics in a quantum world. We stress how reformulating gravity in this language brings it closer to quantum mechanics: it is a theory of operators (charges) and commutator relationships, and the focus is on the algebraic representation theory.

In the following of this Section, we will initiate the study of representations for the $\ucs$. Clearly, we are at an early stage of development of the corner proposal. It is evident that this is a long and difficult road, and success is by no means guaranteed. On the other hand, this proposal is based on very few assumptions, and one could say that it is just a deep appreciation of Noether's second theorem. This is why we believe that it is a solid avenue of research that set a proper stage to address questions in quantum gravity. 

\subsection{UCS, ECS, ACS, and Local Counterparts}\label{sec4.2}

As we discussed, the first step in the corner proposal is the study of representations for the $UCS$ and its physically relevant sub-groups, as described in Subsec. \ref{sec3.2}. 

First, note that the $\mathfrak{gl}(2,\RR)$ algebra in the $\ucs$ is a direct sum of $\mathfrak{sl}(2,\RR)$ and an Abelian factor, that we call $\mathfrak{w}$ (for Weyl) because it is a trace generator:
\beq
\ucs=(\dS\loplus (\mathfrak{w}\oplus{\mathfrak{sl}}(2,\RR))\loplus \RR^2.
\eeq
There are two main sub-algebras that we have identified. The first is the $\ecs$
\beq
\ecs=(\dS\loplus {\mathfrak{sl}}(2,\RR))\loplus \RR^2,
\eeq
which is the maximal algebra at finite distance corners. The second is $\mathfrak{bmsw}$, although we observe two facts. First, more generally we could consider an algebra for asymptotic corners that has both supertranslations turned on.\footnote{The reduction of $\acs$ to $\mathfrak{bmsw}$ is always possible on the coadoint orbits.} Secondly, one can show that the $\RR$ factor in $\mathfrak{bmsw}$ is exactly the trace generator $\mathfrak{w}$. Thus, we introduce the sub-algebra
\beq
\acs=(\dS\loplus \mathfrak{w})\loplus \RR^2,
\eeq
where $\acs$ stands for ``asymptotic corner symmetry". 

Therefore, the $\ecs$ and $\acs$, that are the maximal symmetry groups at finite and asymptotic corners, realize complementary sectors of the $\mathfrak{gl(2,\RR)}$ sub-algebra
\beqn
&\ucs=(\dS\loplus (\mathfrak{w}\oplus{\mathfrak{sl}}(2,\RR))\loplus \RR^2&\\
&\ecs=(\dS\loplus (\qquad  {\mathfrak{sl}}(2,\RR))\loplus \RR^2&\\
&\acs=(\dS\loplus (\mathfrak{w}\qquad \qquad \ \ 
)\loplus \RR^2&.
\eeqn
Although we do not have a Lagrangian theory that realizes the full $\ucs$ at a single corner, we see that studying this algebra allows us to understand both asymptotic and finite-distance corners. 

We can introduce a basis for the $\ucs$ given by the $\dS$ basis $\pa_\alpha$, a $2x2$ matrix $t^a{}_b$ for $\mathfrak{gl}(2.\RR)$, and a  $2-$dimensional vector $t_a$ for the $\RR^2$. The algebra structure is then obtained through the brackets
\beqn\label{comm}
[\pa_\alpha,\pa_\beta]=0 \quad [\pa_\alpha,t^a{}_b]=0=[\pa_\alpha,t_a]\quad [t^a{}_b,t^c{}_d]=\delta^c_b t^a{}_d-\delta^a_d t^c{}_b\quad [t^a{}_b,t_c]=-\delta^a_c t_b\quad [t_a,t_b]=0.
\eeqn
The semi-direct action of $\dS$ on the other sectors of the algebra is entirely given by the action of the $\dS$ basis on the components of elements of the $\ucs$, that are indeed valued on the corner $S$. An element of the algebra $\chi\in\ucs $ is therefore given by\footnote{The algebra is a vector space, so elements of the algebra are vectors that can be expressed in this suitable basis.}
\beq\label{chi}
\chi=\xi^\alpha(\sigma)\pa_\alpha +\theta^a{}_b (\sigma) t^b{}_a + b^a (\sigma) t_a.
\eeq
The restriction to $\ecs$ and $\acs$ can be performed algebraically by imposing that $t^a{}_b$ is traceless, or full trace, respectively. Let us see this concretely. The basis $t^a{}_b$ and $t_a$ are abstract, but we can use a spacetime representation such that the index $a$ is a coordinate index for a $2-$dimensional space and $u^a=(u,\rho)$ are the coordinates. Then, we decompose 
\beq
t^a{}_b \mapsto (w,t_3,t_+,t_-) \qquad w=u\pa_u+\rho\pa_\rho, \quad t_3=u\pa_u-\rho\pa_\rho \quad t_+=u\pa_\rho \quad t_-=\rho\pa_u.
\eeq
Using the spacetime Lie bracket, one shows that $(t_3,t_+,t_-)$ form a $\mathfrak{sl}(2,\RR)$ algebra while $\mathfrak{w}$ is an Abelian ideal. We refer to the latter as a trace operator, because it can be written as $\delta^a{}_bt^b{}_a$, while all $\mathfrak{sl}(2,\RR)$ have vanishing trace. Then, the restriction to $\ecs$ and $\acs$ is achieved by simply disregarding $\mathfrak{w}$ or $(t_3,t_+,t_-)$, respectively.

Therefore, the $\ucs$ generator \eqref{chi} is the central object to inspect. The $\dS$ part acts derivatively on the components, and this creates a technical challenge in the study of representations. It is wiser to start studying the algebra without the $\dS$ part, which can either be seen as a situation in which the corner is $0-$dimensional ($2-$dimensional gravity), or as the algebra analysis point by point on $S$. We thus define the algebras
\beqn
\ucs \to \h \defeq (\mathfrak{w}\oplus{\mathfrak{sl}}(2,\RR))\loplus \RR^2\qquad
\ecs \to \h_s \defeq {\mathfrak{sl}}(2,\RR)\loplus \RR^2 \qquad 
\acs \to \h_w \defeq  \mathfrak{w}\loplus \RR^2.
\eeqn
These algebras are now considerably simpler. They have $6$, $5$, and $3$ generators, respectively. 

We claimed that the $\ecs$ and $\acs$ play a particular role among the $\ucs$ subalgebras, as the algebras associated to finite-distance and asymptotic corners. It is remarkable that these algebras have also a special role from a purely algebraic viewpoint. This can be seen for their local counterparts, $\h_s$ and $\h_w$. Indeed, the reader is invited to show that $\h_s$ and $\h_w$ are the only two non-Abelian ideals of the algebra $\h$. It is the author's personal opinion that this is not a coincidence, and it would be gratifying to have an abstract group theoretical rationale for this pattern. 

We wish now to study representations. These algebras are complicated non-compact nor semisimple Lie algebras, and thus we do not have many mathematical results on their representation theory. We will resort to the coadjoint orbit method, which is much in spirit with these lectures, as it is based on a proper understanding of the geometric structure of these algebras, and their dual coalgebras. 

\subsection{Coadjoint Orbit Method in a Nutshell}\label{sec4.3}

We will first introduce the general coadjoint orbit theory, and then apply it to our algebras. The aim of this Subsection is to give the necessary ingredients for the rest of the notes. This is by far not a scrupulous review, and we re-direct the mathematically-inclined reader to \cite{Kirillov_1962, Kirillov1976ElementsOT, Kirillov_Merits, Kirillov1990, kirillov2004lectures} and \cite{Kostant_2006, ginz, Duistermaat:1982aa, Alekseev1988, ALEKSEEV1989719, Alekseev1990, wildberger_1990, DELIUS1990, Aratyn1990, Brylinski_1994, Alekseev2022} for more details.

Consider a Lie algebra $\g$ over the field of reals $\RR$, and its Lie group $G$. Since $\g$ is a vector space, it admits a dual space $\g^*$ (often called coalgebra) whose elements are maps from $\g$ to $\RR$. Given an element $m\in \g^*$ and a Lie algebra element $\mu\in\g$, we thus have
\beq
m:\g\to \RR,\qquad m(\mu)\in\RR.
\eeq
The adjoint action of $G$ on $G$, given two group elements $g,k\in G$, is defined via
\beq
Ad_g: G\to G \qquad Ad_g(k) \defeq g k g^{-1}.
\eeq
We will use the same symbol for the adjoint action of the group on the algebra
\beq
Ad_g(\mu) \defeq g \mu g^{-1}.
\eeq
In our convention, $Ad$ is a left action:
\beq
Ad_{lg}(k)=lgk(lg)^{-1}=lgkg^{-1}l^{-1}=Ad_l(gkg^{-1})=Ad_l(Ad_g (k)),
\eeq
for all $l,g,k\in G$. From the group adjoint action one derives the adjoint action of the algebra on itself. This should be proven as an exercise, using the Lie group exponential map $exp: \g \to G$. Given $\mu,\nu\in \g$, we have
\beq
ad_\mu:\g\to \g\qquad ad_\mu(\nu)=[\mu,\nu].
\eeq
One can then define the \textit{coadjoint action} of the group $G$ on $\g^*$,
\beq\label{ads}
Ad^*_{g}:\g^*\to \g^*\qquad (Ad^*_g m)(\nu)\defeq m(Ad_{g^{-1}}(\nu)).
\eeq
The appearance of the inverse group element is not conventional. It guarantees that the coadjoint action is also a left action
\beqn
&(Ad^*_{fg} m)(\nu)=m(Ad_{(fg)^1}(\nu))=m(g^{-1}f^{-1}\nu fg)=m(g^{-1}Ad_{f^{-1}}(\nu) g)&\\
&=m(Ad_{g^{-1}}(Ad_{f^{-1}}(\nu)))=(Ad^*_g m)(Ad_{f^{-1}}(\nu))=(Ad^*_f Ad^*_g m)(\nu).&
\eeqn
As for the adjoint action on the algebra, from \eqref{ads} one can derive the coadjoint action of the algebra on the coalgebra 
\beq
ad^*_\mu:\g^*\to \g^*\qquad (ad^*_\mu m)(\nu)=-m(ad_{\mu}(\nu))=-m([\mu,\nu]).\label{coad}
\eeq
The coadjoint orbit at the point $m\in \g^*$ is the group defined via
\beq
O_m\defeq \{g\in G \ \vert \ Ad^*_g m\neq m\},\qquad {\cal O}_m=\{\mu\in \g \ \vert \ ad^*_\mu m\neq m\},
\eeq
where we also introduced the coadjoint orbit algebra ${\cal O}_m$. The stabilizer (or isotropy) group and algebra at $m$ are instead given by those elements that preserve $m$:
\beq
S_m \defeq \{g\in G \ \vert \ Ad^*_g m= m\},\qquad {\cal S}_m=\{\mu\in \g \ \vert \ ad^*_\mu m= m\}.
\eeq

The important result in the theory of coadjoint orbits is that 
\beq
O_m=\bigslant{G}{S_m}.
\eeq
Then, for nilpotent compact Lie groups, the coadjoint orbits $O_m$ are in one to one correspondence with all irreducible representations of $\g$. This is a remarkable result, and allows us to classify all irreducible representations just by studying the orbit geometry of $\g^*$. Unfortunately, this result is not a theorem (that is why it is called the coadjoint orbit ``method''), and it is not clear whether for more complicated groups this correspondence still holds. Nonetheless, it is one the best tools at our disposal to at least find certain representations, and thus we will explore it further it in the rest of these notes. 

The main ingredient to study orbits is the Kirillov-Kostant-Souriau (KKS) symplectic $2-$form\footnote{The $2-$form is, strictly speaking, the map $\Omega_m(.,.)$. In the following, in an abuse of language, we will call the $2-$form both this and the contracted quantity $\Omega_m(\mu,\nu)$.}
\beq\label{KKS}
\Omega_m:{\cal O}_m\otimes {\cal O}_m\to \RR\qquad \Omega_m(\mu,\nu)\defeq \Omega_m(ad^*_\mu m,ad^*_\nu m)\defeq m([\mu,\nu]).
\eeq
By construction, the KKS $2-$form is non-degenerate and invertible on the orbits, because the elements of the stabilizer group have been quotienting out. Understanding the features of this $2-$form is clearly equivalent to understand the coadjoint action, see \eqref{coad}. Mathematically, the coalgebra $\g^*$ is a Poisson manifold, foliated by non-degenerate symplectic leaves, the orbits. Studying orbits is thus a geometric problem on this Poisson manifold with ``metric'' $\Omega_m$, which -- fascinatingly --  provides us information about the properties of the algebra.

\subsection{Applied to \texorpdfstring{$\h$}{}}\label{sec4.4}

The general theory of coadjoint orbit can be applied to our local algebra $\h$, which is a $6-$dimensional algebra. In the $(t^a{}_b,t_c)$ basis described in Subsec. \ref{sec4.2}, an element $\mu\in \h$ is given by
\beq
\mu=\theta^a{}_b t^b{}_a+b^c t_c.
\eeq
The dual basis $(\T^a{}_b,\T^c)$ for the coalgebra is such that
\beq\label{dualba}
\T^a{}_b(t^c{}_d)=\delta^a_d \delta^c_b\qquad \T^a{}_b(t_c)=0\qquad \T^a(t^b{}_c)=0\qquad \T^a(t_b)=\delta^a_b.
\eeq
A generic element $m\in\h^*$ is written as
\beq\label{m}
m=J^a{}_b\T^b{}_a+P_a\T^a,
\eeq
and acts on the algebra as follows
\beq
m(\mu)=\theta^a{}_bJ^b{}_a+b^cP_c.
\eeq
We call the components of $m$ in \eqref{m} $J^a{}_b$ and $P_a$ for analogy with the coadjoint analysis for Poincar\'e, which reveals that they are indeed the angular momentum and the linear momentum of a particle in that case. We will insist on the physics of the coalgebra when we introduce the moment map, but we can already remark at this stage that specifying a point on the coalgebra should be think of as specifying the value of the fields for a theory. Then, moving on an orbit is the algebraic equivalent of a field variation. 

The coadjoint action \eqref{coad} can be computed explicitly for $\h$. Introducing
\beq
\nu=\theta^{\prime a}{}_bt^b{}_a+b^{\prime c}t_c,
\eeq
and using \eqref{comm}, we get
\beq\label{admu}
(ad^*_\mu m)(\nu)=-m([\mu,\nu])=J^a{}_b [\theta,\theta^\prime]^b{}_a+P_a(\theta^a{}_b b^{\prime b}-\theta^{\prime a}{}_b b^b),
\eeq
where we introduced the notation $[\theta,\theta^\prime]^b{}_a=\theta^a{}_c\theta^{\prime c}{}_b-\theta^{\prime a}{}_c\theta^{c}{}_b$.
We want to read $ad^*_\mu m$ from this expression. Since this is a typical procedure, we perform it in detail. Generically, $ad^*_\mu m$ is a new point on the coalgebra.  Therefore, we can decompose it in our basis as
\beq
ad^*_\mu m=\delta_\mu J^a{}_b \T^b{}_a+\delta_\mu P_c \T^c,
\eeq
where we called the components $\delta_\mu J^a{}_b$ and $\delta_\mu P_c$ to emphasize that it is obtained starting from $m$ and acting with $\mu$. This is just a symbolic expression, the symbol $\delta$ is not associated to a field variation at this point.
Acting on $\nu$ we gather
\beq
(ad^*_\mu m)(\nu)=\delta_\mu J^a{}_b \theta^{\prime b}{}_a+\delta_\mu P_c b^{\prime c}.
\eeq
Finally, comparing this expression with \eqref{admu}, we can read off the components of $ad^*_\mu m$,
\beq\label{var}
\delta_\mu J^a{}_b=[J,\theta]^a{}_b-b^aP_b\qquad \delta_\mu P_c=P_b\theta^b{}_c.
\eeq
This means that 
\beq
ad^*_\mu m=([J,\theta]^a{}_b-b^aP_b) \T^b{}_a+P_b\theta^b{}_c \T^c.
\eeq

By computing the coadjoint action, we have already extrapolated the KKS $2-$form \eqref{KKS}
\beq\label{kksh}
\Omega^{\h}_m(\mu,\nu)=-J^a{}_b [\theta,\theta^\prime]^b{}_a-P_a(\theta^a{}_b b^{\prime b}-\theta^{\prime a}{}_b b^b).
\eeq
To study the various orbits, we have to analyse this form. The first observation is that \eqref{var} is a solvable (invertible) system as long as the combination ($\epsilon^{bc}$ is the $2-$dimensional Levi-Civita symbol)
\beq\label{c3}
C_3\defeq P_aJ^a{}_b \epsilon^{bc}P_c,
\eeq
is non-vanishing. This means that \eqref{kksh} can be expressed as a function of $\delta_\mu J^a{}_b$ and $\delta_\mu P_c$ as long as $C_3\neq 0$. Then, we have $6-$dimensional orbits, each labelled by the value of $C_3$. If we are in one of these orbits, it means that we span the whole coalgebra. In fact, as long as $C_3\neq 0$, the isotropy group $S_m$ is trivial, and thus $O_m$ is the full group $H$. In this scenario, the KKS $2-$form gives rise to a Poisson bracket that is exactly the same as the Lie bracket for the algebra $\h$. Let us briefly show this. We compactly denote $Z^A=(J^a{}_b, P_c)$. Then, when $C_3\neq 0$, eq. \eqref{kksh} can be inverted and gives
\beq
\Omega^{\h}_m(\mu,\nu)=\Omega_{AB}(Z)(\delta_\mu Z^A \delta_\nu Z^B-\delta_\nu Z^A \delta_\mu Z^B).
\eeq
Given a symplectic $2-$form, the Poisson bracket are given by the inverse matrix elements:
\beq
\{Z^A,Z^B\}=\Omega^{AB}.
\eeq
Computing this explicitly gives a Poisson bracket algebra which is exactly the same as $\h$.

On the other hand, when the algebra is odd-dimensional, there are no symplectic leaves that span the whole algebra, because a symplectic orbit must be even-dimensional by construction. Instead of classifying the orbits when $C_3=0$, we study the algebras $\h_s$ and $\h_w$, to see how they are realized inside $\h-$orbits. We start with $\h_s$. Then, we must impose that $t^a{}_b$ is traceless. This implies
\beqn
[t^a{}_b,t^c{}_d]=\delta^c_b t^a{}_d-\delta^a_d t^c{}_b\quad [t^a{}_b,t_c]=-\delta^a_ct_b-{1\over 2}\delta^a_bt_c\quad [t_a,t_b]=0.
\eeqn
In particular, also the components $\theta^a{}_b$ are traceless. We leave the computation of the KKS $2-$form as an exercise, but we remark that we can compute the variation of \eqref{c3}. In general, we obtain
\beq
\delta_\mu C_3=Tr(\theta^a{}_b),
\eeq
which thus vanishes for $\h_s$. Therefore, $C_3$ is a cubic Casimir of $\h_s$. Given the existence of a Casimir, there is a constraint among the coadjoint actions on the components, and thus the maximal coadjoint orbit cannot span the full coalgebra. We already anticipated this by noticing that $\h_s$ is $5-$dimensional, and here we confirmed that indeed orbits can be at best $4-$dimensional. 

A very similar computation can be performed for the algebra $\h_w$. The quantity $C_3$ is not a Casimir in this case but, since the algebra is $3-$dimensional, it is easy to compute the KKS $2-$form (exercise) and show that orbits are at best $2-$dimensional. We have not yet done a full classification of orbits for $\h$, $\h_s$, and $h_w$, but more details can be found in \cite{Ciambelli2022a}. We obtained an important result, which is that the maximal orbits of $\h_s$ and $\h_w$ combine into a $6-$dimensional space. One can indeed recognize these orbits as subspaces of the tangent bundle at points in $\h$, confirming that the latter is a unified framework. Recalling that $\h_s$ and $\h_w$ are the local algebras of finite and asymptotic distance corners, it is tempting to speculate that $\h$ is the correct framework to ``add together'' these two corners. Rather than investigating this interesting idea, we study in the next Subsection how to re-introduce $\dS$, which leads us to the theory of algebroids. Although it is natural for us to have in mind applications to corners, and thus to use this part as a warm up to try to reach the full $\ucs$ analysis, this Subsection was a mathematical exercise on its own. It would be interesting to perform the coadjoint analysis to its full extent for these algebras, and link it to their representations.

\subsection{Algebroids and Associated Bundles}\label{sec4.5}

We observed that $\h$ gives rise to a unified framework for finite and asymptotic corners. Can we say the same about the $\ucs$? In other word, can we study the coadjoint orbits of the full $\ucs$? To answer this question, we need to understand how to reinstate diffeomorphisms of the corner. 

\paragraph{Reinstating $\dS$}

This brings back complications, due to the fact that components of the algebra elements are now valued on $S$
\beq
\chi=\xi^\alpha(\sigma)\pa_\alpha+\theta^a{}_b(\sigma)t^b{}_a+b^c(\sigma)t_c,
\eeq
and similarly for the elements of the coalgebra $\ucs^*$. For the latter, the dual basis satisfies
\beq
\D \sigma^\beta(\pa_\alpha)=\delta^\beta_\alpha, 
\eeq
on top of \eqref{dualba}. An element $M\in\ucs^*$ is thus parameterized by
\beq\label{M}
M=\alpha_\beta(\sigma)\D\sigma^\beta+J^a{}_b(\sigma)\T^b{}_a+P_c(\sigma)\T^c,
\eeq
such that (the $\sigma-$dependency of all these components is from now on implicitly assumed)
\beq\label{mchi}
M(\chi)=i_\xi\alpha+\theta^a{}_bJ^b{}_a+b^c P_c,
\eeq
where $\alpha=\alpha_\beta\D \sigma^\beta$ and $\xi=\xi^\beta\pa_\beta$.

This is not yet an invariant pairing, because $M(\chi)$ is still a local expression on $S$ that transforms under $\dS$. The invariant pairing is obtained integrating over $S$:
\beq\label{pairing}
\langle M,\chi\rangle =\int_S Vol_S M(\chi).
\eeq
Note that we could have instead worked with the density
\beq
\tilde{M}=Vol_S M, \qquad \langle \tilde{M},\chi\rangle =\int_S \tilde{M}(\chi).
\eeq
Then, the component $\tilde{\alpha}_\beta$ is also a density, and therefore its coadjoint transformation  contains an extra term. This is the strategy pursued in \cite{Donnelly:2020xgu}. From the intrinsic algebraic viewpoint, the merit of using densities is that $Vol_S$ does not need to be independently introduced. We will continue with \eqref{pairing}, because working with tensors or densities is equivalent for the purposes of these notes.

The coadjoint action \eqref{coad} is now given by (${\cal Y}\in \ucs$)
\beq
\langle ad^*_\chi M,{\cal Y}\rangle=-\langle M,ad_\chi {\cal Y}\rangle,
\eeq
and therefore the KKS $2-$form is 
\beq
\Omega_M(\chi,{\cal Y})\defeq \langle M,[\chi , {\cal Y}]\rangle.
\eeq
Following the same steps as for $\h$, we can compute the coadjoint action on the components of \eqref{M}. We leave as an exercise to prove that the final result is (explicit derivation in \cite{Ciambelli2022a})
\beqn
\delta_{\chi}\alpha &=& \cL_\xi \alpha+J^a{}_b\D \theta^b{}_a+P_c \D b^c\\
\delta_{\chi} J^a{}_b &=& \xi(J^a{}_b)+[J,\theta]^a{}_b-b^aP_b\\
\delta_{\chi}P_a &=& \xi(P_a)+\theta^b{}_a P_b.
\eeqn
From this, we see that the fact that $\h_s$ and $\h_w$ orbits combine into $\h-$orbits transfers to their avatars: the $\ecs$ and $\acs$ are realized inside the $\ucs$ in a complementary way. 

We now face an obstacle. The next step in the orbit analysis is to try to invert the KKS $2-$form, to classify the dimensionality of the orbits and the presence of Casimirs. However, the presence of $\dS$ makes the inversion problem challenging, because it acts as a derivation, and thus inverting it requires non-local integrals on $S$. We here follow a different route, by realizing that the presence of $\dS$ calls for an algebroid's interpretation of these results. This allows us to gain a geometric control on this algebra, and thus better control on representations. Is there a geometric structure whose automorphisms (set of transformations preserving the structure) are given exactly by the $\ucs$?  The answer to this question is the Atiyah Lie algebroid associated to the $H-$principal bundle, which is what we present hereafter. As already stressed, independently from applications to corners, the theory of algebroids is an important mathematical tool in theoretical physics, and thus a useful notion to introduce in these lectures. 

\paragraph{Atiyah Lie algebroids} In a loose sense, an algebroid is a bundle of algebras, where at each point of the base we have a different copy of the algebra defining the bundle. Let us construct it. The $UCS$ group can be written as
\beq
UCS=Diff(S)\ltimes H, \qquad H=GL(2,\RR)\ltimes \RR^2.
\eeq
Starting from $S$, we define the principal $H-$bundle $P$ via the projector $\pi:P\to S$.\footnote{We introduce the algebroid in this specific example, but this construction can be applied to any Lie group.} Locally, this bundle looks like $P\simeq H\times S$. Its tangent bundle is $TP$, which is locally $TP\simeq \h \oplus TS$. Sections of this bundle are closely related to elements of the $\ucs$ algebra, except that $TP$ is a bundle over $P$, not a bundle over $S$, so the components of sections of $TP$ are functions of $P$. To reduce to a bundle over $S$, we need to quotient out one of the two group actions. Indeed, we have a left and a right group action on $TP$. The former gives rise to gauge transformations, while the latter is a mere redundancy. Quotienting out the right action we obtain $TP / H$. We proved in \cite{Ciambelli:2021ujl} that this is now a $H-$bundle over $S$, and thus sections of this bundle are exactly elements of the $\ucs$. This bundle is called the \textit{Atiyah Lie algebroid} associated to the $H-$principal bundle
\beq
A=\bigslant{TP}{H}.
\eeq
This algebroid is defined via the short exact sequence
\beq\label{splitshortExactSeq}
\begin{tikzcd}
0
\arrow{r} 
& 
L
\arrow{r}{j} 
& 
A
\arrow{r}{\rho} 
& 
TS
\arrow{r} 
&
0
\end{tikzcd}.
\eeq

To introduce the concept of short exact sequence, we quickly review the mathematical notion of exact sequences. An exact sequence is a sequence where the image of one morphism equals the Kernel of the next:
\beq
G_0 \stackrel{f_1}{\longrightarrow} G_1 \stackrel{f_2}{\longrightarrow} G_2 \dots \stackrel{f_{n}}{\longrightarrow} G_n
\eeq
with Im$(f_k)=$Ker$(f_{k+1})$. Then:
\begin{itemize}
\item $0\stackrel{f_1}{\longrightarrow} A \stackrel{f_2}{\longrightarrow} B$ is a short exact  sequence if $f_2$ is injective: Ker$(f_2)=0$.
\item $B\stackrel{f_1}{\longrightarrow} C \stackrel{f_2}{\longrightarrow} 0$ is a short  short sequence if $f_1$ is surjective: Im$(f_1)=C$.
\item $0\stackrel{f_1}{\longrightarrow}  A\stackrel{f_2}{\longrightarrow} B \stackrel{f_3}{\longrightarrow} 0$ is a short exact sequence if $f_2$ is an isomorphism: Im$(f_2)=B$, Ker$(f_2)=0$.
\item $0\stackrel{f_1}{\longrightarrow}  A\stackrel{f_2}{\longrightarrow} B \stackrel{f_3}{\longrightarrow} C\stackrel{f_4}{\longrightarrow}0$ is a short exact sequence if $f_2$ is injective, $f_3$ surjective and Im$(f_2)=$Ker$(f_3)$. In other words, $A\subseteq B$ with embedding $f_2$ and $C\simeq B/\text{Ker}(f_3)=B/\text{Im}(f_2)$.
\item A sequence with more than $3$ non-vanishing elements is a long sequence.
\end{itemize}

Therefore, stating that an Atiyah Lie algebroid is defined via the short exact sequence \eqref{splitshortExactSeq} means
\beq
TS\simeq \bigslant{A}{L},
\eeq
where $L$ is by construction the pre-image under the embedding $j$ of the Kernel of $\rho$ 
\beq
Ker(\rho)\defeq \{\chi \in A \vert \rho(\chi)=0\}.
\eeq
The map $\rho$ is called anchor map. An element of $A$ in the Kernel of $\rho$ is thus an element that is not associated to a vector field on $TS$. Therefore, it is associated to an element of the tangent bundle of $H$, which is the Lie algebra $\h$. Indeed, the Kernel of $\rho$ is isomorphic to the adjoint bundle
\beq
L\defeq  P\times_{Ad_H} \h\simeq Ker(\rho),
\eeq
which naturally carries the adjoint representation of $\h$. 

This was a technical and rapid explanation of Atiyah Lie algebroids, in which we proved that this construction is a strikingly perfect description of the $\ucs$. Indeed, what we obtained is a set of bundles ($L$, $A$, and $TS$), each carrying a bit of the $\ucs$, with a built in adjoint action:
\beqn
\text{Element of $\ucs$}&\Rightarrow &\text{Section of $A$} \\
\text{Element of $\h$}&\Rightarrow &\text{Section of $L$} \\
\text{Element of $\dS$}&\Rightarrow & \text{Section of $TS$}.
\eeqn
In this bundle construction, each point on $S$ carries a different copy of the algebra $\h$, which is exactly the intuitive picture of an algebroid, and how \eqref{chi} should be read. To truly complete the geometric description of the $\ucs$, we need to account for the $\ucs$ Lie bracket. This is achieved in the algebroid, because $A$ comes equipped with a bracket structure, satisfying ($f_1,f_2\in C^\infty(S)$)
\beq\label{rhoLeib}
[f_1\chi,f_2{\cal Y}]_{A}=f_1 f_2[\chi,{\cal Y}]_{A}+f_1\rho(\chi)(f_2)\  {\cal Y}-f_2\rho({\cal Y})(f_1)\  \chi,
\eeq
where, since $\rho(\chi)\in TS$, $\rho(\chi)(f_2)\in C^\infty(S)$ is the ordinary derivative of the function $f_2$ along the vector field $\rho(\chi)$. So not only the elements of the $\ucs$ are built in as sections of $A$, but also the $\ucs$ Lie bracket is naturally accounted for:
\beq
\text{$\ucs$ Lie bracket}\Rightarrow \text{$A$ bracket}.
\eeq
This shows how the reinterpretation of the algebra in terms of Atiyah Lie algebroids is very natural, and put the former in a well-defined geometric framework. Without entering too technical details, we mention that, given a bundle structure, there is a privileged semi-direct structure in its autormorphisms. Indeed, in order to preserve the vertical sub-bundle, the base transformations are not allowed to involve the fibre coordinates, while the vertical fibres transform mixing with the base. Putting this together with \eqref{rhoLeib}, one obtains that the automorphisms of $A$ form an algebra which has exactly the semi-direct structure of the $\ucs$. 

Before moving to the study of $\ucs$ representations as associated bundles to $A$, we introduce the last important ingredient, which is the Ehresmann connection. Since $A$ is a bundle, this is similar to the discussion in Subsec. \ref{sec3.1}. An Ehresmann connection on an algebroid is introduced by specifying an inverse short exact sequence
sequence
\beq\label{Ehre}
\begin{tikzcd}
0
\arrow{r} 
& 
L
\arrow{r}{j} 
\arrow[bend left]{l} 
& 
A
\arrow{r}{\rho} 
\arrow[bend left]{l}{\omega}
& 
TS
\arrow{r} 
\arrow[bend left]{l}{\sigma}
&
0
\arrow[bend left]{l} 
\end{tikzcd}.
\eeq
Since $A$ is bigger than $TS$, the anchor map $\rho$ is a projection. Then, the map $\sigma: TS\to A$ is called an Ehresmann connection, and identifies the horizontal sub-bundle $Hor\in A$ as its image, such that
\beq
A=Hor\oplus Ver,
\eeq
where $Ver\simeq Ker(\rho)$ is unambiguously defined, while $Hor=Im(\sigma)$ is ambiguous, and depends on the choice of $\sigma$. This Ehresmann connection is the abstract intrinsic datum that contains the $\RR^2$ connections $a_{(0)}$ and the $\mathfrak{gl}(2,\RR)$ connections $a_{(1)}$. 

\paragraph{Representations}

We showed that $A$ gives a geometric description of the $\ucs$, encapsulating all its properties. This is our starting point for studying representations. The algebroid comes with the adjoint representation on its vertical, which is exactly what we need in the orbit analysis. A thorough study of the coalgebroid $A^*$ is then expected to be viable. We note that, in the theory of group bundles, different representations are given by associated bundles, whose fibres correspond to the representation space. This is well-known in the mathematical formulation of gauge theories on principal bundles, where, for instance, matter fields transforming in the fundamental representation of the gauge group are introduced as sections of the fundamental associated bundle. This is more than an analogy, and indeed by studying associated bundles to the $\ucs$ algebroid we expect to uncover possible representations of it, which are not interpreted as matter fields but rather as gravitational configurations. 

Consider a bundle $E\to S$, sections of the algebroid $A$ act on $E$ as differentiation, via the morphism $\phi_E$:
\beq\label{DerEshortExactSeqGen}
\begin{tikzcd}
& 
L
\arrow{r}{j} 
\arrow{dd}{v_E}
& 
A
\arrow{dr}{\rho} 
\arrow{dd}{\phi_E} 
& 
&
\\
0
\arrow{ur} 
\arrow{dr} 
&&&
TS
\arrow{r} 
&
0
\\
& 
End(E)
\arrow{r}{j_E} 
& 
Der(E)
\arrow{ur}{\rho_E} 
& 
&
\end{tikzcd}.
\eeq
Loosely speaking, the horizontal bundle in $A$ is mapped to gauge-covariant derivatives on $E$, whereas the vertical bundle gives gauge transformations. We will presently identify a relevant associated bundle, which gives rise to a local spacetime representation, but we remark that we reached here the state of the art on the understanding of $\ucs$ representations. We gave all the mathematical tools, and the reader should be at this stage well-equipped to address this problem. Are there unitary and/or irreducible representations?  Is there an associated bundle on which we can immediately recognize quantum gravity features? These are far-reaching questions, to which we do not know the answer, but we believe we now have the right instruments to address. Clearly, the ultimate goal would be to provide a full classification of representations, by carefully studying the coalgebroid $A^*$.

\paragraph{Classical spacetime representation} We can construct an associated bundle in which the $2$ normal coordinates to the corner in a classical spacetime naturally emerge. Along the way, this allows us to discuss features of the dual bundles. Consider an \textit{affine bundle} $\pi_B:B\to S$ of rank $2$. Sections of this bundle can be written as 
\beq
\psi\in B\qquad \psi=\sigma^\beta e_\beta+\psi^a e_a,
\eeq
where the index $a$ is a $2-$dimensional index on the vertical fibres, and ($e_\beta, e_a$) is a basis of $B$. By definition, moving from an open set $U_i\in S$ to another one $U_j\in S$, the components of sections of $B$ undergo on the intersection $U_i\cap U_j$ the affine transformation
\beq\label{aff}
(\sigma_i^\beta, \psi_i^a) = (\sigma_i^\beta(\sigma_j),R_{ij}^a{}_b\psi^b_j+b_{ij}^a),
\eeq
where $R_{ij}^a{}_b$ is a $S-$valued rotational matrix and $b^a_{ij}$ is the affine parameter. Infinitesimally, eq. \eqref{aff} reads
\beq
(\sigma^\beta_i, \psi^a_i) \simeq  (\sigma_j^\beta-\xi^\beta(\sigma_j),\cL_\xi \psi_j^a(\sigma_j)+\theta^a{}_b(\sigma_j) \psi^b_j(\sigma_j)+b^a(\sigma_j)). 
\eeq
This is exactly how the $\ucs$ is expected to act on components of vectors on the normal $2-$plane, see the discussion below \eqref{transform}. So we found that $\psi^a$ can be interpreted as normal coordinates on a $d-$dimensional space (what we called $u^a$ in previous Sections). This associated affine bundle is thus the local classical spacetime representation. It is a local emerging spacetime, in a mathematically rigorous sense. 

We can go a step further by introducing the dual picture of \eqref{DerEshortExactSeqGen}, for the affine bundle $B$,
\beq
\begin{tikzcd}
& 
L^*
\arrow{dl}
& 
A^*
\arrow{l}{j^*} 
& 
&
\\
0
&&&
T^*S
\arrow{dl}{\rho_B^*} 
\arrow{ul}{\rho^*} 
&
0
\arrow{l} 
\\
& 
End(B)^*
\arrow{uu}{v_B^*}
\arrow{ul}
& 
Der(B)^*
\arrow{l}{j_B^*} 
\arrow{uu}{\phi_B^*} 
& 
&
\end{tikzcd}.\label{derBdiagram}
\eeq
Since elements of $A^*$ are ``field configurations'', an element $M_B\in Der(B)^*$ is a field configuration in the local spacetime representation. It thus carries in an abstract manner the metric components near the corner.\footnote{Strictly speaking, this is true in a local trivialization, which we are omitting here, for illustrative purposes.} Given a basis $(\D \sigma^\beta,\tilde{v}^a{}_b,\tilde{v}^a)$ of $Der(B)^*$, we write
\beq
M_B=M_{B\beta}\D\sigma^\beta+M^a_{Bb}\tilde{v}^b{}_a+M_{Bc}\tilde{v}^c.
\eeq
Then, the components $M_{B\beta}$, $M^a_{Bb}$, and $M_{Bc}$ are the classical gravity momenta near the corner. As remarked, this is then a concrete spacetime emergence program, worth pursuing. 

This Subsection gave an overview of how the theory of Atiyah Lie algebroid could help us in shedding light on $\ucs$ representations. It naturally intertwines with the orbit method, as both constructions are anchored to the adjoint representation, and they are two different mathematical tools that we used to address the same problem.

\subsection{Exercise: Moment Maps}\label{sec4.6}

The last notion we wish to convey is that of the \textit{moment map}.\footnote{The term ``moment map'' comes from a wrong French translation (the correct one being ``momentum map''). We prefer to use ``moment map'' to avoid confusion in comparing with previous mathematical works, and because we believe the term ``momentum'' is confusing in this context, since it has many mathematical meanings associated.} We will first present it and then construct it in existing examples, as a brief concluding exercise. 

A moment map relates a classical field space to points of the coalgebra. A classical field space $\Gamma$ comes with a set of symplectomorphisms, that satisfies an algebra, say $\g$. Then, the moment map links $\Gamma$ with the coalgebra $\g^*$:
\beq
\mu: \Gamma \to \g^*. 
\eeq
Given that we have a non-degenerate symplectic $2-$form on $\Gamma$, while $\g^*$ is typically a Poisson manifold, the moment map in an inclusion. It maps to a particular symplectic leaf on $\g^*$. Therefore, classical field spaces are mapped into coadjoint orbits on the coalgebra. This is why we already observed that points on the coalgebra can be though of as field configurations. 

On the extended phase space introduced in Subsec. \ref{sec3.4}, all the $\ucs$ elements are symplectomorphisms, and thus we can construct a moment map to $\ucs^*$. We stress the importance of this result, which is not possible without extending the phase space. Then, we can define the moment map with the pairing \eqref{pairing}. Given $\varphi\in\Gamma$
\beq\label{mom}
\mu(H_\chi)\defeq \langle \mu(\varphi),\chi\rangle=\int_S Vol_S \mu(\varphi)(\chi).
\eeq
We called this expression $\mu(H_\chi)$ because this is how the corner Noether charge (coming from \eqref{inte}) is interpreted from the coalgebra. From this expression it follows that
\beq
\Omega_{\mu(\varphi)}(\chi,{\cal Y})=-\langle ad^*_\chi (\mu(\varphi)),{\cal Y}\rangle=\mu(H_{[\chi,{\cal Y}]})=-\mu(\delta_\chi H_{{\cal Y}})=\mu(I_{V_\chi}I_{V_{\cal Y}}\Omega).
\eeq
This chain of equalities demonstrates that the KKS $2-$form is related to the symplectic $2-$form on $\Gamma$. As a consequence, the field space variation is related to the coadjoint action on the orbit via the moment map
\beq
\mu\circ \delta_\chi = ad^*_\chi\circ \mu.
\eeq
This result is crucial, and it allows us to gather intuition on the coalgebra. Orbits in the latter are inequivalent classical field spaces, labelled by Casimirs. Moving on the orbit with the coadjoint action is equivalent to consider field variations. The brackets coming from the KKS $2-$form are then the Poisson brackets on the field space. 

To familiarize with the moment map, we will explicitly construct it in two simple examples
\begin{enumerate}
\item[Q1)] \textbf{Construct the moment map for the finite-distance charges \eqref{ch}.}

The theory giving rise to \eqref{ch} has asymptotic symmetries given by the $\ecs$. Therefore, we have to construct the intrinsic $\ecs$ invariant pairing
\beq
\langle M,\chi\rangle =\int_S Vol_S M(\chi),
\eeq
where $M\in \ecs^*$ and $\chi\in\ecs$. We do not have to do it from scratch, as we already computed it for the $\ucs$ in \eqref{mchi}. We just need to restrict that expression for traceless $\theta^a{}_b$,
\beq
\langle M,\chi\rangle =\int_S Vol_S \left(i_\xi\alpha+\theta^a{}_bJ^b{}_a+b^cP_c\right).
\eeq
Comparing this expression with \eqref{ch}, and using \eqref{mom}, we derive\footnote{We are still off-shell, whereas the moment map should be constructed on-shell. This is thus a pre-moment map, that gives rise to the coadjoint action on the orbit once the equations of motion are imposed on the field-space momenta.}
\beq
\mu(N^a{}_b)=J^a{}_b\qquad \mu(b_j)=\delta_j^\beta\alpha_\beta\qquad \mu(p_a)=P_a,
\eeq
where the momenta are related to the bulk metric components in the vicinity of the corner as in \eqref{momo}.

\item[Q2)] \textbf{Construct the moment map for the renormalized asymptotic charges of \cite{Freidel:2021fxf}.}

In \cite{Freidel:2021fxf}, the asymptotic symmetry group for corners at null infinity is found to be the $\mathfrak{bmsw}$ group \eqref{bmsw}, generated by vector fields of the form (the index $A$ is an index on $S$ in this exercise, such that $Y^A\pa_A$ is an element of $\dS$)
\beq\label{bmsW}
\xi=T\pa_u+Y^A\pa_A+W(u\pa_u+\rho\pa_\rho).
\eeq
Here, we used the conventions of \cite{Freidel:2021fxf} (see eq. (4.12)), except for the change of coordinates $\rho={1\over r}$ needed to bring this setup in the configuration of Subsec. \ref{sec3.2}. This is why the sign is flipped in the last term of \eqref{bmsW}. The renormalized charges are given in eqs. (9.10)-(9.12) of \cite{Freidel:2021fxf}, and we report them here explicitly ($Vol_S=\sqrt{\overline{q}}\D^2\sigma$)
\beq\label{reno}
Q^R=\int_S Vol_S(T(M-{1\over 2}\overline{D}_AU^A)+4 W\overline{\beta}+Y^A(\overline{P}_A+2\pa_A\overline{\beta})).
\eeq

We must compare this expression to the invariant pairing for $\mathfrak{bmsw}$. This can be computed to be
\beq\label{mchibm}
\langle M,\chi\rangle =\int_S Vol_S \left(i_{Y}\alpha+{1\over 2} W {\cal J}+T P_u\right),
\eeq
where we used a vector field of the form \eqref{bmsW} and an element $M\in\mathfrak{bmsw}^*$ parameterized by
\beq
M=\alpha_\beta\D\sigma^\beta+{\cal J}\tilde{t}+P_u\T^u,
\eeq
with $\tilde{t}$ the trace basis element (the factor ${1\over 2}$ comes from normalization). 

So comparing \eqref{mchibm} with \eqref{reno}, and using \eqref{mom}, we eventually read the moment map
\beq
\mu(M-{1\over 2}\overline{D}_AU^A)=P_u\qquad \mu(16\overline{\beta})={\cal J}\qquad \mu(\overline{P}_A+2\pa_A\overline{\beta})=\delta^\beta_A\alpha_\beta.
\eeq
\end{enumerate}

These two concrete examples allowed us to gather familiarity with the moment map, which was the last notion we wished to introduce in this Section. It would be interesting to apply the coadjoint orbit analysis detailed here to other examples of asymptotic and finite-distance corners in classical gravity. It is important to build intuition on this, to then understand how a quantum representation could differ. Some quantum setup may not have a well-defined classical limit, and thus there could be more convoluted maps linking the operator spectrum of the theory to the coalgebra. 

\section{Conclusions}\label{sec5}

Since we have been verbose both in the Introduction and the main body of these notes, we offer here a brief recap of the main topics discussed and the open questions mentioned.

These lectures had the main goal of presenting the theory of asymptotic symmetries, linking it to the recent corner proposal, and discussing the latter in detail. We especially focussed on the mathematical structures behing these theories, and gave self-consistent digressions on the tools employed. 

After a quick introduction to the covariant phase space, we enunciated Noether's theorems, focussing in particular to Noether's second theorem, which deals with gauge symmetries. Our main question was: how do we distinguish local redundancies from physical symmetries? The answer is via the Noether charges. If the latter are zero, then the symmetry associated is pure gauge. If the charge is non-vanishing, than the symmetry in question maps different field-space configurations, and thus it is a physical symmetry. This applies both to global and local symmetries, and -- as we just explained -- it is a misconception to believe that local symmetries are all redundant, especially in the presence of boundaries. We then proved an important result for diffeomorphisms: the Poisson bracket among charges projectively represents the Lie algebra of vector fields, that is, it is the same algebra, modulo central extensions. We then applied this machinery to electromagnetism, where as an exercise we proved that the total electric charge of the system stems from Noether's second theorem. One of the key features of this theorem is that the charge is a codimension$-2$ object, and thus a surface charge. This is one of the main take-home messages. 

We then stated the theory of asymptotic symmetries, which consists in a series of steps that, given a theory, allows us to extrapolate the Noether charges and the charge algebra. It is a systematic approach that can be applied to a variety of different setups. When applied to gravity, certain features cease to hold. In particular, the canonical charges are no-more always conserved, integrable, and finite. Each of these three properties is understood and cured, but we insisted that integrability is the main problem, for without it one cannot define the Poisson bracket. Anticipating what we discussed afterwards, the Poisson bracket of charges at corners is one of the main ingredient of the corner proposal, so it is crucial that it is well-defined in gravity, and understanding this led us to the extended phase space, recapitulated below. We concluded Section \ref{sec2} with an exercise, where we applied the general theory of asymptotic symmetries to AdS$_3$ Einstein gravity with Brown-Henneaux boundary conditions. There, we computed  the residual symmetries and the charge algebra, explicitly founding that the Poisson bracket represents projectively the Lie bracket, because of the appearance of the  celebrated Brown-Henneaux central extension. We finished mentioning how this example is a precursory realization of the AdS/CFT correspondence. 

We then moved to the corner analysis. Noether charges for local symmetries are evaluated on codimension$-2$ surfaces, which are generically called corners. Corners are manifold of their own, which are embedded into the bulk. The embedding map was the main actor of Section \ref{sec3}, usually disregarded in the previous literature. After a brief introduction to this mathematical concept, we proved that the Lie bracket algebra of diffeomorphisms gives rise to a maximally finitely-generated sub-algebra at the vicinity of corners. This universal algebra, called the $\ucs$, is believed to be one of the most robust features of gravity. We discussed how it contains all known symmetry algebras, both for finite-distance and asymptotic corners. For the former, one can perform a full off-shell and gauge-free analysis, and retrieve that the biggest algebra that can be realized is the extended corner symmetry, in which only the traceless generators of the normal plane symmetries are turned on. Contrarily, only the trace-full generator pertains asymptotic corners, and thus the $\ucs$ can be seen as an unified treatment, applicable to both instances. 

After this, we returned to the problem of integrability and showed that there exists an enlarged field space, called extended phase space, on which all diffeomorphisms are canonical generators. This leads to integrable charges, thanks to the introduction of embeddings as dynamical fields. By tracking their variation under diffeomorphisms we are able to find canonical charges from Noether's second theorem, even when the system dissipates. This is a very powerful result, whose consequences are still unfolding. On such an extended phase space, we have Poisson brackets among all $\ucs$ generators. This is instrumental for the corner proposal, in which the basic ingredients are charges, corners, and Poisson brackets. We then proposed an exercise, to end Section \ref{sec3}, in which we derived the $\mathfrak{bms}_3$ asymptotic symmetry algebra at the black hole horizon, from our general finite-distance discussion. 

Equipped with the $\ucs$ and the extended phase space, we then enunciated the corner proposal in Section \ref{sec4}. The latter refocuses our attention from a Lagrangian and a metric to corners and charges, as the atomic constituents of gravity. Thinking of corners, charges, and their algebra as the building blocks is a far-reaching shift of paradigm, mainly because these concepts are amenable to quantization. This is not the case for a bulk Lagrangian or a full bulk metric, due to many pathologies. Among them, the presence of diffeomorphisms, and the non-renormalizability of the Lagrangian. Loosing ties with these features is an important step. The corner proposal is a concrete program, made of  solid results, and we believe it will be able to address unanswered questions in quantum gravity. Indeed, given corners and algebras, one can study representations of the latter. While we found representations that are suitable to describe a classical emergent spacetime, there exist other representations, whose properties are more suitable to a quantum interpretation. 

We are at a primitive stage in the study of $\ucs$ representations. Therefore, rather than specializing to some particular aspects, we continued Section \ref{sec4} with an overview of the mathematical tools that we need to address this problem. We first studied the coadjoint orbit method developed by Kirillov, and discussed how it allows to understand the asymptotic and finite-distance corner orbits as embedded in the full universal corner symmetry. The study of orbits of the coalgebra is a formidable mathematical problem, which renders more geometrical the study of representations. We then observed how the $\dS$ part of the $\ucs$ algebra complicates the analysis, and calls for an interpretation of the $\ucs$ as the set of automorphisms of an Atiyah Lie algebroid over the corner. After a mathematical introduction to algebroids, we identified an affine associated bundle to the $\ucs$ algebroid, whose fibres can be thought of as giving rise to the normal-to-the-corner coordinates in an emergent spacetime. We closed Section \ref{sec4} with a brief exercise, in which we familiarized with the concept of moment map. The latter identifies field space configurations with points in the coalgebra. Then, the symplectic structure on the field space is related to the KKS structure on coadjoint orbits. This is how classical gravity is embedded in the abstract orbit analysis. 

We recollect here the various open questions and avenues of investigation that we touched upon. Fascinatingly, the general theory of symmetries for gravity is  still unravelling new ideas. The asymptotic symmetry theory is a paradigm that could be applied to many instances. Two main roads are important to pursue. First, we should relax as much as possible boundary conditions. This brought new challenges to the community, and we expect that there are new features to be understood in this direction. Secondly, in order to build toward a quantum gravity theory, it is fundamental to abandon the gauge-fixing approach, and construct a gauge-free analysis. In particular, there is more and more evidence that the symmetries used to gauge-fix could be charged, and thus setting the associated fields to zero is a physical constraint on the system, which should be avoided. 

Another urgent direction to pursue is the application and understanding of the extended phase space. How can embeddings lead to integrable charges for all dissipative systems? While this is true mathematically, we need to test this theory in specific examples, to gather more intuition on this fact. Another application of the extended phase space worth exploring is in the asymptotic charge analysis, where charges diverge, and renormalization is needed. How does the latter influence embeddings? This is still a relevant open question. 

Finally, the most important pending question is how far the corner proposal can lead us in addressing quantum gravity. We believe that the $\ucs$ is present in quantum gravity as the algebra organising observables. Thus, the first step in this path is the study of the $\ucs$ representations. Is it possible to find unitary and irreducible representations? Can we apply the theory of geometric quantization to the corner proposal? Can a discrete pattern emerge, suitable for describing non-commutative geometries? These are some profound questions that we can tackle using the developed tools. It is the author's personal opinion that we are in a fascinating moment in the developments of the corner proposal, in which corners are permeating various fields, as a unified underlying framework. We have gathered a lot of evidence suggesting that this framework has wide-ranging consequences, some of which yet to be unveiled. 

\subsection*{Acknowledgements}

I am grateful to the organisers of the XVIII Modave summer school for the invitation, and to all participants for the stimulating atmosphere and discussions during the school. I am indebted to Glenn Barnich and Laurent Freidel for precious discussions. My deepest thanks to Rob Leigh, for developing ideas presented here together.
I would like to acknowledge Nicolas Cresto, 
Arnaud Delfante, Gerben Oling, 
Sabrina Pasterski, 
 James Read, 
and Simone Speziale 
for discussions and feedbacks on the manuscript, and Fran\c{c}ois Gieres for pointing out important historical papers.  Thanks to the participants of the conference Quantum Gravity Around the Corner, held in Perimeter Institute, for interesting exchanges among communities. Research at Perimeter Institute is supported in part by the Government of Canada through the Department of Innovation, Science and Economic Development Canada and by the Province of Ontario through the Ministry of Colleges and Universities.

\bibliographystyle{uiuchept}
\bibliography{LucaModave.bib}

\end{document}